\RequirePackage{fix-cm}
\documentclass[twocolumn]{svjour3}          % twocolumn
\smartqed  % flush right qed marks, e.g. at end of proof
\usepackage{graphicx}
\usepackage{amsfonts}
\usepackage{array}
\usepackage{amsmath}
\usepackage{amsfonts}
\usepackage{amssymb}
\usepackage{amsxtra}
\usepackage{enumerate}
\usepackage{epsfig}
\usepackage{mathtools}
\usepackage{setspace}
\usepackage{multirow}
\usepackage{natbib}
\usepackage{hyphenat}
\usepackage{authblk}
\usepackage{microtype}

\usepackage{doi}
\usepackage{xcolor}
\hypersetup{
    colorlinks,
    linkcolor={red!50!black},
    citecolor={blue!50!black},
    urlcolor={blue!80!black}
}
\setcitestyle{aysep={}} 
\newcommand{\bvu}{\ensuremath\boldsymbol{u}}

\newcommand{\bvx}{\ensuremath\boldsymbol{x}}
\newcommand{\bvX}{\ensuremath\boldsymbol{X}}
\newcommand{\bvf}{\ensuremath\boldsymbol{f}}
\newcommand{\bvF}{\ensuremath\boldsymbol{F}}

 \journalname{Biomechanics and Modeling in Mechanobiology}

\usepackage{hyperref}

\usepackage{etoolbox}
\newcommand*{\affaddr}[1]{#1} % No op here. Customize it for different styles.
\newcommand*{\affmark}[1][*]{\textsuperscript{#1}}
\definecolor{myred}{rgb}{0, 0, 0}
\begin{document}

\title{Pump efficacy in a \textcolor{myred}{two-dimensional,} fluid-structure interaction model of a chain of contracting lymphangions%\thanks{Grants or other notes
}


\author{%
Hallie Elich {\protect\affmark[1]} \and Aaron Barrett \affmark[1] \and Varun Shankar\affmark[2] \and Aaron L. Fogelson\affmark[1,3]}


\institute{
              Corresponding author: Hallie Elich \\
              elich@math.utah.edu\\
              \\
%              Tel.: +123-45-678910\\
%              Fax: +123-45-678910\\
%              \email{elich@math.utah.edu}           %  \\
%              second address\\
              \affaddr{\affmark[1]Department of Mathematics}\\
\affaddr{\affmark[2]School of Computing}\\
\affaddr{\affmark[3]Department of Biomedical Engineering}\\
\\
\affaddr{University of Utah, Salt Lake City, UT, USA}%
}

\authorrunning{
Hallie Elich \and 
Aaron Barrett \and 
Varun Shankar \and 
Aaron L. Fogelson
}

\date{Received: date / Accepted: date}

\maketitle
%\onehalfspacing
%\doublespacing
\begin{abstract}
The transport of lymph through the lymphatic vasculature is the mechanism for returning excess interstitial fluid to the circulatory system, and it is essential for fluid homeostasis.  Collecting lymphatic vessels comprise a significant portion of the lymphatic vasculature and are divided by valves into contractile segments known as lymphangions.  Despite its importance, lymphatic transport in collecting vessels is not well understood.  We present a computational model to study lymph flow through chains of valved, contracting lymphangions. We used the Navier\hyp{}Stokes equations to model the fluid flow and the immersed boundary method to handle the two\hyp{}way, fluid\hyp{}structure interaction \textcolor{myred}{in 2D, non-axisymmetric simulations}. We used our model to evaluate the effects of chain length, contraction style, and adverse axial pressure difference (AAPD) on cycle-mean flow rates (CMFRs).  In the model, longer lymphangion chains generally yield larger CMFRs, and they fail to generate positive CMFRs at higher AAPDs than shorter chains.  Simultaneously contracting pumps generate the largest CMFRs at nearly every AAPD and for every chain length.  Due to the contraction timing and valve dynamics, non\hyp{}simultaneous pumps generate lower CMFRs than the simultaneous pumps; the discrepancy diminishes as the AAPD increases.  Valve dynamics vary with the contraction style and exhibit hysteretic opening and closing behaviors.  Our model provides insight into how contraction propagation affects flow rates and transport through a lymphangion chain.  

\end{abstract}

\section{Introduction}
\label{sec:1}
The lymphatic vasculature is a structural component of the lymphatic system, and its primary function is to return around \textcolor{myred}{8 L/day \citep{Renkin1986} of interstitial fluid to the venous circulation via \textcolor{myred}{lymph node microvessels} \citep{Adair_Guyton_1983,Knox_Pflug_1983} and the thoracic duct.}  Once interstitial fluid enters the blind\hyp{}ended, initial lymphatic vessels, it is referred to as lymph.  \textcolor{myred}{For a nice schematic of the lymphatic system, we refer the reader to Fig. 2 in a review by \citet{Moore2018}.}  Lymphatic transport is crucial for fluid homeostasis, but it is also important more broadly in immunology and cancer biology \citep{SHIELDS2007526, Swartz2012, WiigSwartz2012}.  Impaired lymph transport results in an accumulation of fluid, or lymphedema.  \textcolor{myred}{Lymphatic filariasis infection, lymph node removal in cancer\hyp{}related surgeries \citep{ROCKSON2001288}, and congenital lymphatic valve defects \citep{Bazigou2009,cancers12082297} are each associated with lymphedema.}  Lymphedema can be painful and disfiguring; mainstream treatments are limited to compression garments.  

Despite its importance, lymphatic transport is poorly understood.  This is somewhat surprising given that lymphatic studies, like those for the cardiovascular system, date back hundreds of years.  However, compared to blood vessels, lymphatic vessels were long\hyp{}regarded as passive drainage routes, they were difficult to find and dissect, and they lacked specific cell markers until the turn of the 21st century.  Given advances in science and technology, lymphatic vascular biology has undergone a resurgence in the past 20\hyp{}40 years.  

Numerous biological experiments have been performed, especially on isolated rat mesenteric collecting lymphatic vessels \citep{Davis2009a,Davis2009,Davis2011,Davis2012,Scallan2012,Scallan2013,Zawieja2009}.  Collecting lymphatic vessels are segmented by valves into functional units known as lymphangions which can actively contract in a cyclical manner to propel lymph against an adverse pressure gradient.  \textcolor{myred}{Lymphangions can be highly contractile with 10\hyp{}20 contractions per minute \citep{Zawieja2009} and undergo contractions with amplitudes around 40\hyp{}50\% of their diastolic diameter \citep{Davis2012,Zawieja2009}.}  \textcolor{myred}{Additionally, these contractions can propagate in the forward (orthograde) or backward (retrograde) directions along a chain of lymphangions \citep{Zawieja1993}.  Mechanisms of this directionality are not well understood \citep{Bertram2018a,Kunert2015}.}  Considering the fragility and size of most collecting lymphatics, impressively technical biology experiments have been performed to better understand their contractile properties \textcolor{myred}{(for example, see \citet{Crowe1997, Davis2012, Gashev2002, Scallan2012, Zawieja1993, Zhang2013}).  We refer the reader to beautiful videos (S1 and S2) of contractile lymphangions in isolated vessel experiments by \citet{Davis2012}.}  Complementary to biological studies, computational models have been developed with aims of better understanding lymph flow through these intricate vessels.

Lumped\hyp{}parameter models of collecting lymphatics first appeared in the literature in the 1970s \citep{Reddy1975,REDDY1977181}.  Since then, most of the additional quantitative models that have been developed are also lumped\hyp{}parameter and generally feature flow through one or more contractile lymphangions with valves modelled as resistors \citep{Bertram2011,Quick2007,Venugopal2007}.  Lumped\hyp{}parameter models from the Bertram group comprise most of the recent quantitative lymphatic literature.  Various generations of their 2011 model have been developed \citep{Bertram2014a,Bertram2014,Bertram2017,Jamalian2013} and used to simulate pumping in chains of lymphangions \citep{Bertram2016,Bertram2018a,Jamalian2016, Jamalian2017}.  The model \citep{Jamalian2013} was also used with an experimentally based constitutive relation for vessel wall mechanics \citep{Caulk2016,Razavi2017}.  Zero\hyp{}dimensional (lumped) models are tractable and especially useful in simulating flow through vessel networks \citep{Jamalian2016,REDDY1977181}.  

\textcolor{myred}{With regards to the fluid flow, lumped models can only provide information on the bulk flow.  They are also unable to explicitly include the lymphangion geometry, particularly the valves whose leaflets influence flow dynamics.  Moreover, the valves and variable \textit{in vivo} lymph flow patterns \citep{Gashev2002} additionally complicate the lumped\hyp{}model Poiseuille\hyp{}flow assumption.  Based on the velocity and diameter measurements reported for \textit{in vivo} rat mesenteric lymphatics \citep{Dixon2006} and using the density and viscosity of water, we obtain a Reynolds number of 0.08 (using a mean velocity of 0.87 mm/s \citep{Dixon2006}) and a Reynolds number of 0.82 (using a peak velocity of 9.0 mm/s \citep{Dixon2006}).  Also, Reynolds numbers of 16-160 (mean vs. peak) have been estimated for flow through the human thoracic duct (a final outflow tract of the collecting lymphatic vasculature) \citep{Moore2018}.  Thus, Reynolds numbers for flow through collecting lymphatic vessels vary depending on the species, anatomical location, and lymph velocities.  Higher-dimensional, fluid\hyp{}structure interaction (FSI) models are needed to resolve flow details, to visualize the velocity field, and to elucidate valve behaviors among adjacent, contractile lymphangions.}  

\textcolor{myred}{Higher\hyp{}dimensional models of collecting vessels have been developed, but they all make restrictive assumptions on the flow, the FSI, the vessel or valve geometry, or the number of lymphangions.  \citet{Macdonald2008} modified the \citet{Reddy1975} model to study 1D flow and contractile wave-propagation inside a single lymphangion whose valves were modelled as resistors.  A 2D model of flow through a lymphangion with contractions modulated by oscillations in calcium and nitric oxide (NO) concentrations was developed and simulated in chains of up to eight lymphangions \citep{Kunert2015}.  The valves were described as numerical check valves, and their model did not feature any sinus geometry.  In later work \citep{Li2019}, this group incorporated leaflet structures into their FSI model.  The model geometry featured an initial lymphatic vessel upstream of a single, contractile lymphangion with no valve-sinus regions.  Both vessels were embedded in a porous tissue.  The vessel failed to yield forward flow and featured retrograde flow through leaky valves when the adverse axial pressure difference was larger than 0.025 cmH$_2$O.}  

\textcolor{myred}{\citet{RAHBAR20111001} constructed a 3D model of flow through a non-valved portion of a single, contractile, cylindrical lymphangion whose contractions were prescribed.  The FSI was not modelled, but they examined the velocity profile and quantified shear stress.  \citet{Wilson2013} simulated incompressible, Navier\hyp{}Stokes flow through a 3D static fluid geometry featuring one valve that was based on confocal images of a rat mesenteric lymphatic vessel.  They studied NO advection, diffusion, and degradation in the fluid region.  Later work by this group \citep{WILSON20153584} examined the open-valve resistance in steady flow simulations through a 3D, single lymphangion with a rigid vessel wall and open, rigid valve leaflets.  They used an idealized geometry with parameters based on confocal images of rat mesenteric lymphatics.  They indicate that the sinus facilitates valve opening and reduces resistance.  The FSI was uncoupled; they present a fully coupled FSI model of a valve in a single lymphangion in later work \citep{Wilson2018}, although they still assume a rigid vessel wall and study only one valve.  Work by a related group \citep{Watson2017} featured a reconstructed geometry of a portion of a 3D lymphangion with a single valve and an elasto-plastic model.  There was no fluid flow or contractions; mechanical valve\hyp{}closure tests were performed for a variety of leaflet shear moduli with rigid or flexible vessel walls.  They indicate that valve behavior is poorly understood.  \citet{Bertram2020} developed a 3D finite-element, FSI model of a portion of a lymphangion featuring one valve with a flexible leaflet and vessel wall.  There were no contractions; valve opening and closing were driven by pressure differences.  A quarter of the geometry was simulated, and the leaflets did not completely close.  However, this was the first 3D, FSI study with both flexible leaflets and a flexible vessel wall.  \citet{Ballard2018} report a 3D, fully coupled, FSI model of portion of a rigid lymphangion with one valve.  Their model features no sinus geometry.  They performed FSI simulations and studied how forward-flow valve resistance and backflow valve conductance changed with valve aspect ratio and bending stiffness.  They also simulated dynamic valve behavior in the rigid vessel via time-varying imposed axial pressure gradients; they generated plots exhibiting hysteretic behavior of the valve gap distance.}  

In the current work, we present a novel, two\hyp{}way\hyp{}FSI, computational model that we used to study flow through various\hyp{}length chains of contracting lymphangions in two spatial dimensions.  Simultaneous, forward\hyp{}propagating (orthograde), and backward\hyp{}propagating (retrograde) contractions are considered.  For all chain lengths and at nearly all adverse axial pressure differences (AAPDs), the simultaneously contracting pumps yield the largest cycle\hyp{}mean flow rates (CMFRs).  Lengthening the chain generally leads to increased CMFRs and higher AAPDs at which the pump fails to yield positive CMFR.  In addition to performing pump\hyp{}function studies, we used simulation videos \textcolor{myred}{(see Online Resources 1\hyp{}19)} and velocity\hyp{}field, pressure, diameter, valve\hyp{}dynamics, flow\hyp{}rate, and tracer\hyp{}transport plots to assess pump behaviors.  The model provides insight into lymphatic transport and contributes to the quantitative lymphatic biology field.

The paper is organized as follows.  In Section 2, we present the model.  \textcolor{myred}{In Section 3, we provide and discuss results from benchmark simulations, pump\hyp{}function studies, and pump\hyp{}behavior analyses.  We also compare our results to the literature.  In Section 4, we conclude our findings.}  

\section{Model}
\label{sec:2}
We implement the immersed boundary (IB) method to simulate lymph flow through contracting lymphangions.  The IB method was developed by Charles Peskin in the 1970s \citep{Peskin1972,Peskin1977} and numerically implemented to study the fluid dynamics of blood flow around moving valves in a computational model of a contracting heart.  Since then, this two\hyp{}way, fully coupled method has been applied in a wide variety of FSI problems \textcolor{myred}{and has advantages over traditional computational approaches, particularly for problems with complex and highly deformable geometries where re\hyp{}meshing is not required.}  Some of the biofluid applications involve swimming organisms \citep{DILLON1995325,FAUCI1995679,FAUCI198885,Hamlet2018,HOOVER201513,Lim2012}, cochlear dynamics \citep{BEYER1992145}, biofilm processes \citep{DILLON199657}, insect flight \citep{Miller3073,Miller195}, esophageal transport \citep{KOU2017433}, platelet aggregation \citep{FOGELSON1984111,FOGELSON20082087}, arteriolar flow \citep{ARTHURS1998402}, heart development \citep{Battista2018}, and bioprosthetic heart valves \citep{Lee2020}. 
%Text with citations \cite{RefB} and \cite{RefJ}.

\begin{figure*}
    \includegraphics[width=1\textwidth]{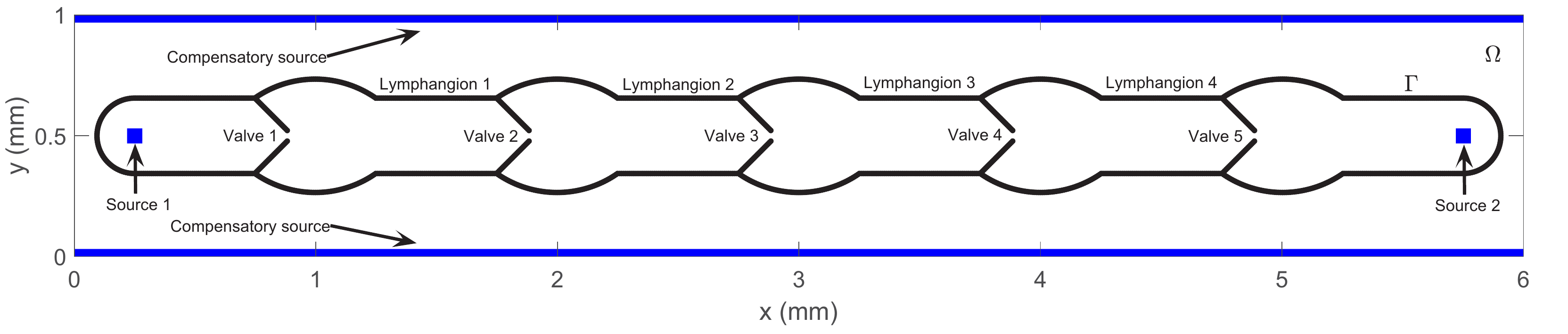}
%    \includegraphics[width=1\textwidth]{Fig1.eps}
\caption{Geometry of a 4-lymphangion chain in its initial configuration with complete lymphangions and valves numbered from left to right and fluid\hyp{}source locations shown in blue.  The entire rectangular region is labelled as $\Omega;$ the IB is labelled $\Gamma$}
\label{fig:2.1.1}       % Give a unique label
\end{figure*}

\subsection{Model geometry}
\label{sec:2.1}
\textcolor{myred}{Based on the diameter range for rat mesenteric lymphatics \citep{Dixon2006}, the lymphangion valve spacing every 3-10 diameters \citep{SchmidSchonbein1990}, the sinus-to-root ratio \citep{WILSON20153584}, the valve-sinus length \citep{WILSON20153584}, and the valve leaflet depth \citep{Lapinski2017},} we define an \textit{in silico}, \textcolor{myred}{idealized} model geometry representative of a portion of a collecting lymphatic vessel.  \textcolor{myred}{This is similar to the idealized geometry used by \citet{WILSON20153584}.  We refer the reader to images of lymphangions showing valve-sinus regions \citep{Bohlen2009,Margaris2016} (see their Figs. 5A and 3, respectively)  and to the image of four lymphangions in \citet{Bertram2016} (see their Fig. 9).}     

The model vessel comprises a chain of one to four contractile lymphangions separated by valves.  We define a rectangular region, $\Omega = [0,aL]\times [0,bL]$ in $\mathbb{R}^2,$ filled entirely with fluid and define a subset $\Gamma \subset \Omega$ that is a capsular\hyp{}shaped, closed curve with semicircular endcaps, arched protrusions, and interior processes; see Fig. \ref{fig:2.1.1} for an illustration of a 4\hyp{}lymphangion chain.  We call $\Gamma$ the immersed boundary (IB).  Away from the semicircular endcaps, the IB corresponds to the vessel wall and valves; the arched protrusions correspond to the valve\hyp{}sinus regions, and the line\hyp{}segment interior processes represent the valve leaflets.  \textcolor{myred}{We refer to the portion of $\Gamma$ without the valve leaflets as the capsule.}

The semicircular endcaps enclose regions where fluid can enter or leave $\Omega.$  The fluid source\footnote{We use the term ``source'' liberally throughout this work to obviate the need for repeating ``source or sink.''} supports are shown as blue squares in Fig. \ref{fig:2.1.1}.  \textcolor{myred}{By source support, we mean a subset of $\Omega$ in which fluid can enter or leave $\Omega.$}  For each of the two sources inside the vessel, an Ohm's\hyp{}law equation governs flow rates based on the difference in pressure between a virtual\footnote{By virtual, we describe something that is implicitly rather than explicitly included in the physical model domain.} reservoir (in which pressure is prescribed) and the source support (over which the computational pressure is averaged).  Each reservoir and source support are connected virtually with tubing and a cannulation pipette.  \textcolor{myred}{Our approach to setting upstream and downstream pressures is similar to what is done in cannulated, isolated vessel experiments.}  To allow for volume changes inside the lymphangion chain, a source is placed exterior to the capsule; see Fig. \ref{fig:2.1.1}.  Flow through it is set equal and opposite to the net inflow through the endcap sources.  The source setup is similar to that used by \citet{ARTHURS1998402}.  We simulate a vessel filled with lymph and immersed in an interstitial fluid/tissue mix comprising the interstitium.  We model the interstitium as a fluid\hyp{}filled, porous medium.  
   
\subsection{Model equations}
\label{sec:2.2}
As in \citet{Peskin1977}, we assume $\Omega$ is filled with a viscous, incompressible, Newtonian fluid of constant viscosity and density and that $\Gamma$ is neutrally buoyant.  Thus, lymph and interstitial fluid are assumed Newtonian, and the vessel wall and valves are assumed neutrally buoyant in lymph and in the interstitium.  
The IB formulation of the model consists of the system of equations:
\begin{equation}\label{eq:2.2.1}
\rho \left( \bvu_t + \nabla \cdot \left(\bvu\bvu^T\right) \right) = -\nabla p + \mu\Delta \bvu + \bvf - \kappa\sigma\bvu,
\end{equation}

\begin{equation}\label{eq:2.2.2}
\nabla \cdot \bvu = \sum_{i=1}^2 Q_i(t)\psi_i(\bvx),
\end{equation}

\begin{equation}\label{eq:2.2.3}
Q_i(t) = \dfrac{1}{R_i}\left\lbrace P_i^{\text{ext}}(t)-\int_{\Omega_{3D}}p(\bvx,t)\psi_i(\bvx)\,d\bvx \right\rbrace,
\end{equation}

\begin{equation}\label{eq:2.2.4}
\bvf(\bvx,t) = \int_{\Gamma} \bvF(s,t)\delta(\bvx-\bvX(s,t))\,ds,
\end{equation}
\begin{equation}\label{eq:2.2.5}
\dfrac{\partial \bvX}{\partial t}(s,t) = \bvu(\bvX(s,t),t) = \int_{\Omega_{3D}}\bvu(\bvx,t)\delta(\bvx-\bvX(s,t))\,d\bvx,
\end{equation}
\begin{equation}\label{eq:2.2.6}
\bvu(\bvx,0) \equiv \boldsymbol{0},
\end{equation} where $\bvu$ is the fluid velocity, $p$ is the pressure, $\bvf$ is the Eulerian force density, and $\sigma$ is an indicator function that is 1 outside the vessel and 0 inside the vessel; each is a function of $\bvx$ and $t.$  The functions $Q_i$, $\psi_i,$ and $P_i^{\text{ext}}$ are described below.  The IB configuration at time $t$ is $\bvX(s,t)$ with $0 \leq s \leq l$, and $\bvF$ is the Lagrangian force density.  The Dirac delta distribution appears in Eqs. \ref{eq:2.2.4}-\ref{eq:2.2.5}.  Also, $\rho$ and $\mu$ are the constant fluid density and dynamic viscosity, respectively; $\kappa$ is a Brinkman damping constant; and $R_i$ for $i = 1,2$ are resistances associated with cannulation pipettes and tubing.  

Because the model geometry lies in a 2D subspace of $\mathbb{R}^3$, we assume nothing varies in the $z$\hyp{}direction (which is orthogonal to the plane in which $\Omega$ lies) at least for some distance $L_z,$ and we assume vector $z$\hyp{}components are all 0.  \textcolor{myred}{Thus, the model is a 2D channel model; we use a characteristic depth, $L_z,$ taken to be the diameter, to convert area\hyp{}metric flow rates to volumetric flow rates.}  Define $\Omega_{3D} := [0,aL]\times[0,bL]\times[0,L_z]$.  We use periodic boundary conditions in the $x$\hyp{} and $y$\hyp{}directions; all dependent variables in Eqs. \ref{eq:2.2.1}\hyp{}\ref{eq:2.2.6} with $\bvx$\hyp{}dependence are periodic functions on $\Omega.$          

Eq. \ref{eq:2.2.1} is the Navier\hyp{}Stokes momentum equation with the advective term written in conservative form, with a body force\hyp{}density term (the Eulerian force density), and with a Brinkman (porous media) term.  The body force density is present due to the IB forces, and the Brinkman term acts to damp the velocity only outside the vessel due to the interstitium.  

The usual incompressibility constraint, $\nabla\cdot \bvu = 0$, is modified in Eq. \ref{eq:2.2.2} to account for the fluid sources.  The volumetric flow rates at the left and right fluid sources are denoted $Q_1$ and $Q_2,$ respectively.  The index $i$ is equal to 1 or 2, unless values are specified otherwise.  We define $\psi_i(\bvx) = \psi_i^r(\bvx)-\psi^c(\bvx),$ with $\psi_i^r, \psi^c \geq 0,$ and require      
\begin{equation}\label{eq:2.2.7}
\int_{\Omega_{3D}}\psi_i^r(\bvx)\,d\bvx = 1  \hspace{10pt} \text{ and} \hspace{10pt} \int_{\Omega_{3D}}\psi^c(\bvx)\,d\bvx = 1.
\end{equation}  We also require these functions to not vary in the $z$\hyp{}direction.  Here, $\psi_i^r$ corresponds to the (real) source labelled $i$ inside the vessel segment, and $\psi^c$ corresponds to the (compensatory) source outside the vessel segment.  In multiplying $Q_i$ and $\psi_i$ in Eq. \ref{eq:2.2.2}, we distribute the $Q_i$ volumetric flow rate over the support of $\psi_i$ in a weighted manner according to the $\psi_i$ function values.  Our treatment of sources and flow rates is similar to that in \citet{ARTHURS1998402} and in \citet{Peskin1977}.  

In Eq. \ref{eq:2.2.3}, we assume $R_i$ and $P_i^{\text{ext}}$ are given resistances and reservoir pressures, respectively.  We set $P_i^{\text{ext}}(t)=P_i^{\text{coeff}}\tanh(t);$ thus, we ramp up the pressure gradually at the start of each calculation.  The computational pressure is averaged over the source support, but in the integral, $\psi_i = \psi_i^r-\psi^c$ appears due to the need for a reference pressure.  We average the computational pressure over the compensatory source and use this as the reference pressure.  We consider this equal to the pressure with respect to  which $P_i^{\text{ext}}$ is measured.  We give more details on this in Sect. 1.2 of the Supplementary Information (SI); see Online Resource 0.

In Eqs. \ref{eq:2.2.4}-\ref{eq:2.2.5}, $\bvX(s,t)$ appears.  In the IB method, a Lagrangian representation of $\Gamma$ is used.  We parameterize each curve comprising $\Gamma$ with respect to arc length.  Namely, there is a closed, capsular curve for the vessel wall and an open\hyp{}curve line segment for each valve leaflet.   We parameterize the capsule using notation $\bvX_c(s_c,t)$ for $0\leq s_c \leq l_c$ and a valve leaflet using $\bvX_{v}(s_{v},t)$ for $0\leq s_{v} \leq l_{v},$ where $l_c$ is the arc length of the capsule in its initial configuration, and $l_{v}$ is the arc length of the valve leaflet initial configuration.  In practice, additional notation is used to distinguish top and bottom leaflets and multiple valves, but for ease of notation and unless stated otherwise, we write $\bvX(s,t)$ with $0\leq s \leq l$ to indicate the collection of curves comprising $\Gamma$ (each of whose $s$\hyp{}parameters vary within their individual ranges).  The Lagrangian variable $s\in [0,l]$ labels a material point, namely the point with (Eulerian) coordinates $\bvX(s,t)$ on $\Gamma$ at time $t.$  Throughout this work, we generally use capital letters to denote functions of the Lagrangian variable and lowercase letters to denote Eulerian variables and functions thereof.

Eqs. \ref{eq:2.2.4}-\ref{eq:2.2.5} are fluid\hyp{}structure interaction equations.  The kernel is a scaled, 2D Dirac delta distribution, i.e., $\delta(\bvx) = \dfrac{1}{L_z}\delta(x)\delta(y),$ where $\delta(x)$ and $\delta(y)$ are 1D Dirac delta distributions.  In Eq. \ref{eq:2.2.4}, the Lagrangian force density, $\bvF,$ is recast as an Eulerian force density, $\bvf,$ which acts on the fluid in Eq. \ref{eq:2.2.1}.  In Eq. \ref{eq:2.2.5}, the boundary moves with the fluid and satisfies a no\hyp{}slip condition.  The integral part of the equation comes from the defining property of the Dirac delta distribution.  \textcolor{myred}{In the IB method, the boundary moves at the local fluid velocity, and forces are generated in association with the deformed boundary.  These forces are then spread to the nearby fluid.}
   
 \begin{figure*}
        \includegraphics[width=1\textwidth]{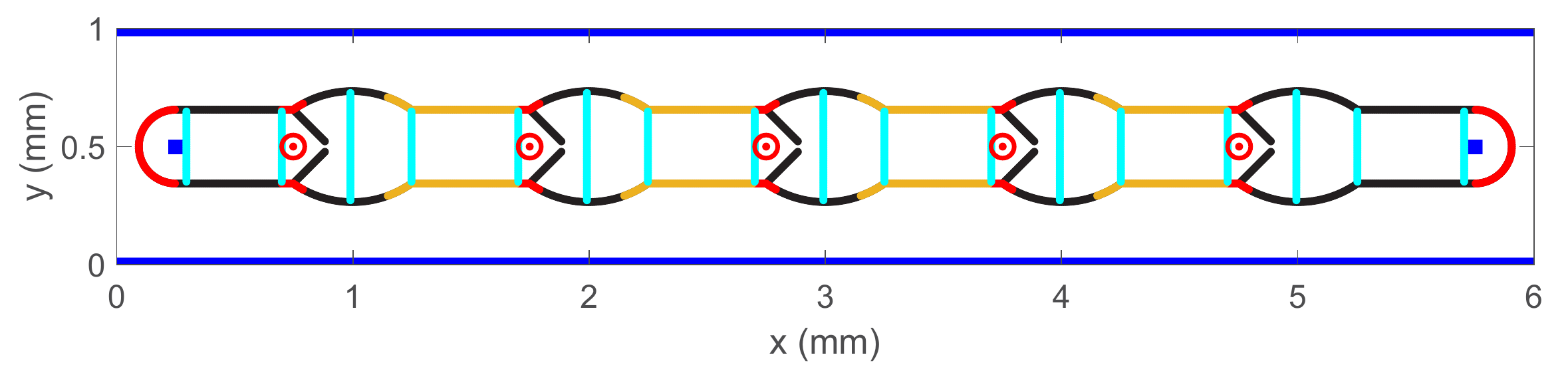}
%        \includegraphics[width=1\textwidth]{Fig2.eps}
\caption{Color\hyp{}coded force regions in a 4-lymphangion chain in its initial configuration.  The entire capsular structure and valves resist tension and bending.  Red regions are strongly tethered to their initial locations.  All other portions of the capsule including those along the \textcolor{myred}{dark yellow} contractile regions are weakly tethered to their initial locations.  Flow-meter locations are indicated in cyan.  Tracers are displayed as red dots with initial locations circled in red.  No color\hyp{}coding is shown for buttress forces at leaflet endpoints.  See Sect. \ref{sec:2.3} for a description of flow meters and tracers (which are numbered in increasing order from left to right).  Fluid-source locations are shown in blue as in Fig \ref{fig:2.1.1}}
\label{fig:2.2.1}       % Give a unique label
\end{figure*}
 
It remains to describe the boundary forces.  As is often done, the Lagrangian force density is defined as negative the first variational derivative of an energy functional.  The energy functional depends on the boundary configuration, $\bvX(s,t).$  The choice of the energy functional is motivated by the desired material properties of the IB; it is common to use more than one energy functional, each giving rise to a different type of Lagrangian force density.  In our model, the vessel wall and valve leaflets resist stretching, compression, and bending.  Thus, we consider tension and bending elastic energy functionals \citep{PeskinNotes}:
\begin{equation}\label{eq:2.2.8}
E_T[\boldsymbol{X}(\cdot,t)] = \dfrac{1}{2}\int_0^l k_t\left(\left\Vert \dfrac{\partial\boldsymbol{X}(s,t)}{\partial s}\right\Vert-l_0(s) \right)^2 \,ds,
\end{equation}
and 
\begin{equation}\label{eq:2.2.9}
E_B[\boldsymbol{X}(\cdot,t)] = \dfrac{1}{2}\int_0^l k_b\left\Vert \dfrac{\partial^2\boldsymbol{X}(s,t)}{\partial s^2} - \dfrac{\partial^2\boldsymbol{X}^0(s)}{\partial s^2}\right\Vert^2 \,ds,
\end{equation}
 where $\boldsymbol{X}^0(s)$ is a reference configuration for the curve $\bvX(s,t)$ with $0 \leq s \leq l$; $l_0(s) = \left\Vert \dfrac{\partial \boldsymbol{X}^0}{\partial s}(s)\right\Vert;$ and $k_t$ and $k_b$ are tension and bending constants, respectively.  We use the initial capsule configuration as the reference configuration.  The same functionals are used for valve leaflets but with $k_t$ and $k_b$ replaced with $k_{t_v}$ and $k_{b_v}$, respectively.  Each leaflet's initial configuration is used as its reference configuration.  In the numerical implementation of the IB method, the energy functionals are discretized, and then finite differences are taken to approximate the Lagrangian force densities.  The details and equations for the tension and bending Lagrangian force densities are provided in Sect. 2 of the SI.     
  
In addition to tension and bending\hyp{}resistant forces, we use spring forces to tether portions of the vessel wall or valve leaflets to fixed physical locations.  Tether forces act along the semicircular endcaps and aim to keep them relatively fixed in space; \textcolor{myred}{this is similar to vessels attached to stationary cannulating pipettes in experiments.}  Additionally, since regions of the vessel wall near the valve\hyp{}insertion sites are fortified with collagen in rat \citep{Rahbar2012} and in cat \citep{Vajda1971} mesenteric lymphatics, we tether these ``valve\hyp{}stiffness'' regions to their initial locations so they are less mobile; see Fig. \ref{fig:2.2.1} which shows regions along the IB where different forces are modelled.  We also weakly tether the remaining parts of the vessel wall to their initial locations to represent forces from circumferentially oriented components of the vessel that act to maintain vessel shape.  

For strong tether forces used along the semicircular endcaps and valve\hyp{}stiffness regions, we use a Hookean spring model 
\begin{equation}\label{eq:2.2.10}
\bvF_{\text{teth},k} = K_{\text{teth}}\left(\bvX_k^{\text{teth}}-\bvX_k\right)
\end{equation}
for the Lagrangian force density at the IB point $\bvX_k$ arising due to its connection to the tether point $\bvX_k^{\text{teth}}.$  We refer to the collection of points comprising the discretized boundary as IB points.  See Sect. 1 of the SI for discretization information and notation.  We use a similar equation for capsule IB points that are weakly tethered to their initial locations but with a smaller spring constant, $K_{\text{teth,weak}};$ see Table \ref{tab:2.2.1}.  

To implicitly model the downstream valve\hyp{}insertion points that are located outside our 2D model plane, we include conditional tether spring forces (buttress forces) that resist leaflet endpoints from opening too wide or moving upstream of their initial locations.  For example, the vertical force at an upper valve leaflet endpoint with $y$\hyp{}coordinate $Y_k$ is given by 
\begin{equation}\label{eq:2.2.11}
G_{\text{bo}} =   \left\{
\begin{array}{lll}
      K_{\text{bo}}\left(Y_{\text{bo}}-Y_k\right) & \text{\hspace{0.25 in}}& Y_k > Y_{\text{bo}}  \\
      0& \text{\hspace{0.25in}}& Y_k \leq Y_{\text{bo}}\\
\end{array} 
\right., 
\end{equation}
where 
$Y_{\text{bo}}$ is the buttress\hyp{}opening target point.  We set $Y_{\text{bo}}$ to be one\hyp{}half the initial\hyp{}configuration tube radius above the vessel midline.  With slight abuse of notation, we provide the force in Eq. \ref{eq:2.2.11} rather than a force density, as it is a force acting only at the leaflet endpoint.  A similar conditional force acts at the endpoint of each bottom leaflet.  The horizontal force at a valve leaflet endpoint with $x$\hyp{}coordinate $X_k$ is
\begin{equation}\label{eq:2.2.12}
F_{\text{bu}} =   \left\{
\begin{array}{lll}
      K_{\text{bu}}\left(X_{\text{bu}}-X_k\right) & \text{\hspace{0.25 in}} & X_k < X_{\text{bu}}  \\
      0& \text{\hspace{0.25 in}} & X_k \geq X_{\text{bu}}\\
\end{array} 
\right., 
\end{equation} 
where the initial\hyp{}configuration leaflet endpoint is $X_{\text{bu}}.$  A similar conditional force acts at the endpoint of each bottom leaflet.  
%	in the Watson paper, was leaflet tip movement backwards needed for a tighter valve seal? 

\begin{figure}
\includegraphics[width=0.5\textwidth]{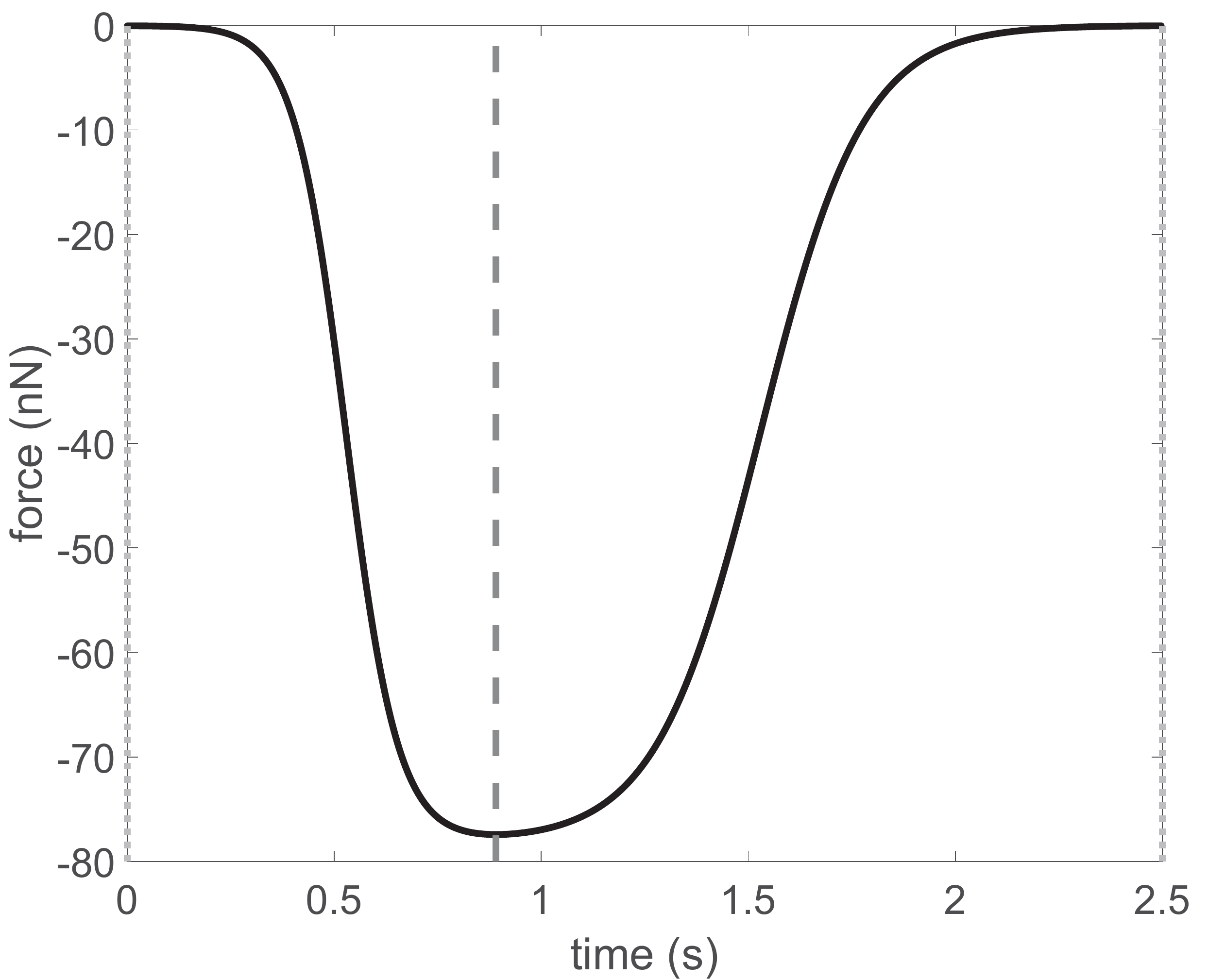}
%   \includegraphics[width=0.5\textwidth]{Fig3.eps}
\caption{One period of the contraction force acting along the upper half of the vessel wall in the contractile regions of a simultaneously contracting lymphangion chain.  Dark gray dashed line segment indicates the time (approximately 0.89 s) at which the force magnitude is maximal; light gray dotted line segments indicate times (0 s and 2.5 s) at which the force magnitude is minimal}
\label{fig:2.2.2}       % Give a unique label
\end{figure}

In the model, flow occurs via pressure\hyp{}driven sources and lymphangion contractions. We drive contraction by prescribing diameter-reducing forces along regions of each lymphangion where the muscle cells are predominantly located \citep{Bridenbaugh2013,Zawieja2018} and circumferentially oriented \citep{Bridenbaugh2013}.  Within a model lymphangion, the contractile region extends along the downstream portion of the sinus region and terminates adjacent to the downstream valve\hyp{}stiffness region; see Fig. \ref{fig:2.2.1}.  The contraction force pointing downward at each upper contractile\hyp{}region IB point in a simultaneously contracting lymphangion chain is given by a periodic replication (in time) of the function

\begin{align}
G_c(t)\Delta s &= -C_{\text{amp}}\left\lbrace\tanh(-(t + 1.25 - \tau)/0.25)\right.\\ \nonumber
& + \tanh((t + 2.25 - \tau)/0.125)\\ \nonumber
& + \tanh(-(t - 1.25 - \tau)/0.25)\\ \nonumber
& + \tanh((t - 0.25 - \tau)/0.125)\\ \nonumber
&  + \tanh(-(t - 3.75 - \tau)/0.25)\\ \nonumber
& + \left.\tanh((t - 2.75 - \tau)/0.125)\right\rbrace \nonumber
\end{align} \label{eq:2.2.13}on the interval $0 \leq t \leq 2.5$ s; see Fig. \ref{fig:2.2.2}.  The negative of this force acts at each lower contractile\hyp{}region IB point.  The force density is $G_c(t).$  The constant, $C_{\text{amp}},$ is set to yield contractions with diameter changes similar to those reported experimentally, and the constant, $\tau,$ is a time shift so that the contraction force is minimal at $t = 0$ s and $t = 2.5$ s.  Also, $\Delta s$ is the uniform increment in Lagrangian parameter space; see the discretization in Sect. 1.1 of the SI.

In each cycle, the contraction force magnitude increases throughout systole until it peaks and subsequently decreases throughout diastole.  We refer to maximal lymphangion contraction and relaxation as end\hyp{}systole and end\hyp{}diastole, respectively.  They occur at approximately the same times as maximal and minimal contraction force magnitudes, respectively.  

The period of our contraction force function is consistent with the literature \citep{Bertram2016,Zawieja2009,Zweifach1975}.  The peak contraction force occurring approximately 0.89 s after contraction onset is similar to that in \citet{Bertram2016} and to the experimental reports by \citet{Zweifach1975} on rat mesenteric lymphatics.  The steeper descent than ascent of the curve in Fig. \ref{fig:2.2.2} yields brisk contractions and slower relaxations which are physiological \citep{Macdonald2008,Zweifach1975}.

We use time\hyp{}delay shifts of $G_c(t)\Delta s$ (on the interval $0 \leq t \leq 2.5$ s) and replicate them periodically in time to construct \textcolor{myred}{different style} contraction forces (for the non\hyp{}simultaneous pumps).  \textcolor{myred}{In the non\hyp{}simultaneous pumps, all points in the contractile region of a single lymphangion are uniformly driven to contract, but adjacent lymphangions contract at different time delays.  These time delays are approximately 0.26 s or 0.52 s and correspond to travelling waves with velocities of 2 mm/s or 4 mm/s, respectively, that propagate along the vessel wall and signal uniform contraction within each lymphangion when they pass through the valve regions where pacemakers are likely located (see e.g., \citet{Hald2018}).  The contraction force within each lymphangion is given by a periodic replication of $G_c(t-t_{\text{delay}})\Delta s$ with $t_{\text{delay}}$ values differing among neighboring lymphangions by 0.26 s or 0.52 s and whose overall values are provided in Table \ref{tab:2.2.1}.  We provide figures in Sect. 4 of the SI to illustrate the contraction forces at various times throughout a cycle in the simultaneous, orthograde, and retrograde pumps.}

Contractions in rat mesenteric lymphatics \textit{in situ} propagate at 4-8 mm/s \citep{Zawieja1993}; in isolated rat mesenteric lymphatics, they propagate at around 6-10 mm/s \citep{Akl2011,Scallan2013}.  \textcolor{myred}{Based on the former measurements and unpublished observations where we found little difference in pumps whose contractions propagated at 8 mm/s or simultaneously (infinitely fast),} we set the travelling\hyp{}wave velocity to 2 mm/s or 4 mm/s.  We refer to the former as slow or just orthograde or retrograde, depending on the direction, and the latter as fast\hyp{}orthograde or fast\hyp{}retrograde, equivalently orthograde x2 or retrograde x2, depending on the direction.  \textcolor{myred}{Also, contractions are described as simultaneous in bovine mesenteric lymphatics \citep{McHale1976} and nearly synchronous in mouse popliteal lymphatics \citep{Scallan_2016}, so the simultaneous contractions are relevant.}            
    
% Note-to-self: eye towards refinment in your description, remember your discrete energy functional constant values and written work?  Fogelson continuum of tethers to avoid springs?).  tethers, buttress, vertical links, initial tethers, imposed forces, time-varying links, moving targets, etc.  Also recall the continuous approach and confusion, RBF, FD, etc. 
% Note to self: Why just says elastic material vs. viscoelastic, confusion.  Does viscoelastic property not apply to surfaces?
% Does the initial condition need to be divergence-free or match modified incompressibility?
% Do we have to write periodic BCs for pressure, forces?

\subsection{Parameters, flow meters, and tracers}
\label{sec:2.3}
All parameters are listed in Table \ref{tab:2.2.1} or described in the paragraphs that follow.  Whenever possible, they are based on those in the rat mesenteric lymphatic literature.  The radius measured at the valve sinus is 1.5 times the tubular radius; this ratio is consistent with those reported by \citet{WILSON20153584} for rat mesenteric lymphatics, although their vessel diameters were 1/3 as large as those in the present study.  \textcolor{myred}{The leaflet length is consistent with the leaflet depth reported by \citet{Lapinski2017}.}  The sinus length (measured horizontally) is 0.5 mm which is similar to that reported in \citet{WILSON20153584}.  We estimated and varied other parameters (e.g., $\kappa$ and tension and bending constants) empirically to yield simulations with benchmark output data in the physiological range.  The valve\hyp{}stiffness regions (see Fig. \ref{fig:2.2.1}) extend a distance of approximately 0.049 mm upstream and 0.041 mm downstream of each valve\hyp{}insertion point (with Euclidean distances reported and measured only in the $x$\hyp{}direction).  Each contractile region (shown in dark yellow in Fig. \ref{fig:2.2.1}) between two consecutive valves begins 0.1 mm to the left of the sinus\hyp{}wall junction and ends at the IB point immediately left of the valve\hyp{}stiffness region.  
The left\hyp{}endcap tether region extends from the left up to an $x$\hyp{}coordinate of 0.2468 mm; the right\hyp{}endcap tether region has a farthest-left $x$\hyp{}coordinate of 2.753; 3.753; 4.753; or 5.753 mm depending on the number of lymphangions; see red endcaps in Fig. \ref{fig:2.2.1}.

Flow meters are shown in cyan in Fig. \ref{fig:2.2.1} and are vertical line segments where we measure flow rates, velocities, and pressures.  One is located 0.048 mm inward from each red\hyp{}endcap terminus, and there are three flow meters placed around each sinus region: one at the upstream end of the valve\hyp{}stiffness region, one at the sinus apex, and one at the downstream end of the sinus.  The flow meters are numbered from left to right.  When we refer to interior flow meters, we mean the set of flow meters excluding the first and last.  See Sect. 3 of the SI for details on how flow rates, velocities, and pressures are measured at flow meters.   

We seed the flow with tracers (circled red dots in Fig. \ref{fig:2.2.1}) centered at each valve\hyp{}insertion $x$\hyp{}coordinate.  These tracers move at the local fluid velocity but do not affect the flow in any way.  They are used to gauge velocities and evaluate transport.  In simulation videos (see Online Resources 1-6 and 8-19), ten evenly spaced tracers are shown within each lymphangion and are used to visualize passive particle transport through a pumping lymphangion chain. 

\begin{table}
\caption{Model parameters with semicolons used to order those that vary with increasing lymphangion\hyp{}chain length}
\label{tab:2.2.1}       % Give a unique label
\begin{tabular}{ll}
\hline\noalign{\smallskip}
Fluid parameters\\
\hline\noalign{\smallskip}
$\rho$ & $10^3 \text{ kg/m}^3$\\
$\mu$ & $10^{-3} \text{ kg/(ms)}$\\
$\kappa$ & $2.0\times 10^7 \text{ kg/sm}^3$ \\ 
$U_c$ & $10^{-3} \text{ m/s}$\\
Re & 1\\
\hline\noalign{\smallskip}
Geometry parameters (mm)\\
\hline\noalign{\smallskip}
$a$ & 3; 4; 5; 6\\
$b$ & 1\\
$L$ & 1\\
$L_z$ & 0.3125\\
tube radius, r & 0.15625\\
tube length & 2.5; 3.5; 4.5; 5.5\\
sinus radius & $\approx 0.2344$\\
sinus length & $\approx 0.5$\\
left endcap center & $(0.25,0.5)$\\
right endcap center & $(2.75; 3.75; 4.75; 5.75,0.5)$\\
top valve\hyp{}insertion pts. & $(0.75; 1.75; 2.75; 3.75; 4.75,0.6562)$\\
bottom valve\hyp{}insertion pts. & $(0.75; 1.75; 2.75; 3.75; 4.75,0.3438)$\\
leaflet initial arc length & $\sqrt{2}r-2h\approx 0.1897$\\ 
leaflet initial angle & $-\pi/4$ radians\\
$X_{\text{bu}}$ & $\approx 0.88; 1.88; 2.88; 3.89; 4.89$\\
$Y_{\text{bo}}$ & 0.578125\\
\hline\noalign{\smallskip}
Misc. parameters\\
\hline\noalign{\smallskip}
$P_1^{\text{coeff}}$ & $0.076$ cmH$_2$O \\
$P_2^{\text{coeff}}$ & $0.076$ cmH$_2$O and higher \\
$R_i$ & $1.6\times 10^{9}\text{ kg/sm}^4$ \\
\hline\noalign{\smallskip}
Discretization parameters\\
\hline\noalign{\smallskip}
$h$ & $1/64$ mm\\
$n_x$ & 192; 256; 320; 384\\
$n_y$ & 64\\
$k = \Delta t$ & $1/32768$ s\\
$T$ & 25 s (10 cycles)\\
$\Delta s \approx \Delta s_v$ & 0.008 mm\\
$N$ & 756; 1012; 1268; 1524 \\
$N_v$ & 25 \\
CFL constant $C$ & 0.1 \\
\hline\noalign{\smallskip}
Force parameters\\
\hline\noalign{\smallskip}
$k_t = k_{t_v}$ & $1.5625\times 10^{-6} \text{ N}$\\
$k_b$ & $3.125\times 10^{-17} \text{ Nm}^2$\\
$k_{b_v}$ & $6.25\times 10^{-16} \text{ Nm}^2$\\
$K_{\text{teth}}$ & $\approx 387\text{ N/m}^2$\\
$K_{\text{teth,weak}}$ & $\approx 77\text{ N/m}^2$\\
$K_{\text{bo}}$ & $3.125\times 10^{-4} \text{ N/m}$\\
$K_{\text{bu}}$ & $3.125\times 10^{-4} \text{ N/m}$\\
$C_{\text{amp}}$ & $39.0625$ nN\\
$\tau$ & $\approx 0.2784$ s \\
$t_{\text{delay}}$ (for 2 mm/s) & $\approx 0.28; 0.80; 1.31; 1.83$ s\\
$t_{\text{delay}}$  (for 4 mm/s) & $\approx 0.14; 0.40; 0.66; 0.91$ s\\
\noalign{\smallskip}\hline
\end{tabular}
\end{table}

\subsection{Non-dimensionalization, discretization, and numerical implementation}
\label{sec:2.4}  
We non\hyp{}dimensionalize the system by scaling length in the $x$\hyp{} and $y$\hyp{}directions by $L$, length in the $z$\hyp{}direction by $L_z$, time by $L/U_c$, velocity by $U_c$, pressure by $\mu U_c/L$, the Eulerian force density by $\mu U_c/L^2$, the volumetric flow rate by $U_cLL_z$, the resistance by $\mu/(L^2L_z),$ $\psi_i$ and $\delta$ by $1/(L^2L_z)$, the Lagrangian parameter $s$ by $L$, the Lagrangian force density by $L_z\mu U_c/L$, and $\kappa$ by $\mu/L^2.$  The specific parameter values are displayed in Table \ref{tab:2.2.1}.  We set the characteristic length scale $L = 10^{-3}$ m as the distance between two consecutive valve\hyp{}insertion points \citep{Zawieja1993} and set $L_z = 3.125\times10^{-4}$ m as the initial tube diameter which is within physiological range \citep{Dixon2006}.  Using approximate values for the density and viscosity of water, we have $\rho = 10^3 \text{ kg/m}^3$ and $\mu = 10^{-3} \text{ kg/(ms)}.$  We take $U_c = 10^{-3} \text{ m/s}$ based on experimental measurements \citep{Dixon2006}.  These parameter values yield a Reynolds number Re = $\dfrac{\rho U_cL}{\mu} = 1.$  

For the dimensionless system we obtain equations that appear the same as Eqs. \ref{eq:2.2.1}-\ref{eq:2.2.6} except with the momentum equation replaced with:

\begin{equation}\label{eq:2.2.14}
\text{Re}\left(\bvu_t + \nabla \cdot \left(\bvu\bvu^T\right)\right) = -\nabla p + \Delta\bvu + \bvf - \kappa\sigma\bvu,
\end{equation}
$\Omega_{3D} = \left[0, a\right]\times \left[0, b\right]\times \left[0,1\right],$ and the IB configuration written as $\bvX(s,t)$ with $0 \leq s\leq \dfrac{l}{L}.$  Since we assume all vector third components are identically zero and nothing varies in the $z$\hyp{}direction, we reinterpret these dimensionless equations in 2D.  Additionally, because $p$ and $\psi_i$ are constant in $z,$ and the $z$\hyp{}integration limits are from 0 to 1, the integral in the dimensionless version of Eq. \ref{eq:2.2.3} is equivalent to a 2D integral over $\Omega$ with the integrand unchanged.  Similarly, the integral in the dimensionless version of Eq. \ref{eq:2.2.5} is equivalent to a 2D integral over $\Omega.$  In order to simulate the 2D dimensionless system, we discretize the model in space and time. 

We discretize $\Omega$ using a uniform grid of meshwidth $h$ and let $n_x = a/h $ and $n_y = b/h$ denote the number of grid cells in the $x$\hyp{} and $y$\hyp{}directions, respectively.  We use a marker\hyp{}and\hyp{}cell, or MAC, grid \citep{HarlowWelch1965} upon which scalar functions are defined only at cell centers, and vector\hyp{}valued functions have $x$\hyp{} and $y$\hyp{}components defined only at cell left\hyp{}edge centers and bottom\hyp{}edge centers, respectively.  

We specify a number of evenly spaced Lagrangian nodes in parameter space so they correspond to IB points on the initial $\Gamma$ configuration that are evenly spaced and approximately a Euclidean distance of $h/2$ apart (as commonly used in the IB method).    

A discrete version of the Dirac delta distribution, $\delta_h,$ is necessary for numerical implementation of the IB method.  As in \citet{Peskin1977, Peskin2002}, we set
$\delta_h(\bvx) = \delta_h^{1D}(x)\delta_h^{1D}(y),$ where 
\[\delta_h^{1D}(r) =\begin{cases} 
      \left(\dfrac{1}{4h}\right)\left(1+\cos\left(\dfrac{\pi r}{2h}\right)\right) & |r|< 2h \\
      0 & |r|\geq 2h. 
   \end{cases}
\]  
This function is defined on all of $\Omega$ (on all of $\mathbb{R}^2$ due to periodicity), and its support is an interval of length $4h.$  We define the dimensionless functions $\psi_i$ based on $\delta_h$, taking $\psi_i^r$ equal to a shifted $\delta_h$ with center at the $i$th source center and $\psi^c$ equal to a scaled (multiplied by $1/a$) horizontal tiling of $\delta_h^{1D}$ functions, each oriented in the vertical direction with center along $y = 0.$ Due to the periodic boundary conditions, $2h$ of the support lies just above $y = 0$ and $2h$ of the support lies just below $y = b;$ see Fig. \ref{fig:2.1.1}.   
 
We use centered differencing and averaging, the latter when quantities are not defined at requisite grid locations, to spatially discretize the 2D dimensionless system and obtain semi\hyp{}discrete approximations to the system equations.  We provide details in Sect. 1.1 of the SI.    
 
We numerically integrate the semi\hyp{}discrete system using a second\hyp{}order\hyp{}accurate Runge\hyp{}Kutta method based on the midpoint rule as in \citet{Peskin2002}.  In both the preliminary and main substeps of the time\hyp{}stepping scheme, we compute the discrete divergence of the fully discretized momentum equation to obtain a pressure\hyp{}Poisson problem.  Similar to what is done by \citet{ARTHURS1998402} and \citet{Peskin1977}, we decompose the pressure as a linear combination of a source\hyp{}free pressure and two pressures related to the interior fluid sources.  We generate three auxiliary Poisson problems and use MATLAB's built-in direct solver to solve them.  The flow rates appear in the last two terms in the pressure decomposition and are determined by solving a pair of linear equations.  Once the pressure is computed, the velocity is then updated by a direct-solve of a Helmholtz equation.  Details of the temporal discretization and numerical-implementation scheme are provided in Sects. 1.2\hyp{}1.3 of the SI.    

The model was coded in MATLAB and run on clusters at the Center for High\hyp{}Performance Computing (CHPC) at the University of Utah.  Spatial and temporal refinement studies were performed to assess convergence.  Results \textcolor{myred}{(see Sect. 1.4 of the SI) indicate approximate first\hyp{}order convergence for velocity and the IB structure in space and time (in the 2\hyp{}norm for grid functions \citep{leveque2007finite}).}
    
\section{Results \textcolor{myred}{and discussion}}
\label{sec:3}
We investigated lymphangion chains of various lengths with different contraction styles that pumped against a range of adverse axial pressure differences (AAPDs).  Our aims were to assess and compare the pumps' abilities to generate positive CMFR and transport lymph (pump efficacy).  We used velocity plots, pressure plots, and simulation videos (see Online Resources 1-6 and 8-19) to glean insight into pump behaviors.  We analyzed diameter, valve\hyp{}state, and pressure data in timecourse plots similar to those appearing in the literature.  Additionally, we examined trans\hyp{}valve pressure differences and flow\hyp{}rate data.  Pump\hyp{}function plots were also constructed and analyzed.  \textcolor{myred}{The AAPDs are driven by connections with the virtual reservoirs in which pressures are controlled.}  All pumps had $P_1^{\text{coeff}}= 0.076$ cmH$_2$O and $P_2^{\text{coeff}}= 0.076$ cmH$_2$O or higher as described in the sections that follow.

\subsection{Velocity and pressure in different valve states}
\label{sec:3.1}
As representative model output, we provide velocity and pressure plots for a 4\hyp{}lymphangion chain with fast\hyp{}orthograde contractions operating at an AAPD of 0.051 cmH$_2$O.  All valves in this pump opened and closed during each 2.5\hyp{}s contraction cycle; \textcolor{myred}{the opening and closing dynamics were not prescribed in any of our simulations but were a product of the flow.}  The pump was simulated for 25 s or equivalently, 10 contraction cycles.  The difference in CMFR averaged among interior flow meters from cycle 9 to cycle 10 was less than 1\%; thus, cycle\hyp{}10 ($T = 22.5$\hyp{}$25$ s) results were considered to be quasi\hyp{}steady state. 

\begin{figure}
%  \includegraphics[width=0.5\textwidth]{vel_comp_pcolor_times_2_zoom.pdf}
     \includegraphics[width=0.5\textwidth]{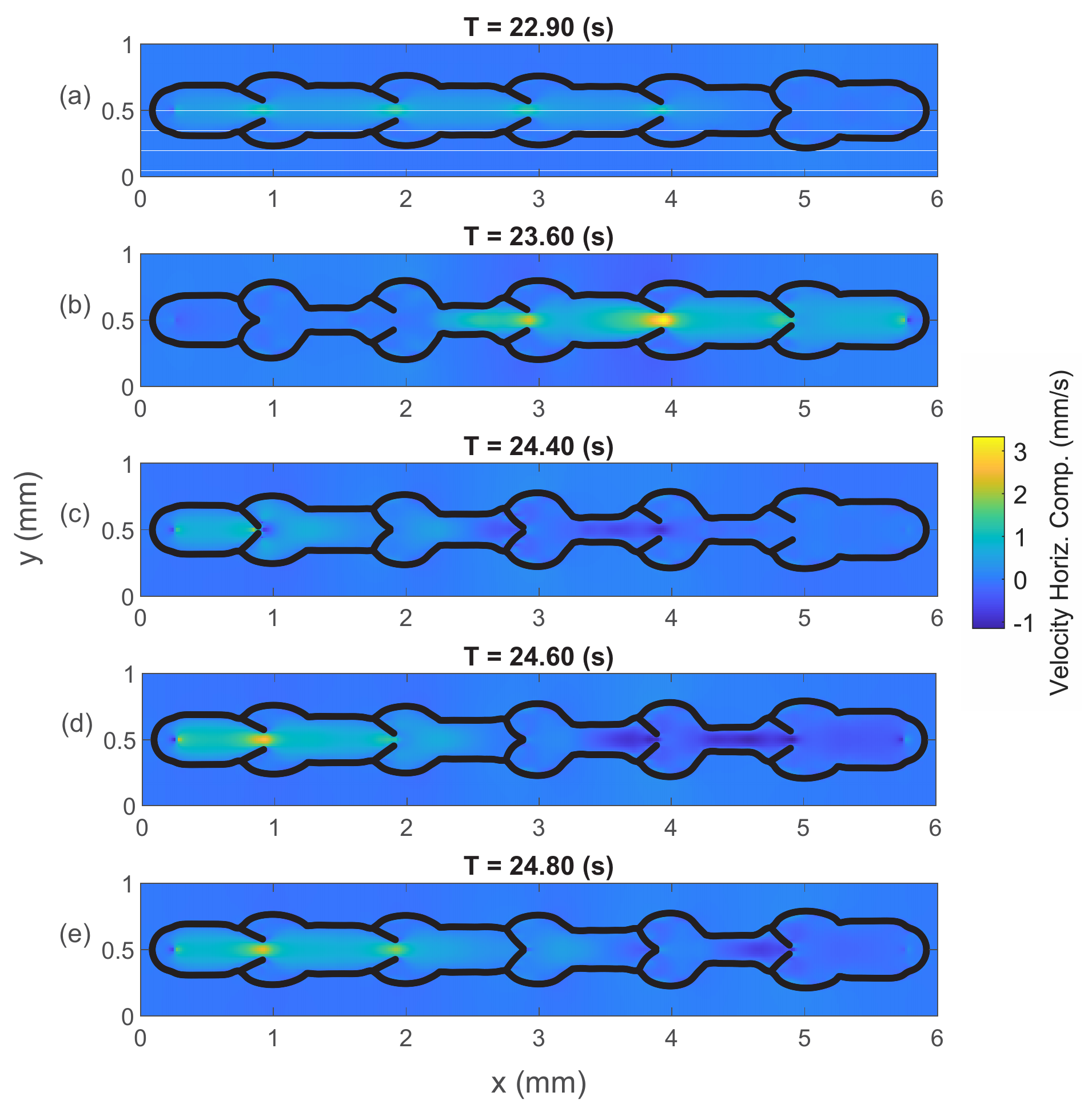}   
%    \includegraphics[width=0.5\textwidth]{Fig4.eps}
\caption{Horizontal velocity component at select times during cycle 10 in a 4\hyp{}lymphangion, fast\hyp{}orthograde pump with an adverse axial pressure difference (AAPD) of 0.051 cmH$_2$O}
\label{fig:3.1.1}       % Give a unique label
\end{figure}

\begin{figure}
%  \includegraphics[width=0.525\textwidth]{press_time_course_2_zoom.pdf}
%    \includegraphics[width=0.525\textwidth]{Figures/Fig6}
        \includegraphics[width=0.525\textwidth]{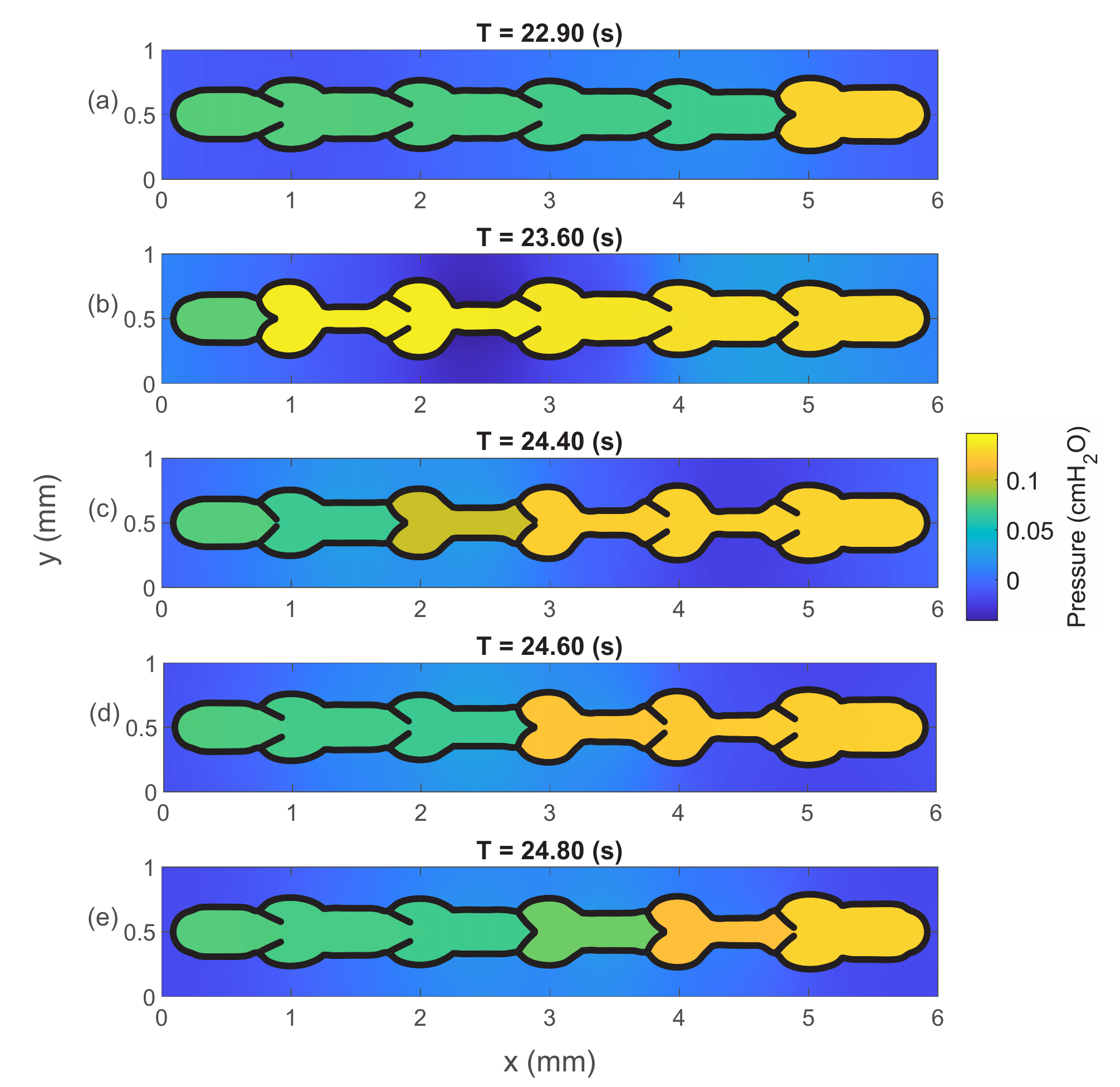}
%        \includegraphics[width=0.525\textwidth]{Fig5.eps}
\caption{Pressure at select times during cycle 10 in a 4\hyp{}lymphangion, fast\hyp{}orthograde pump with an AAPD of 0.051 cmH$_2$O}
\label{fig:3.1.2}       % Give a unique label
\end{figure} 

Velocity and pressure plots at times exhibiting valve\hyp{}state changes are shown in Figs. \ref{fig:3.1.1}\hyp{}\ref{fig:3.1.2}.  At 22.9 s, valves 1\hyp{}4 are open with positive, horizontal fluid velocity between the leaflets.  Valve 5 is shut and has a negative trans\hyp{}valve pressure (see panels (a) of both figures).  At 23.6 s, the first two lymphangions are contracting, \textcolor{myred}{and lymphangion 3 is just beginning to contract.}  Valves 2\hyp{}5  are open with large flow velocities through valves 3 and 4; valve 1 is shut and bears a negative trans\hyp{}valve pressure (see panels (b)).  At 24.4 s, valve 1 is opening, valve 2 is shut, valve 3 is closing, and valves 4\hyp{}5 are open (see panels (c)).  At this time, contraction is prominent for lymphangions 3 and 4, though there is backflow through valve 4; \textcolor{myred}{lymphangions 1 and 2 are relaxing, and lymphangion 3 is beginning to relax.}  After an additional 0.2 s, valve 3 is shut and all other valves are open with prominent positive velocity through valve 1 and backflow through valves 4 and 5 (see panels (d)).  \textcolor{myred}{Lymphangions 1-3 are relaxing, and lymphangion 4 is beginning to relax.}  Lastly at 24.8 s, valve 4 is closed, valve 3 is opening, and valves 1, 2, and 5 are open (see panels (e)).  \textcolor{myred}{At this time, all lymphangions are relaxing.}  

\subsection{Vortices in the valve-sinus region}
\label{sec:3.2}

\begin{figure}
      \includegraphics[width=0.45\textwidth]{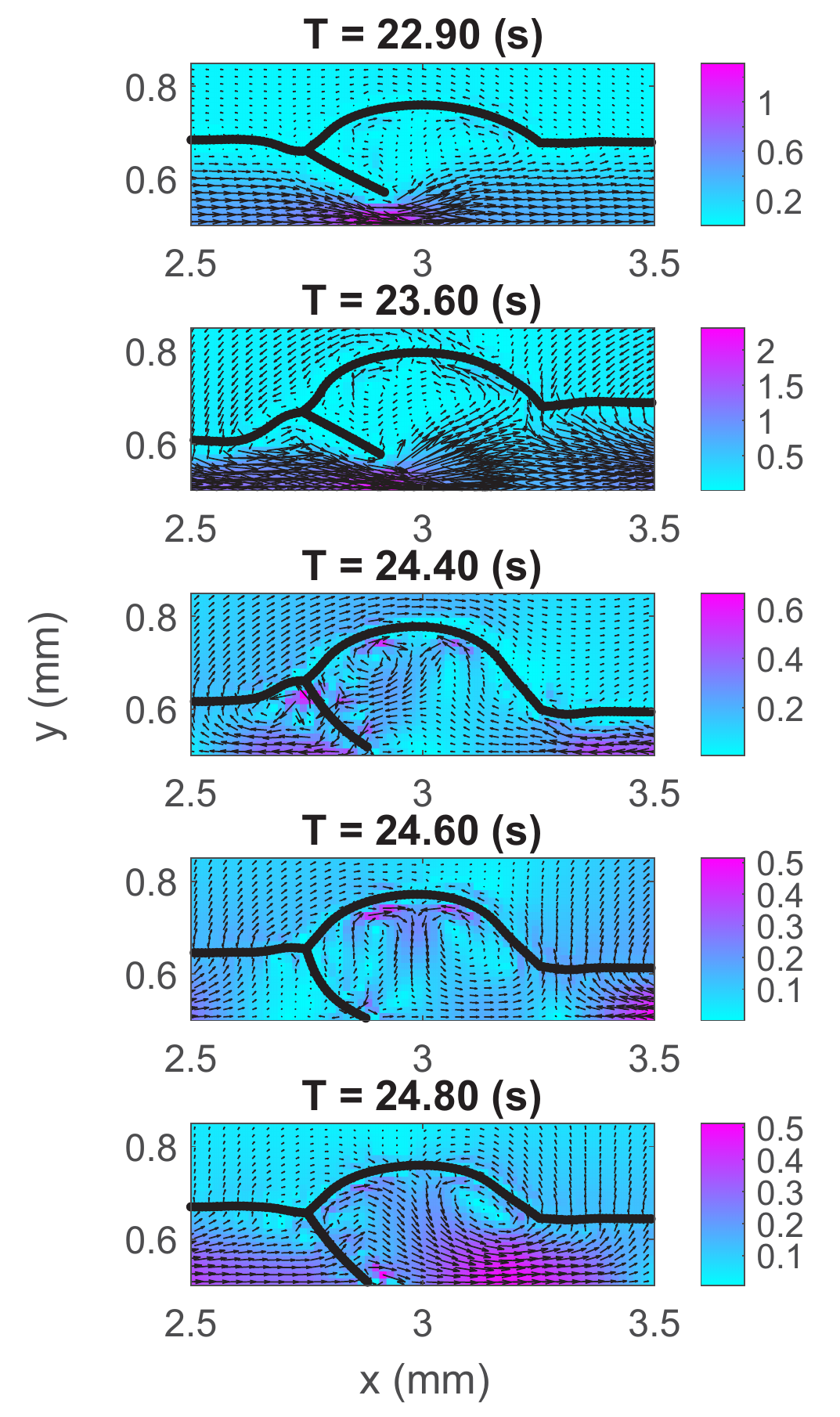}
%      \includegraphics[width=0.45\textwidth]{Fig6.eps}
\caption{Velocity field surrounding the central valve (valve 3) at select times during cycle 10 in a 4\hyp{}lymphangion, fast\hyp{}orthograde pump with an AAPD of 0.051 cmH$_2$O.  Every other velocity vector is included in the plot, the vectors are scaled by a factor of 0.125, \textcolor{myred}{and the (unscaled) velocity magnitude is plotted with color bars measured in mm/s}}
\label{fig:3.2.1}       % Give a unique label
\end{figure}

Flow patterns around valve 3 of the same 4-lymphangion, fast-orthograde pump are provided in Fig. \ref{fig:3.2.1}.  The plots display a portion of the overall domain and velocity field, although the entire model geometry and fluid region were simulated.  The central valve is bounded on both sides by contractile lymphangions, so its motion and flow patterns are considered most native to the model.  Notably, the sinus region features counter\hyp{}rotating vortices at nearly all time snapshots which include periods of forward flow and \textcolor{myred}{valve opening and} closure.  \textcolor{myred}{In their microparticle image velocimetry experiments, \citet{Margaris2016} observed eddies in valve\hyp{}sinus regions (with favorable or adverse 0.5 cmH$_2$O axial gradients in rat mesenteric lymphatics).  \citet{Wilson2018} also reported persistent eddies throughout a pressure\hyp{}driven contraction cycle in the (rigid) valve\hyp{}sinus region that were most prominent during peak forward flow.  They report velocities up to 4 mm/s in the eddy regions during peak systole.  Our simulations feature peak velocity magnitudes in the sinus region around 0.40 mm/s (T = 23.60 s in Fig. \ref{fig:3.2.1}); the trans-valve pressure at this time is around 0.0036 cmH$_2$O, and the AAPD is 0.051 cmH$_2$O.  Our trans-valve pressure differs from the peak trans-valve pressure of 0.61 cmH$_2$O in \citet{Wilson2018}, and our AAPD is an order of magnitude smaller.  \citet{Wilson2018} also report a peak overall velocity of 44 mm/s which seems large compared to peak velocities reported in the literature.  Our peak velocity at 23.6 s is 3.33 mm/s which is more consistent with the literature \citep{Dixon2006}.  The difference in models (2D vs. 3D), vessel-wall rigidity, and pressure regimes likely influence these discrepancies.}

% old one shown was velocity_fields_coarsen.pdf

\subsection{Wall, valve, \textcolor{myred}{pressure, and flow} dynamics}
\label{sec:3.3}
\begin{figure*}
   \includegraphics[width=1\textwidth]{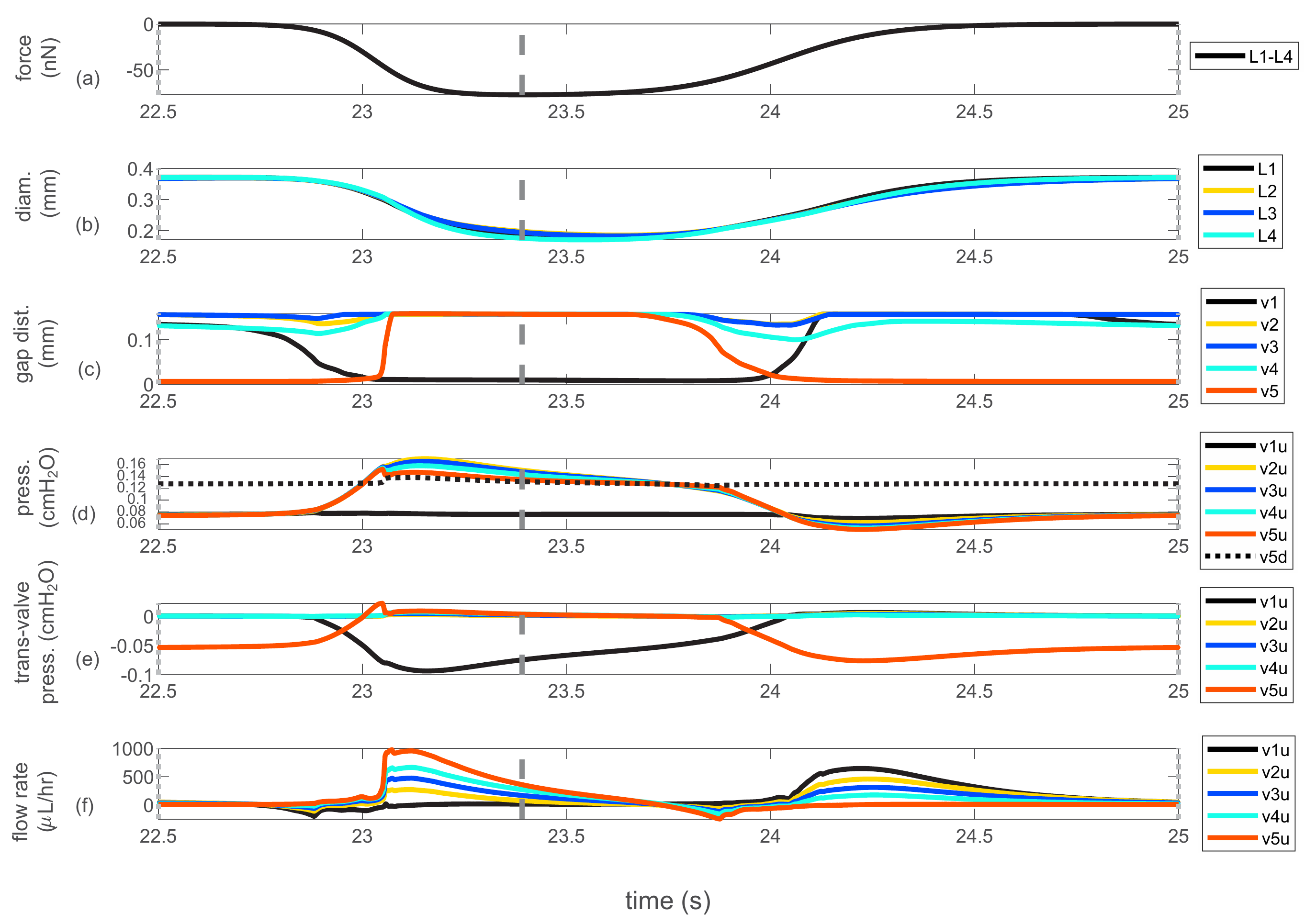}
%   \includegraphics[width=1\textwidth]{Fig7.eps}
\caption{Timecourse plots for a 4-lymphangion, simultaneous pump with an AAPD of 0.051 cmH$_2$O \textcolor{myred}{during cycle 10 (T = 22.5\hyp{}25 s)}.   Panel (a) is the inward contraction force acting simultaneously along all lymphangion upper contractile regions; the negative of this force acts along the lower contractile regions.  \textcolor{myred}{Dark gray dashed and light gray dotted (the latter at 22.5 s and 25 s)} line segments have the same meaning as in Fig. \ref{fig:2.2.2}.  Panel (b) is the lymphangion diameter measured at axial locations approximately 0.25 mm to the left of each downstream valve of lymphangions 1-4 (L1-L4).  Panel (c) is the distance between valve leaflet endpoints for each of the five valves.  Panel (d) is the pressure measured at flow-meter centers (vessel midline) just upstream of each valve (v1u-v5u) and downstream of valve 5 (v5d).  Panel (e) displays the trans-valve pressure for each valve; it is the pressure measured at the flow-meter center just upstream of each valve minus the pressure at the center of the flow meter located at the sinus-wall junction.  Panel (f) shows the flow rate measured at flow meters just upstream of each valve}
\label{fig:3.3.1}       % Give a unique label
\end{figure*}

We provide \textcolor{myred}{contraction force}, diameter, valve gap\hyp{}distance, pressure, and flow\hyp{}rate timecourse plots for a 4\hyp{}lymphangion, simultaneously contracting chain at an AAPD of 0.051 cmH$_2$O during cycle 10 in Fig. \ref{fig:3.3.1}.  \textcolor{myred}{The concurrent diameter, valve\hyp{}state, and pressure plots are similar to those generated from isolated lymphatic vessel experiments \citep{Bertram2018a,Davis2011,Scallan2012,Scallan2013}.  We devote this subsection to analysis of the results in Fig. \ref{fig:3.3.1} for this 4\hyp{}lymphangion, simultaneous pump.}

\subsubsection{\textcolor{myred}{Physiological diameter changes and ejection fraction}}
\label{sec:3.3.1}
The diameter was measured at an axial location approximately 0.25 mm to the left of the downstream valve in each lymphangion.  The diameter tracings for all four lymphangions are shown in panel (b) of Fig. \ref{fig:3.3.1}.  They exhibit shapes similar to those in the literature (e.g., Fig. 1 in \citet{Davis2011}) with brisk contractions and slower relaxations.  In cycle 10, the lymphangion 1 maximal diameter (end\hyp{}diastolic diameter, EDD) was 0.372 mm, and its minimal diameter (end\hyp{}systolic diameter, ESD) was 0.180 mm.  \textcolor{myred}{The minimal diameter occurred 0.988 s after the maximal diameter; the velocity of shortening is estimated as 0.194 mm/s or 0.52 L/s (dividing 0.194 mm/s by the EDD), where L is the EDD.  Values for velocity of shortening in the literature are 103.9 $\mu$m/s or 0.48 L/s \citep{Zhang2013} and 140 $\mu$m/s or 2 L/s \citep{Benoit1989}, though the experimental setups differ.}  In our work, the ESD was 48.3\% of EDD; this is a physiological contraction amplitude \citep{Davis2012,Zawieja2009}.  Assuming no backflow and a cylindrical geometry, these ESD and EDD values yield an ejection fraction (EF) of 76.6\% ($EF = ((\text{EDD}^2-\text{ESD}^2)/\text{EDD}^2)\times 100\%$) which agrees with the literature \citep{Scallan2012}.  \textcolor{myred}{Thus, this pump exhibits physiological diameter changes.}  

\subsubsection{Valve gap-distance dynamics}
\label{sec:3.3.2}
\textcolor{myred}{In our model, the valve dynamics occur as a result of the leaflets' interactions with the fluid; their motion is not prescribed.}  For each valve, the distance between the top and bottom leaflet endpoints was measured throughout a simulation and is referred to as the gap distance.  Gap\hyp{}distance plots for all five valves are shown in panel (c) of Fig. \ref{fig:3.3.1}.  Notably, only valves 1 and 5 close.  The three interior valves remain open throughout a contraction cycle with only small gap\hyp{}distance fluctuations when valves 1 and 5 change state.  Moreover, the gap\hyp{}distance curves for valves 1 and 5 are out of phase; when one is closed, the other is open.  Valve 5 is closed longer and opens more abruptly than valve 1.  Starting at 22.5 s around end\hyp{}diastole, valve 5 is closed, and valve 1 is open.  As contraction commences, the valve\hyp{}1 gap distance decreases as it begins to close.  Around 23 s, valve 1 closes, and valve 5 abruptly opens.  Valve 5 remains open until about 0.25 s after end\hyp{}systole, and then its gap distance decreases.  As the lymphangions are relaxing and the diameters are increasing, valve 5 closes around 24 s, and valve 1 opens.  These valves remain in these states for the remainder of the cycle through 25 s.  The simultaneous\hyp{}pump interior valve behaviors differ from those in the fast\hyp{}orthograde pump. Because only the first and last valves open and close each cycle in the simultaneous pump, the 4\hyp{}lymphangion chain is essentially functioning as a single, long lymphangion.  This interior\hyp{}valve behavior was also reported in the model by \citet{Bertram2016} for their simultaneous pumps.     

\subsubsection{Cyclical intraluminal pressure variation and periods of forward flow}
\label{sec:3.3.3}
The pressure at the vessel midline was measured at each flow meter and plotted over time in panel (d) of Fig. \ref{fig:3.3.1}.  Starting at 22.5 s and as contraction commences, pressures measured at flow meters just upstream of valves 2\hyp{}5 gradually rise above the (nearly constant) pressure just upstream of valve 1.  They then rise in tandem briskly and surpass the pressure downstream of valve 5 around 23 s.  During the time between 22.5 s and 23 s, valve 1 is closing and valve 5 is starting to open.  The rising pressures measured upstream of valves 2\hyp{}5 peak around the time valve 5 is fully open and prior to end\hyp{}systole.  During the time period between 23 s and just before 24 s, these pressures decrease moving down the chain.  During this time period, valve 1 is shut and valve 5 is open; there is a period of forward flow through the tube.  \textcolor{myred}{As relaxation ensues, pressures drop in tandem; coincident with the pressure drop is the closing of valve 5 followed by the opening of valve 1 (around 24 s and shortly thereafter, respectively).}  During the time period just after 24 s and until 25 s, interior pressures again decrease moving down the chain; there is a period of forward flow through the tube.  Thus, significant forward flow occurs twice per cycle in the simultaneous pump: once during systole and again during diastole. 

\subsubsection{Trans-valve pressure differences}
\label{sec:3.3.4}
Next, trans\hyp{}valve pressures were computed as the difference in midline pressures measured at flow meters upstream and downstream of each valve near the sinus\hyp{}wall junctions and plotted over time in panel (e) of Fig. \ref{fig:3.3.1}.  Consistent with their lack of closure, trans\hyp{}valve pressures for the three interior valves are nearly 0 and exhibit little fluctuation.  The shapes of the gap\hyp{}distance plots in panel (c) and the trans\hyp{}valve pressure plots in panel (e) are similar in spirit; we relate time periods of valve openness or closure to time periods during which the trans\hyp{}valve pressure difference is positive or negative, respectively.  Notice that the trans\hyp{}valve pressure for valve 5 features a positive spike, but the same is not true for valve 1.  \textcolor{myred}{Also, valve 1 bears a larger adverse trans\hyp{}valve pressure than valve 5.}          

\subsubsection{Strong forward flow during both systole and diastole}
\label{sec:3.3.5}
Flow rates measured at the flow meters just upstream of each valve are plotted in panel (f) of Fig. \ref{fig:3.3.1}.  During each cycle, there are two pronounced periods of forward flow and two periods of backflow.  The forward flow rates are much larger than the backward flow rates, and forward flow occurs over a longer time period than the backflow.  Recall the pressure drops that occur moving down the chain in panel (d) between approximately 23\hyp{}24 s with valve 1 shut and valve 5 open and again between approximately 24\hyp{}25 s with valve 1 open and valve 5 shut.  These pressure\hyp{}drop periods correspond to periods of forward flow during both systole and diastole.  The backflow occurs when valves 1 and 5 are closing. 

\begin{figure}
%  \includegraphics[width=0.5\textwidth]{hysteresis_zoom2.pdf}
%    \includegraphics[width=0.5\textwidth]{Figures/Fig9}
    \includegraphics[width=0.5\textwidth]{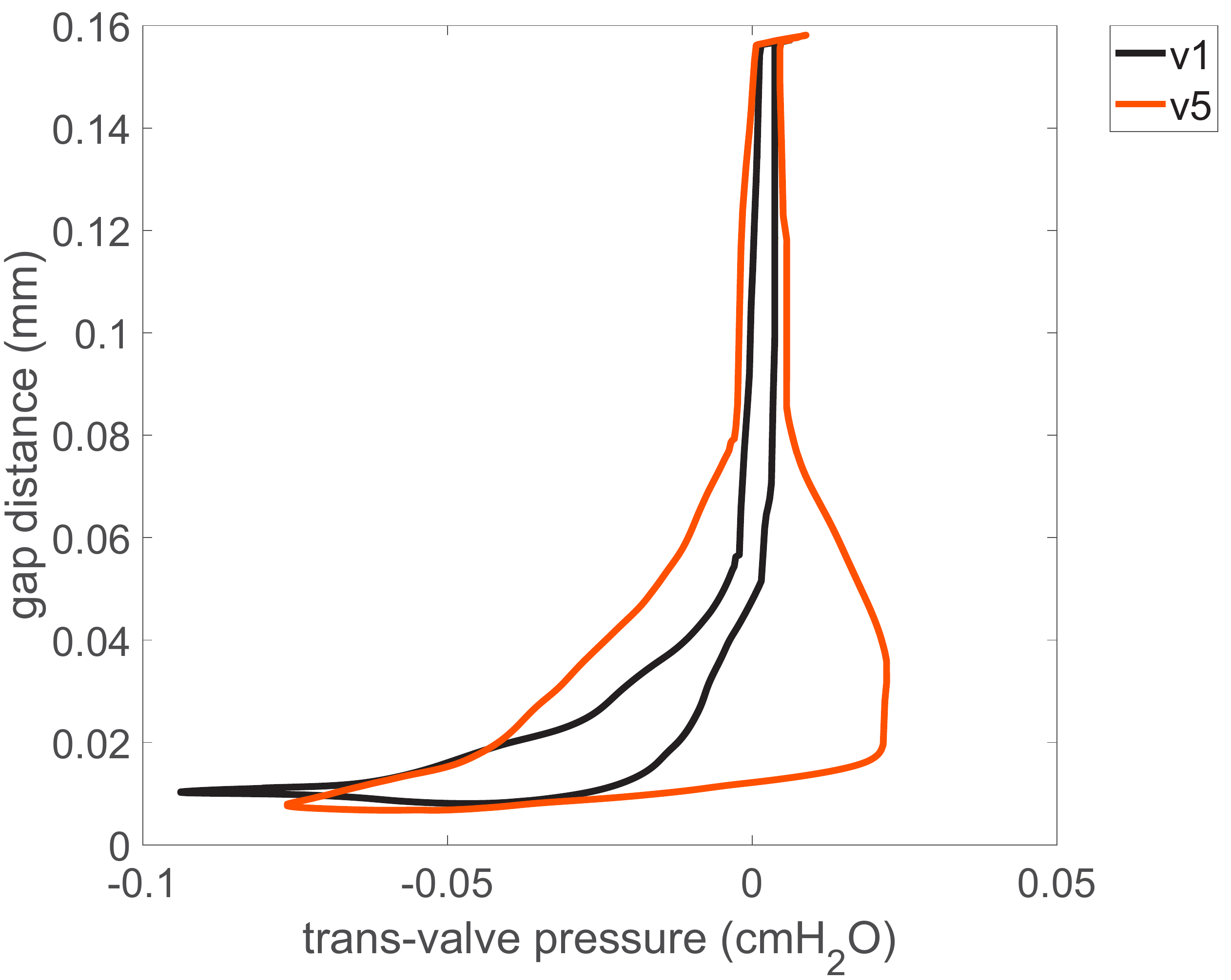}
%    \includegraphics[width=0.5\textwidth]{Fig8.eps}
\caption{Gap distance vs. trans-valve pressure for valves 1 and 5 in a 4-lymphangion, simultaneous pump with an AAPD of 0.051 cmH$_2$O over cycles 9 and 10.  These plots are traversed in the counter-clockwise direction}
\label{fig:3.3.2}       % Give a unique label
\end{figure}

\subsubsection{\textcolor{myred}{Valve gap distance exhibits hysteresis}}
\label{sec:3.3.6}
Next, the gap distance vs. trans\hyp{}valve pressure is plotted for valves 1 and 5 in Fig. \ref{fig:3.3.2}; both variables are plotted at the same times from 20\hyp{}25 s (cycles 9 and 10).  Plots for valves 2 through 4 are omitted because they remained open each cycle and exhibited little gap\hyp{}distance variation.  As time advances, these plots are traced out in the counter\hyp{}clockwise direction.  Animations of dual curve\hyp{}traversal over cycle 10 are provided (Online Resource 7).  These plots are multivalued; for each valve, a given trans\hyp{}valve pressure is associated with two gap distances whose values depend on whether the valve is opening or closing.  Viewed another way, specifying a threshold gap distance beyond which a valve is considered open, a given valve opens and closes at different trans\hyp{}valve pressures.  This phenomenon is oft\hyp{}cited in the literature as valve hysteresis \citep{Ballard2018,Bertram2014,Wilson2018}. 

In Fig. \ref{fig:3.3.2}, starting from the state of a tightly closed valve (minimal gap distance), as time advances, the trans\hyp{}valve pressure increases, but the gap distance initially does not change much.  As the trans\hyp{}valve pressure increases further, the gap distance increases, and the valve opens.  For an open valve, as the trans\hyp{}valve pressure begins to decrease, the gap distance initially does not change much.  This is likely related to fluid inertia; flow\hyp{}reversal is not instantaneous.  Further decreases in trans\hyp{}valve pressure are associated with decreases in gap distance and valve closure.  \textcolor{myred}{Based on the largest average (cycle-mean) and instantaneous tracer velocities measured during cycles 9 and 10, we report Reynolds (Re) numbers of 0.59 and 7.68, respectively.  The Womersley number is 1.59.  Differences between these Re and those we computed based on \citet{Dixon2006} (Re 0.08 and 0.82) stem from differences in the diameter and length scales.}      

When closed, valve 1 bears a larger trans\hyp{}valve pressure than valve 5; also, its gap distance is more responsive to initial increases in trans\hyp{}valve pressure.  In contrast, valve 5 is harder to open; it experiences a large, positive trans\hyp{}valve pressure prior to opening.  This is likely related to the fact that valve 1 opens while the vessel is relaxing and valve 5 is shut.  Compared to valve 5, valve 1 is less exposed to the downstream reservoir pressure; its opening does not require overcoming as large a pressure as for valve 5.  \textcolor{myred}{In fact, it starts to open when the trans\hyp{}valve pressure is adverse, see Fig. \ref{fig:3.3.2}; valves opening at negative trans\hyp{}valve pressures have been reported experimentally \citep{Davis2011}}.  In order for valve 5 to open, the increased pressure during systole must overcome the high downstream pressure.  Also, valve 1 is easier to close than valve 5; to achieve the same gap distance in valve 5, it generally takes a larger trans\hyp{}valve adverse pressure.  Similar to experiments by \citet{Davis2011} indicating transmural\hyp{}pressure\hyp{}dependent hysteresis (though in non\hyp{}contractile, single\hyp{}valved lymphangion segments), the variation in trans\hyp{}valve pressures for opening and closing valves 1 and 5 indicate that hysteretic \textcolor{myred}{behaviors} may vary along a lymphangion chain (in valves of uniform construction) due to variations in pressure and flow.

\textcolor{myred}{The plots in Fig. \ref{fig:3.3.2} exhibit similar shapes to hysteresis plots generated from 3D computational models \citep{Ballard2018,Wilson2018}.  Both of these models assume a rigid vessel wall and feature a single valve with a contraction cycle simulated via pressure waveforms.  
The \citet{Ballard2018} model does not incorporate the valve\hyp{}sinus geometry, and snapshots and videos of the valve at different times in an opening\hyp{}closing cycle do not exhibit much leaflet motion or bulge\hyp{}back \citep{Zawieja2009} in the vicinity of the upstream valve\hyp{}insertion locations.  Based on expected symmetry, only 1/4 the full vessel geometry was simulated in \citet{Wilson2018}, and simulations were stopped to avoid leaflet contact and numerical issues.  Despite the model and geometry differences, the leaflet gap distance vs. trans\hyp{}valve pressure plots in these references exhibit similar shapes to ours.}       

\subsubsection{Assessing transport with tracers}
\label{sec:3.3.7}
\begin{figure}
     \includegraphics[width=0.5\textwidth]{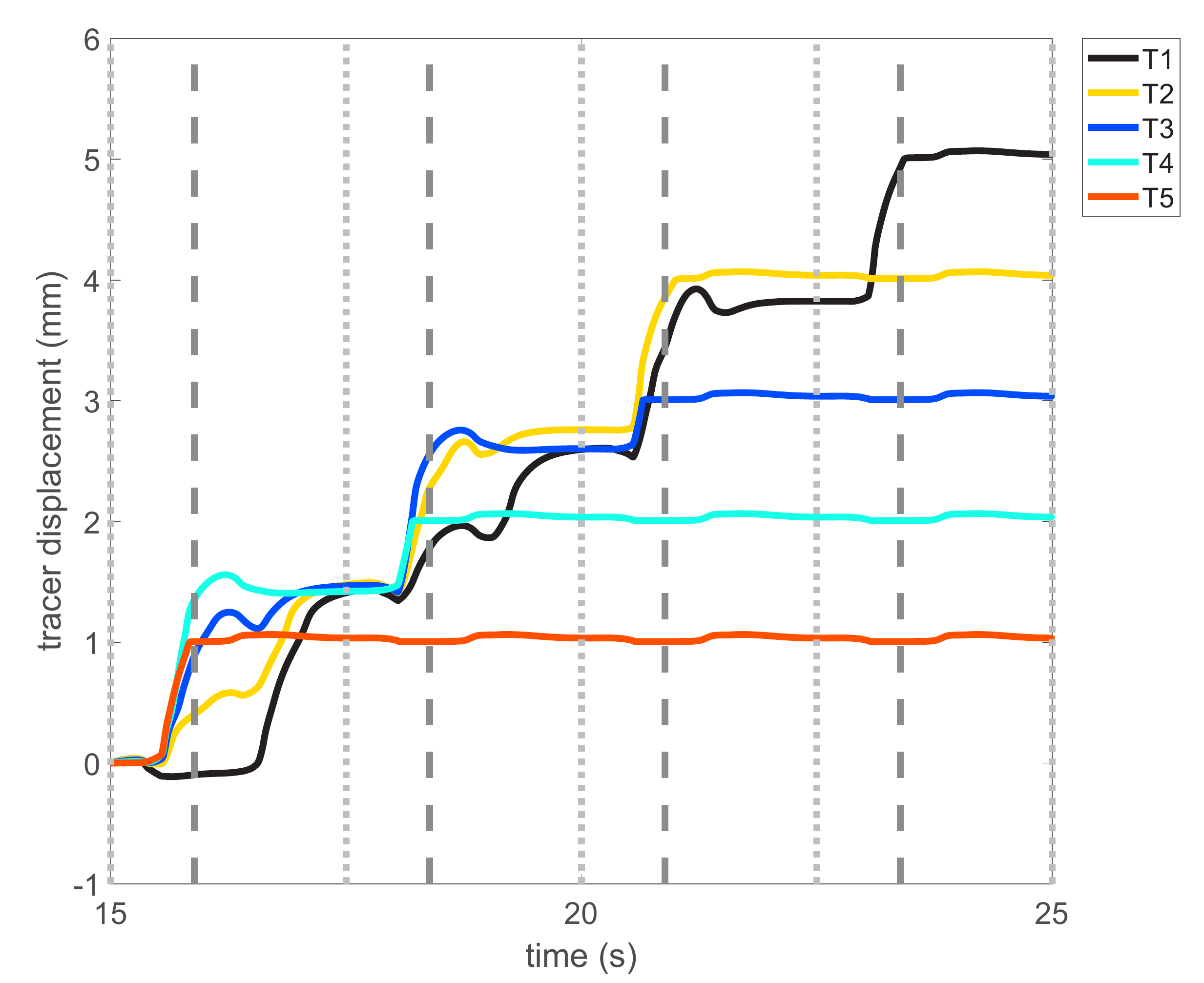}
%     \includegraphics[width=0.5\textwidth]{Fig9.eps}
\caption{Tracer displacement for tracers seeded at valve-insertion centers at the start of cycle 7 (T = 15 s) in a 4-lymphangion, simultaneous pump with an AAPD of 0.051 cmH$_2$O.  Tracers are numbered from left to right according to valve number.  \textcolor{myred}{Dark gray dashed and light gray dotted} lines as in Fig. \ref{fig:2.2.2} }
\label{fig:3.3.3}       % Give a unique label
\end{figure}

We seeded the flow in the 4\hyp{}lymphangion, simultaneous pump with passive tracers initially centered at each valve; see Fig. \ref{fig:2.2.1}.  Figure \ref{fig:3.3.3} illustrates the tracer displacement timecourse for five tracers seeded at the start of cycle 7 ($T = 15$ s) and numbered according to the valve at which they are originally centered.  By the end of the simulation ($T = 25$ s), each tracer was located at the downstream source.  It took tracer 1 just under 3.5 contraction cycles to traverse the tube and reach the downstream source where it stagnated.  \textcolor{myred}{Tracer cycle\hyp{}mean velocities of up to 0.59 mm/s and instantaneous velocities of up to 9.69 mm/s were observed.  These values are close to those reported by \citet{Dixon2006} for \textit{in situ} rat mesenteric prenodal lymphatics (average lymph velocity of 0.87 mm/s with peaks from 2.2-9.0 mm/s).}  Tracers generally moved the fastest just prior to end\hyp{}systole (near dark gray dashed vertical lines in Fig. \ref{fig:3.3.3}).  \textcolor{myred}{We are unaware of any results in the literature comparable to our tracer\hyp{}transport plots.}  Our model enables the study of tracer movement in the flow, much like microparticle velocimetry experiments but without confounding factors that arise from particle\hyp{}insertion.

\subsubsection{Open-valve resistance}
\label{sec:3.3.8}
By virtue of the interior valves remaining open throughout each contraction cycle in the simultaneous, 4\hyp{}lymphangion pump, we estimated the open\hyp{}valve resistance.  This resistance estimate is for the vessel and valve in a contractile setting.  We assumed an Ohm's\hyp{}law relationship and divided the trans\hyp{}valve pressure drop by the flow rate to estimate the resistance.  Based on cycle\hyp{}mean pressures measured at the centers of flow meters bounding each valve\hyp{}sinus region and cycle\hyp{}mean flow rates, we obtained an open\hyp{}valve resistance of approximately $4.02\times10^4$ dyn$\cdot$s/cm$^5$ (averaged among the three interior valves).

\textcolor{myred}{The flow resistance associated with an open\hyp{}valve is an important parameter when valves are modelled as resistors in lumped\hyp{}parameter models.  Open\hyp{}valve resistance parameters in the literature vary: 0.25 dyn$\cdot$s/cm$^5$ was used by \citet{Venugopal2007}; 0.6$\times 10^6$ dyn$\cdot$s/cm$^{5}$ was estimated by \citet{Bertram2014a}, though in earlier work they used a value of 600 dyn$\cdot$s/cm$^5$ \citep{Bertram2011}; 2.68$\times 10^6$ dyn$\cdot$s/cm$^5$ was estimated by \citet{Wilson2018}; and 2.59$\times 10^6$ dyn$\cdot$s/cm$^5$ and 3.26$\times 10^6$ dyn$\cdot$s/cm$^5$ were estimated by \citet{Bertram2020} in the case of a flexible or rigid vessel wall, respectively.  Hence, our open\hyp{}valve resistance estimate is one to two orders of magnitude smaller than the \textit{largest} ones in the literature.  This may be attributable to differences in the models, experimental setups, and definitions of open\hyp{}valve resistance.  The 2D nature of our model might also reduce the open\hyp{}valve resistance.  In our model, the valve opening is effectively a channel in the $z$\hyp{}direction, and there is no resistance offered by the walls of the lymphangion in this direction.}  

\textcolor{myred}{For the sake of comparison, pipette resistances in our model are set to $1.6\times10^4$ dyn$\cdot$s/cm$^5$ and are about 2.5 times smaller than our open\hyp{}valve resistance estimate.  In \citet{Bertram2016}, pipette resistances were approximately four times larger than the open\hyp{}valve resistance.  Pipette resistances in \citet{Venugopal2007} were nearly three orders of magnitude larger than the open\hyp{}valve resistance.  If cannulated vessel experiments and computational simulations are designed to emulate what occurs \textit{in vivo} and for inlet and outlet pressures to be meaningful, it seems desirable for pipette resistances to be the same order of magnitude as the open\hyp{}valve resistance.}      

\begin{figure*}
\includegraphics[width=1\textwidth]{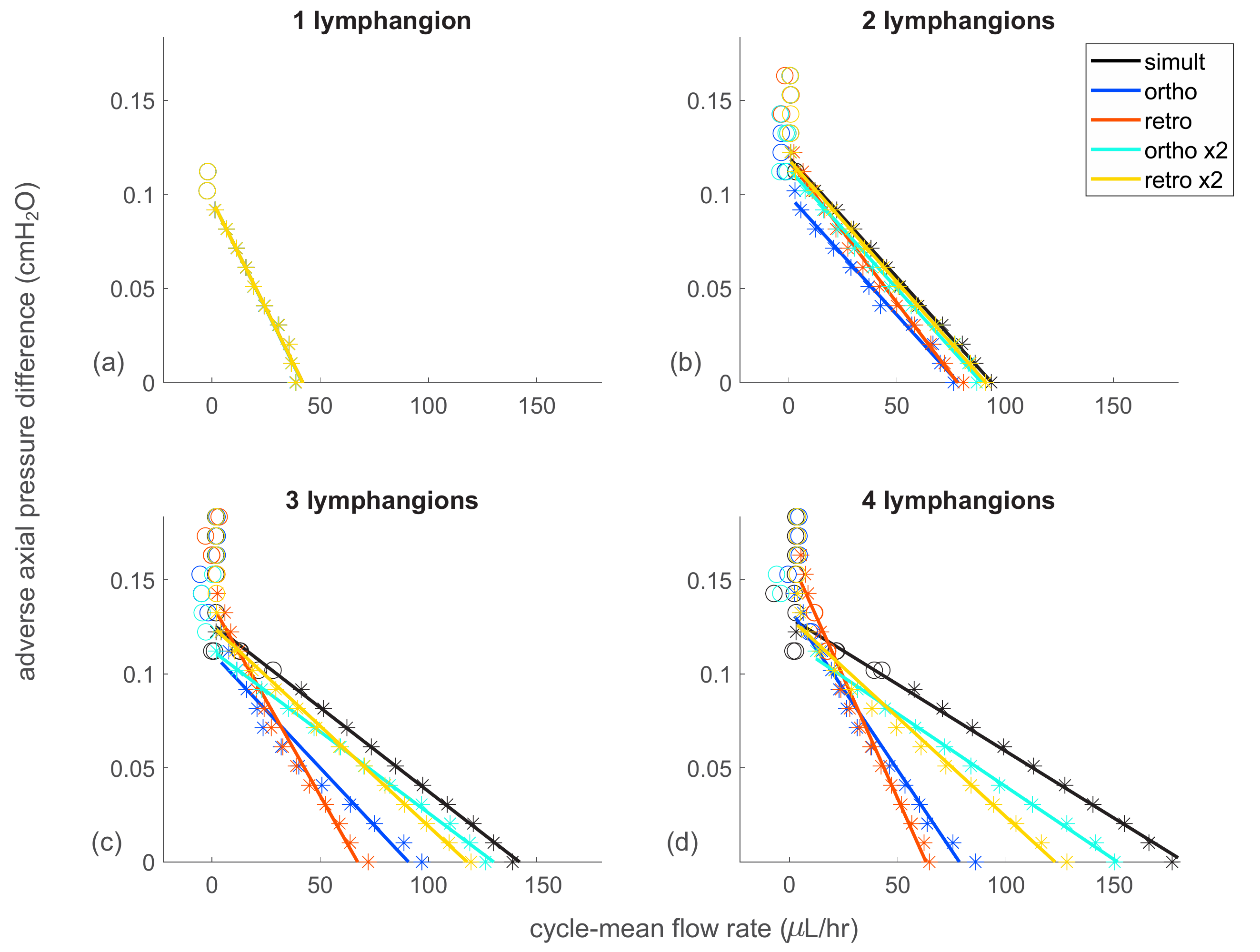}
%\includegraphics[width=1\textwidth]{Fig10.eps}
\caption{Pump-function plots for chains of one to four lymphangions with different contraction styles.  Plots are grouped into panels (a) through (d) according to chain length.  Linear regression was performed for the \textcolor{myred}{non-vertically clustered} data points for which CMFRs were positive (asterisk data points), and lines of best fit are displayed.  The circled data points were not included in the regression due to vertical \textcolor{myred}{clustering} or negative CMFRs.  The legend applies to panels (a)-(d).  The data for 1-lymphangion pumps are overlapping}
\label{fig:3.4.1}       % Give a unique label
\end{figure*} 
 
\subsection{Pump-function behavior}
\label{sec:3.4}
Pump\hyp{}function plots are useful in assessing pump performance.  Chains of one to four lymphangions operating at various AAPDs were simulated for 25 s, or 10 contraction cycles.  The AAPDs between the left and right virtual reservoirs had approximate values of 0, 0.010, 0.020, 0.031, 0.041, 0.051, 0.061, 0.071, 0.082, 0.092, 0.102, and 0.112 cmH$_2$O for all chain lengths; AAPDs of 0.122, 0.133, 0.143, 0.153, and 0.163 cmH$_2$O were also simulated for chains with more than one lymphangion; and AAPDs of 0.173 and 0.184 cmH$_2$O were also simulated for 3\hyp{} and 4\hyp{}lymphangion chains.  Moreover, each chain was simulated with simultaneous, orthograde, retrograde, fast\hyp{}orthograde, or fast\hyp{}retrograde contractions.  Thus, 20 types of pumps operating at various AAPDs were considered.  

\textcolor{myred}{These AAPDs are smaller than those typically used in isolated\hyp{}vessel experiments (1-10 cmH$_2$O, e.g., see \citet{Scallan2012}).  However, lymphangions exhibit variable axial pressure differences \textit{in vivo} which may oppose or favor forward flow; forward flow occurs during periods of increased lymph formation or over short time periods due to contractions of upstream lymphangions \citep{Gashev2002}.  Based on the variable pressure and flow conditions lymphangions encounter \textit{in vivo} and experiments by \citet{Gashev2002} indicating the relevance of slow changes in pressure gradients and adjustment periods whose timescales last minutes, we reason that as AAPDs change from adverse to favorable, periods during which AAPDs of the magnitudes we have listed above and simulated are physiologically relevant (at least over time periods of 25 s, which is what we generally consider).  In the current work, when we refer to low, moderate, and high AAPDs, this is only within the context of the AAPDs we consider in the present study.}

The CMFR data averaged over the interior flow meters from cycle 10 was considered quasi\hyp{}steady state (QSS) and plotted as the CMFR if the percent change among the average cycle\hyp{}10 CMFR and average cycle\hyp{}9 CMFR differed by no more than 1\%.  If QSS was not attained by cycle 10, additional cycles were run.  The CMFR data from the first cycle after the 10th featuring $\leq 1\%$ change from the previous cycle was considered QSS.

\subsubsection{Effect of chain length}
\label{sec:3.4.1}
For each type of pump (specified chain length and contraction style), the QSS data points were plotted and exhibited a negatively sloped, linear trend among low and moderate AAPDs and a vertical trend in the high\hyp{}AAPD regime; see Fig. \ref{fig:3.4.1}.  Linear regression was performed for the non\hyp{}vertically \textcolor{myred}{clustered} data points for which CMFRs were positive.  These lines of best fit are interpreted as pump\hyp{}function curves and plotted in Fig. \ref{fig:3.4.1}.  The asterisk data points were used in the regression, and the circled data points were not (due to vertical \textcolor{myred}{clustering} or negative CMFR).  The slope, intercept, and $R^2$ values are provided in Table \ref{tab:3.4.1}.
  
Among the 335 pumps that were simulated, 12 pumps featured CMFRs that repeated every two cycles or every four cycles and attained QSS among the alternating cycles.  Each of these pumps was operating in the high\hyp{}AAPD regime; their data points are included in the pump-function plots as circles, but they were not included in any regression.  This behavior is under current investigation.     

As seen in panel (a) of Fig. \ref{fig:3.4.1}, all five 1\hyp{}lymphangion pumps yielded nearly identical QSS data sets and thus pump\hyp{}function curves.  \textcolor{myred}{This is expected for a 1\hyp{}lymphangion chain given the way we drive contractions.  The only difference in results occurs for the orthograde and fast\hyp{}orthograde pumps; this is due to a slight difference in the initial values of the contraction functions (while the vessel is pressurizing) compared to those for the three other pumps.  The orthograde contraction signals traverse the lymphangion from left to right prior to reaching the contraction-initiation location near the downstream valve.}  \textcolor{myred}{\citet{Dixon2006} reported an average volumetric flow rate of 13.95 $\mu$L/hr for individual mesenteric collecting lymphatics.  It is unclear how many lymphangions were involved.  Our 1-lymphangion pumps at 0.051 cmH$_2$O yielded CMFRs around 19 $\mu$L/hr, and the CMFRs increased with the chain length.  It is unclear if the \citet{Dixon2006} flow rate is a cycle-mean flow rate like ours or a different type of average.  Also, pressures in the \textit{in situ} setting were not reported, so the comparable AAPD is unclear.}  
  
Panel (b) shows pump\hyp{}function curves and data for 2\hyp{}lymphangion chains.  These curves are extremely similar for the simultaneous and fast, non\hyp{}simultaneous pumps.  At a given pressure difference, the largest CMFRs were generated by the simultaneous pump and then by the fast, non\hyp{}simultaneous pumps.  Also, the disparity among the trio of the simultaneous and fast, non\hyp{}simultaneous pumps vs. the retrograde pump shrinks with rising AAPDs.  As seen in Table \ref{tab:3.4.1}, slopes are quite similar for all 2-lymphangion pumps except the retrograde pump.  Its larger\hyp{}magnitude slope indicates that a one\hyp{}unit increase in CMFR requires a larger pressure drop.  The 2-lymphangion intercepts are similar and largest for the simultaneous and both retrograde pumps and lowest for both orthograde pumps.  Thus, the orthograde pumps are expected to fail (i.e., produce CMFR $\leq 0$) at lower AAPDs than the other pumps.  

Panels (c) and (d) of Fig. \ref{fig:3.4.1} show pump\hyp{}function curves and data for the 3\hyp{} and 4\hyp{}lymphangion chains, respectively.  In both panels, the simultaneous pumps yielded the largest CMFRs except at the highest pressures.  Also, in the low\hyp{} and moderate\hyp{}AAPD regimes, the fast\hyp{}orthograde and fast\hyp{}retrograde pumps yielded larger CMFRs than the orthograde and retrograde pumps.  For both chain lengths, the fast\hyp{}orthograde and fast\hyp{}retrograde pump\hyp{}function curves intersect; the same is true for the orthograde and retrograde curves.  Also, as reflected in the differences in the slopes, the advantage of the simultaneous pumps in producing the largest CMFRs diminishes as the AAPD increases. 

%	6 decimal places included
\begin{table}
{\scriptsize
\caption{Pump-function linear regression slope, intercept, and R$^2$ data.  Slope units are cmH$_2$O/($\mu$L/hr), intercept units are cmH$_2$O, and LA denotes lymphangion}
\label{tab:3.4.1}       % Give a unique label
\begin{tabular}{lrrrrrr}
\hline\noalign{\smallskip}
\text{ } & simult & ortho & retro & ortho x2 & retro x2\\
\noalign{\smallskip}\hline\noalign{\smallskip}
1 LA\\
\hline\noalign{\smallskip}
slope & -0.002312 & -0.002323 & -0.002312 & -0.002319 & -0.002312\\
int. & 0.096951 & 0.097003 & 0.096951 & 0.096998 & 0.096951\\
R$^2$ & 0.985473 & 0.985475 & 0.985473 & 0.985426 & 0.985473\\ 
\noalign{\smallskip}\hline
2 LA\\
\hline\noalign{\smallskip}
slope & -0.001277 & -0.001270 & -0.001516 & -0.001273 & -0.001301\\
int. & 0.119847 & 0.099403 & 0.118162 &	 0.113602 & 	0.118482\\
R$^2$ & 0.997029 & 0.989047 &	0.990261 & 0.987739 & 0.997734\\ 
\noalign{\smallskip}\hline
3 LA\\
\hline\noalign{\smallskip}
slope & -0.000892 & -0.001227 & -0.002034 & -0.000864 & -0.001065\\
int. & 0.126676 & 0.111377 & 0.136890 & 0.112472 & 0.125534\\
R$^2$ &  0.998350 & 0.948896 & 0.979237 & 0.998102 & 0.994456\\ 
\noalign{\smallskip}\hline
4 LA\\
\hline\noalign{\smallskip}
slope & -0.000711 & -0.001716 & -0.002573 & -0.000775 & -0.001057\\
int. & 0.129907 & 0.134558 & 0.162173 & 0.117511 & 0.129696\\
R$^2$ &  0.995006 & 0.972854 & 0.981860 & 0.997314 & 0.978646\\ 
\end{tabular}}
\end{table}

As seen from the 3\hyp{} and 4\hyp{}lymphangion slope information in Table \ref{tab:3.4.1}, compared to the orthograde and retrograde pumps, the simultaneous and fast, non\hyp{}simultaneous pumps have CMFRs that are more responsive to pressure changes.  Based on the intercept information, the retrograde pumps are expected to fail at higher AAPDs than the other pumps; the orthograde and fast\hyp{}orthograde pumps are expected to fail at lower AAPDs than the other 3\hyp{} and 4\hyp{}lymphangion pumps, respectively.   \textcolor{myred}{Also, the pump\hyp{}function data are more spread out among the different contraction styles at low and moderate AAPDs as the chain lengthens.} \textcolor{myred}{To examine the intercept and some of the data\hyp{}spread observations, we decomposed the flow\hyp{}rate data into positive and negative parts and averaged each over a contraction cycle among the interior flow meters to obtain cycle\hyp{}mean forward and backward flow rates, CMFFRs and CMBFRs, respectively.  Pump\hyp{}function data plots analogous to those in panels (c) and (d) of Fig. \ref{fig:3.4.1} are provided in Fig. 5 of Sect. 5 of the SI but with CMFFRs and CMBFRs replacing CMFRs.  These plots reveal that the orthograde and retrograde pumps have similar CMFFRs in the high\hyp{}AAPD regime (see panels (a) and (c)) but that the retrograde pump has smaller\hyp{}magnitude CMBFRs (see panels (b) and (d)).  Hence, the resulting CMFRs for the 3\hyp{} and 4\hyp{}lymphangion retrograde pumps exceed those for the orthograde pumps in this pressure regime, and the retrograde pumps fail at higher AAPDs than the orthograde pumps.  This provides insight into the larger intercepts for the retrograde pumps compared to the orthograde pumps.  In the lowest AAPD regimes, the orthograde pumps yield larger CMFFRs and smaller\hyp{}magnitude CMBFRs than the retrograde pumps, and this is consistent with the CMFR data shown in panels (c) and (d) of Fig. \ref{fig:3.4.1}.}

\textcolor{myred}{To gain further insight into the retrograde high\hyp{}AAPD CMFR advantage, we analyze the video showing the trio of the simultaneous, orthograde, and retrograde pumps at an AAPD of 0.092 cmH$_2$O (see Online Resource 5).  The orthograde pump features a slow pressure increase within the lymphangion chain that eventually opens the downstream valve and is achieved via sequential contraction of lymphangions 1\hyp{}4 with backflow (evidenced by tracer motion) through each of the interior valves as each immediate\hyp{}downstream lymphangion contracts.  We see that some of the forward flow generated by an upstream lymphangion is negated when its neighboring downstream lymphangion contracts and the valve between them remains open.  In contrast, in the retrograde pump, individual lymphangion contractions build sufficiently high pressures to open closed valves.  As the contraction propagates upstream, the interior valves are leveraged, resulting in little backflow (compared to that in the orthograde pump) and summated lymphangion efforts that drive forward flow.  In the retrograde pump, we see that the forward flow generated by a lymphangion contraction is accommodated by downstream lymphangions relaxing.  This is consistent with the description by \citet{GMcHale1992} of pressure increases in contracting segments driving flow into relaxing segments behind the contractile wave.  Additionally,  \citet{Bertram2018a} reported antegrade contraction propagation when the inlet pressure was equal to or exceeded the outlet pressure and retrograde contraction propagation when the inlet pressure was less than the outlet pressure in isolated vessel experiments on rat mesenteric lymphatics.  Combined with our observation (retrograde high-AAPD advantage over orthograde), this suggests that lymphangion chains attempt to optimize their pumping style for ambient AAPDs.}

\textcolor{myred}{Returning to Fig. 5 in the SI, as the chain lengthens from three to four lymphangions, the simultaneous pump significantly increases its CMFFR (see panels (a) and (c)) and slightly decreases the magnitude of its CMBFR (see panels (b) and (d)) at every AAPD.  For the orthograde and retrograde pumps, in going from three to four lymphangions, the CMFFRs decrease and the CMBFRs increase in magnitude at low AAPDs; on the other hand, the CMFFRs increase and the CMBFRs decrease in magnitude for these pumps at high AAPDs.  Thus, the orthograde and retrograde CMFRs decrease as the chain lengthens from three to four lymphangions in the low\hyp{}pressure regime, but they increase in the high\hyp{}pressure regime.  However, the increases are modest in comparison to those for the simultaneous pump.  The increase in the pump\hyp{}function curve spread among the 3\hyp{} and 4\hyp{}lymphangion simultaneous, orthograde, and retrograde pumps is attributable to these differences in CMFFR and CMBFR trends.}            
     
\begin{figure*}
  \includegraphics[width=1\textwidth]{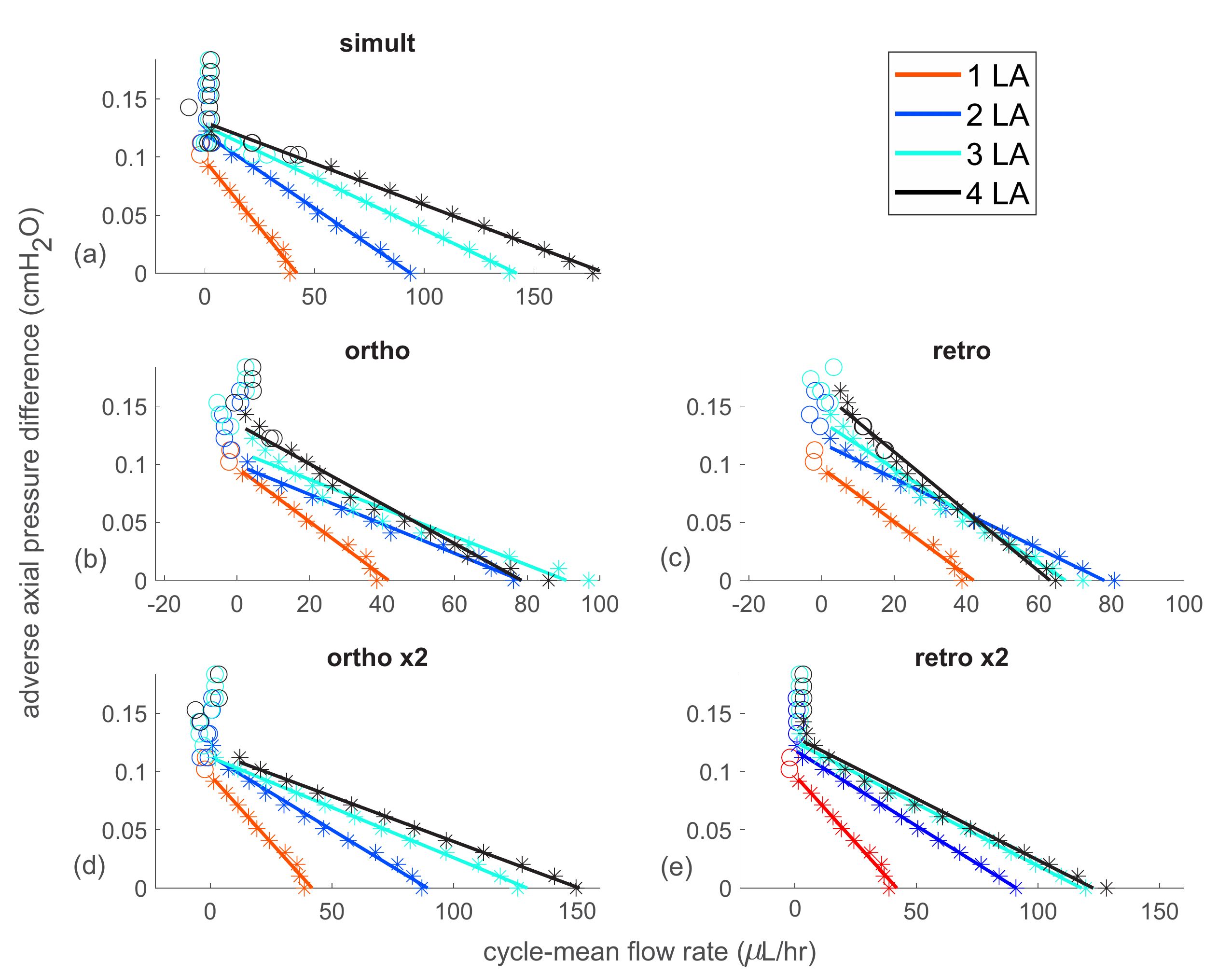}
%  \includegraphics[width=1\textwidth]{Fig11.eps}
\caption{Pump-function plots for chains of one to four lymphangions with different contraction styles.  Plots are grouped into panels (a) through (e) according to contraction style.  The data-point labelling and linear regression are as in Fig. \ref{fig:3.4.1}.  The legend applies to panels (a)-(e), and LA in the legend denotes lymphangion}
\label{fig:3.4.2}       % Give a unique label
\end{figure*}  
   
\subsubsection{Effect of contraction style}
\label{sec:3.4.2}
Lengthening the chain by adding a lymphangion generally yielded larger CMFRs, smaller pump\hyp{}function slopes, and larger intercepts-- though there were a few exceptions; see Fig. \ref{fig:3.4.2}.  For the retrograde pump, there was a reduction in CMFR when lengthening the chain from two to three lymphangions at low AAPDs; there was an increase in CMFR when lengthening the chain at moderate and high AAPDs.  For orthograde and retrograde pumps, there was a slight reduction in CMFR when lengthening the chain from three to four lymphangions at low AAPDs; there was a slight increase in CMFR when lengthening the chain from three to four lymphangions at moderate and high AAPDs.  Also, the orthograde and retrograde pump\hyp{}function curves steepened as the chain lengthened from three to four lymphangions; the retrograde pump\hyp{}function curves also steepened as the chain lengthened from two to three lymphangions.  Lastly, the intercept for the fast\hyp{}orthograde pump decreased in going from a 2\hyp{} to a 3\hyp{}lymphangion chain.

\subsubsection{Effect of additional lymphangions in simultaneous pumps}
\label{sec:3.4.3}
Since the simultaneous pumps of all chain lengths generated the largest CMFRs, we examined them further.  Simultaneous pumps operating at a moderate AAPD of 0.051 cmH$_2$O generated approximate CMFRs of 19, 51, 85, and 113 (all $\mu$L/hr) for 1\hyp{} to 4\hyp{}lymphangion chains, respectively.  The CMFRs increased by factors of approximately 2.7, 1.7, and 1.3.  Thus, the added value of an additional pumping lymphangion decreased with chain length.  As seen from Table \ref{tab:3.4.1}, the slope magnitude decreased by approximately 45\%, 30\%, and 20\% as the chain lengthened to two, three, and four lymphangions, respectively.  The decreasing slope magnitude indicates the CMFR was more responsive to pressure changes as the chain lengthened, but the percent\hyp{}change trend indicates diminishing CMFR responsivity to pressure changes as the chain lengthened.  Lastly, the intercepts increased by approximately 24\%, 6\%, and 3\% as the chain lengthened to two, three, and four lymphangions, respectively.  Thus, the pumps are predicted to fail at higher AAPDs as the chain lengthens but to a diminishing extent.

\subsubsection{\textcolor{myred}{Pump-function behavior: literature comparison}}
\label{sec:3.4.3}
Pump\hyp{}function plots appear in the lumped\hyp{}parameter literature \citep{Bertram2011,Bertram2016,Razavi2017}.  We are unaware of any \textcolor{myred}{extensive} pump\hyp{}function plots from higher\hyp{}dimensional models.  \citet{Bertram2011} report pump\hyp{}function plots for chains of three to five lymphangions whose lymphangions contracted at 0.5\hyp{}s orthograde lags.  \textcolor{myred}{Their pump\hyp{}function curves (see their Fig. 8(b)) are concave\hyp{}down with a pronounced bend in their moderate\hyp{}AAPD (around 300 dyn/cm$^2$ or 0.31 cmH$_2$O) regime.  At their low AAPDs, all three pumps yield similar cycle-mean flow rates (CMFRs) around 0.18 ml/hr which is comparable to our CMFRs.  In their higher AAPD regime, the pumps exhibit variation with longer chains yielding a given CMFR at larger AAPDs than shorter chains.  Similar shaped pump\hyp{}function plots (when the axes are interchanged in their Fig. 5A) are provided by \citet{Razavi2017} for chains of 8 to 36 lymphangions in a related lumped model.}      

\textcolor{myred}{The contractions in our orthograde pumps correspond to approximate lag times of 0.5 s between successive lymphangions, so we compare these pumps to those in the lumped models.  Our pump\hyp{}function data are linear and negatively sloped for positive CMFRs and nearly vertical in the high\hyp{}pressure/low\hyp{}CMFR regime.  This is in direct contrast to the concave\hyp{}down, nonlinear pump\hyp{}function curves in \citet{Bertram2011} but is similar to their 1-lymphangion pump\hyp{}function plot which becomes vertical at sufficiently high downstream pressures.  Also, our orthograde pump\hyp{}function data are quite similar for chains of two to four lymphangions (see Fig. \ref{fig:3.4.2}), and this is qualitatively similar to what \citet{Bertram2011} report.  The observation that longer chains yield 0 CMFR at higher AAPDs than shorter chains \citep{Bertram2011,Razavi2017} is consistent with our findings.}   

\textcolor{myred}{Pump\hyp{}function plots are also provided by \citet{Bertram2011} for a 4\hyp{}lymphangion chain whose lymphangions contract at various lags (simultaneous; 0.25\hyp{}s retrograde; or 0.25, 0.5, 0.75, or 1\hyp{}s orthograde, \textcolor{myred}{where we have converted their angular phase lags to time lags}); see their Fig. 13.  At all AAPDs, the largest CMFRs are generally attained in pumps with longer lags, and the lowest CMFRs are produced by the simultaneous pump.  This suggests larger CMFRs occur in pumps with slower propagating contractions than for faster ones.  Additionally, compared to the lagged pumps, the simultaneous pump fails to yield positive CMFR at the lowest AAPD.  These are opposite our results, \textcolor{myred}{however, we explored a comparatively limited range of contraction propagation speeds (time lags of 0.26 or 0.52 s; equivalently, 37 or 75 degree phase lags, respectively).}  The \citet{Bertram2011} results indicate an advantage for 0.25\hyp{}s orthograde pumps over 0.25\hyp{}s retrograde pumps in terms of large CMFRs, and the advantage diminishes as the downstream pressure increases; this is consistent with our findings for the fast, non\hyp{}simultaneous pumps.  Similar to our data, their pump\hyp{}function curve appears nearly linear for the simultaneous pump, but in contrast, their other pump\hyp{}function curves are concave\hyp{}down with a bend that sharpens with lag length.}

\textcolor{myred}{With further developments to the 2011 model, in particular with regards to hysteretic and transmural\hyp{}pressure\hyp{}dependent valve opening and closing, \citet{Bertram2016} simulated a 5\hyp{}lymphangion pump at various AAPDs with different refractory periods (delayed successive contractions) and different orthograde and retrograde lags.  In direct contrast to their 2011\hyp{}model results, they found the highest CMFRs occurred at each AAPD with simultaneous pumping and a positive refractory period.  In each case, only the first and sixth valves were operational, and all interior valves remained open each cycle.  These results are fully consistent with our observations.  In their non\hyp{}simultaneous pumps, different valve behaviors occurred; some valves remained open during a cycle and never closed. We observed no such behaviors in our non\hyp{}simultaneous pumps except at the lowest AAPDs for the fast\hyp{}orthograde and fast\hyp{}retrograde pumps.  Generally, their retrograde pumps yield lower CMFRs than their orthograde counterparts (at the same pressure and with the same refractory period), and this is similar to our results.}

\textcolor{myred}{Additional investigations of the effects of contraction delays on CMFRs appear in other low\hyp{}dimensional models \citep{Macdonald2008,Venugopal2007}.  \citet{Venugopal2007} created a lumped\hyp{}parameter model corresponding to bovine mesenteric lymphatics and report little variation in CMFR when contractions among adjacent lymphangions in a 3\hyp{}lymphangion chain occur in orthograde or retrograde lags with delays ranging from 0 up to 1 s.  The authors indicate the CMFR is slightly greater for orthograde than retrograde delays.  They state that the vessel behavior (presumably CMFR trends) did not change as they increased the number of lymphangions to four or more, but results are not shown.  They generated pump\hyp{}function data for a 2\hyp{}lymphangion chain, although contraction delays, if any, among the lymphangions are unclear.  The model flow\hyp{}rate data are linear.  Linear regression of corresponding experimental data has a reciprocal slope magnitude of 0.0005665 cmH$_2$O/($\mu$L/hr) which is an order of magnitude smaller than any of our 2\hyp{}lymphangion pump\hyp{}function slope magnitudes.  Their AAPDs are larger than ours by an order of magnitude.  Although the linearity of their pump\hyp{}function data and larger CMFRs in orthograde vs. retrograde pumps are qualitatively similar to our results, at low pressures we observed marked differences in CMFRs as contraction styles and chain lengths varied.}    

\textcolor{myred}{\citet{Macdonald2008} modified a version of the \cite{Reddy1975} model to develop a 1D-flow model inside a bovine, contractile lymphangion with valves modelled as resistors.  They examined contractile wave propagation along a single lymphangion and report little difference in flow whether the 1 cm/s travelling wave propagates forward or backward.  The best pumping occurs when all sections of the lymphangion wall contract in a quickly propagating wave; large phase differences among neighboring computational cells within the lymphangion shuttles fluid back and forth.  This is generally consistent with our results.}

\subsection{Valve dynamics in pump failure}
\label{sec:3.5}
Various valve dynamics were observed in the pump\hyp{}function simulations.  Only the first and last valves in simultaneous pumps opened and closed during contraction cycles.  All interior valves remained open.  Valve opening and closing in orthograde and retrograde pumps generally occurred in forward and backward sequences, respectively.  This is consistent with the contraction\hyp{}propagation direction.  In general, valves closed more tightly in simulations with higher downstream pressures. At the highest downstream pressures, the last valve remained closed throughout a contraction cycle with the other valves generally opening and closing each cycle.  Also, there was a tendency for the first valve to exhibit hindered opening as the downstream pressure rose.  Generally, other valves also opened less wide as the downstream pressure increased, but the hindrance to the first valve occurred at lower downstream pressures than for the other valves.  It often was observed that at sufficiently high downstream pressures, the first valve remained closed while the other valves generally exhibited some degree of opening and closing throughout each contraction cycle.  At even higher downstream pressures, there was recovery of the first\hyp{}valve opening with further increases in downstream pressure associated with the last valve remaining closed and all other valves generally opening and closing each cycle.  

\textcolor{myred}{For their lumped\hyp{}parameter model, Bertram's group investigated pump\hyp{}failure modes in a single lymphangion featuring a valve that remained closed throughout a contraction cycle \citep{Bertram2017}.  We observed two of their eight\hyp{}listed failure modes in our 1\hyp{}lymphangion simulations (results not shown).  At an AAPD of 0.112 cmH$_2$O, we observed the inlet valve to remain closed each cycle and the outlet valve to open and close each cycle.  At an AAPD of 0.122 cmH$_2$O, the outlet valve remained closed, while the inlet valve opened and closed each cycle.  Interestingly, at 0.133 cmH$_2$O both valves closed and slightly opened each cycle.  Also, unlike the \citet{Bertram2017} results, neither the inlet nor outlet valve remained open each cycle at any pressure in any of our 1\hyp{}lymphangion pumps.  Due to the complex valve treatment in \citet{Bertram2017}, it is unclear if the failure modes are physical though some were also observed experimentally.}

\subsection{Comparison of 4-lymphangion chains with different contraction styles}
\label{sec:3.6}
The pump\hyp{}function plots show that simultaneous pumps yield the largest CMFRs at all pressures below pump failure.  The simultaneous advantage is particularly pronounced for the 4\hyp{}lymphangion chain at low and moderate pressures.  However, the advantage diminishes as the downstream pressure increases.  To better understand these pump characteristics, we compared the 4\hyp{}lymphangion chains with different contraction styles at select AAPDs of 0, 0.051 (moderate), and 0.092 (high) cmH$_2$O. 

\subsubsection{Video comparison}
\label{sec:3.6.1}  
Videos showing the last 5 cycles from T = 12.5\hyp{}25 s are provided for the simultaneous, orthograde, and retrograde pump trio at AAPDs of 0, 0.051, and 0.092 cmH$_2$O (Online Resources 1, 3, 5, respectively).  Similar videos are provided for simultaneous, fast-orthograde, and fast-retrograde pumps at the same AAPDs (Online Resources 2, 4, 6).  These videos show passive tracer transport with tracers color\hyp{}coded according to which lymphangion they originate in and are particularly helpful in visualizing the progression of flow through the lymphangions.

At 0 AAPD, we observe swift tracer propulsion by the simultaneous pump during both systolic and diastolic periods with essentially no backflow.  Due to the slowly propagating contraction and the associated interior\hyp{}valve closure, the orthograde and retrograde pumps feature slower and less coordinated tracer movement with periods of flow stagnation and backflow.  The tracers get shuttled down the lymphangion chain in a back\hyp{}and\hyp{}forth motion.  On the other hand, the fast\hyp{}orthograde and fast\hyp{}retrograde contractions propagate sufficiently fast that they yield more coordinated tracer movement with significant forward flow and less backflow than the orthograde and retrograde pumps.  In the simultaneous pump, only valves 1 and 5 open and close each cycle.  In the orthograde and retrograde pumps, all valves open and close each cycle.  In the fast\hyp{}orthograde pump, valve 2 remains open, and all other valves open and close each cycle.  In the fast\hyp{}retrograde pump, valves 3 and 4 remain open, and all other valves open and close each cycle.

At an AAPD of 0.051 cm H$_2$O, the simultaneous pump still exhibits two significant periods of forward flow, though peak flow rates are smaller than they are at 0 AAPD, and there is slight backflow.  Again, only valves 1 and 5 open and close each cycle; all interior valves remain open.  The orthograde and retrograde pumps again exhibit uncoordinated transport with periods of forward flow, backward flow, and stagnation.  All valves open and close, and valves 1 and 5 generally operate in phase; they both open and close during the same time periods.  The fast\hyp{}orthograde and fast\hyp{}retrograde pumps exhibit more rapid tracer transport than the orthograde and retrograde pumps but slower tracer transport than the simultaneous pump.  All valves open and close in the fast, non\hyp{}simultaneous pumps.  Also, in the fast\hyp{}orthograde pump, pressures in the four lymphangions rise in tandem during the systolic period and drop in sequence (from left to right) during the diastolic period (Online Resource 14).  In the fast\hyp{}retrograde pump, pressures in the four lymphangions rise in sequence (from right to left) during the systolic period and drop in tandem during the diastolic period (Online Resource 15).

At an AAPD of 0.092 cm H$_2$O, tracer transport in all pumps occurs more slowly than at lower AAPDs, but the same general trends observed at a pressure of 0.051 cm H$_2$O persist in terms of tracer transport and valve behaviors.  Also, the systolic diameters are larger than at lower pressures but feature marked constriction just downstream of the sinus for all pump types.  The pressure rise/fall trends in the fast\hyp{}orthograde and fast\hyp{}retrograde pumps also persist (Online Resources 18-19).

\begin{figure*}
        \includegraphics[width=1\textwidth]{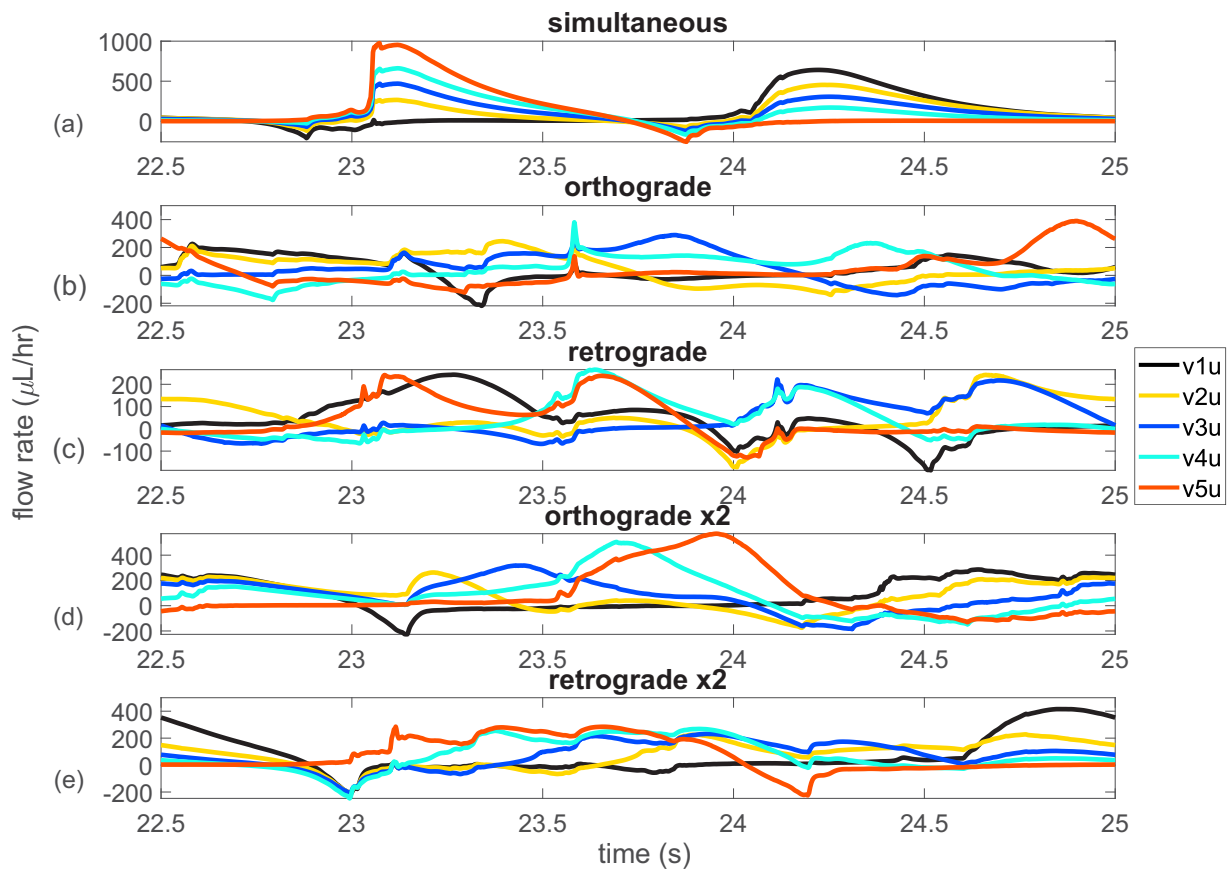}
%        \includegraphics[width=1\textwidth]{Fig12.eps}
\caption{Flow rates in 4-lymphangion chains with different contraction styles pumping at an AAPD of 0.051 cmH$_2$O.  Flow rates were measured at flow meters just upstream of each valve and are displayed over \textcolor{myred}{cycle 10 (T = 22.5\hyp{}25 s)}.  The legend applies to panels (a)-(e), e.g., v1u indicates flow rates measured upstream of valve 1}
\label{fig:3.6.1}       % Give a unique label
\end{figure*}

\begin{figure*}
    \includegraphics[width=1\textwidth]{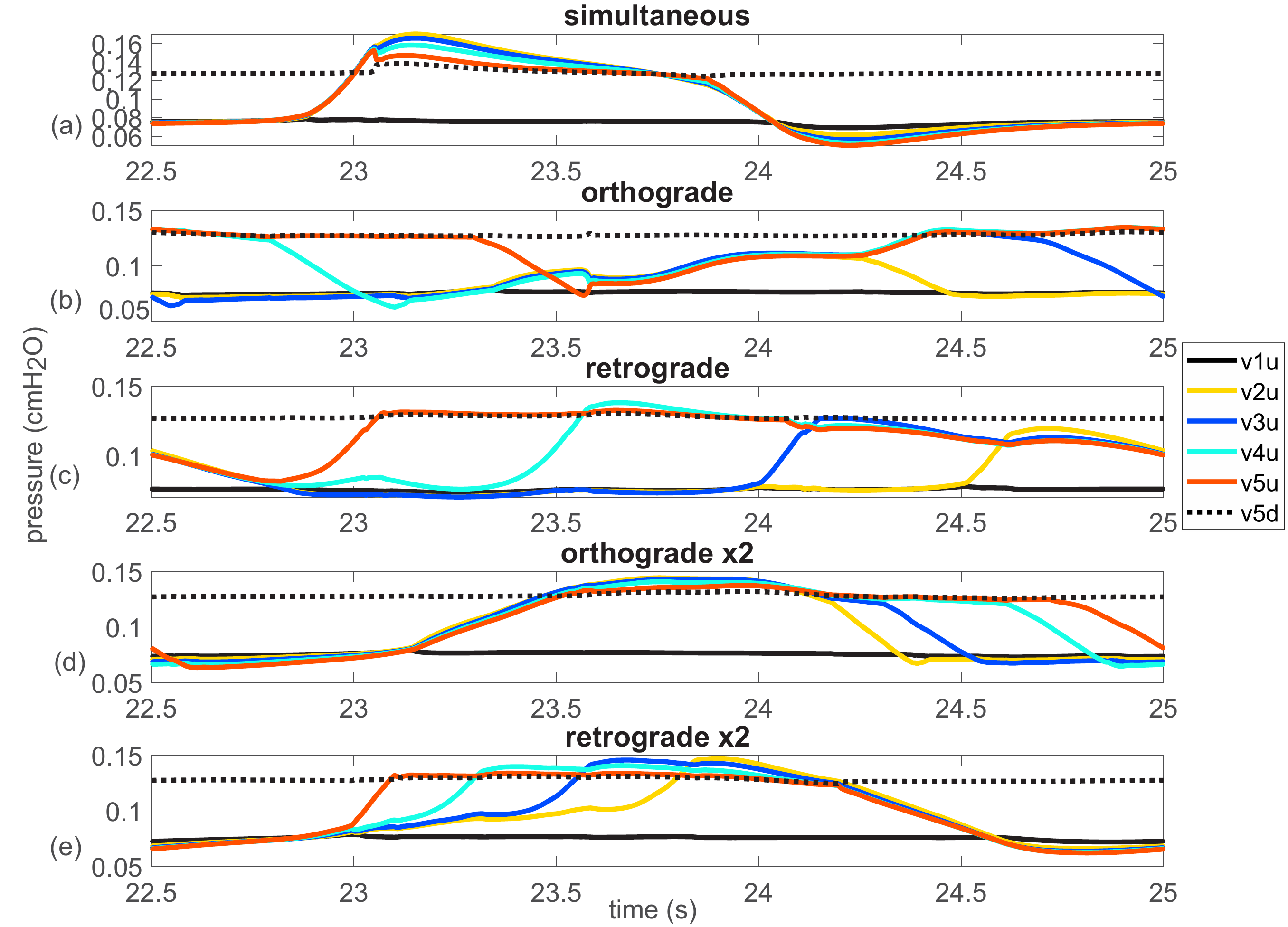}
%    \includegraphics[width=1\textwidth]{Fig13.eps}
\caption{Pressures in 4-lymphangion chains with different contraction styles pumping at an AAPD of 0.051 cmH$_2$O.  Pressures were measured at the vessel midline at flow meters just upstream of each valve and also downstream of valve 5; pressure timecourse plots are displayed over \textcolor{myred}{cycle 10 (T = 22.5\hyp{}25 s)}.  Due to the fluid-source proximity, pressures just upstream of valve 1 (black solid, v1u) and just downstream of valve 5 (black dotted, v5d) are nearly constant for all subplots.  The legend applies to panels (a)-(e)}
\label{fig:3.6.2}       % Give a unique label
\end{figure*}

\begin{figure*}
    \includegraphics[width=1\textwidth]{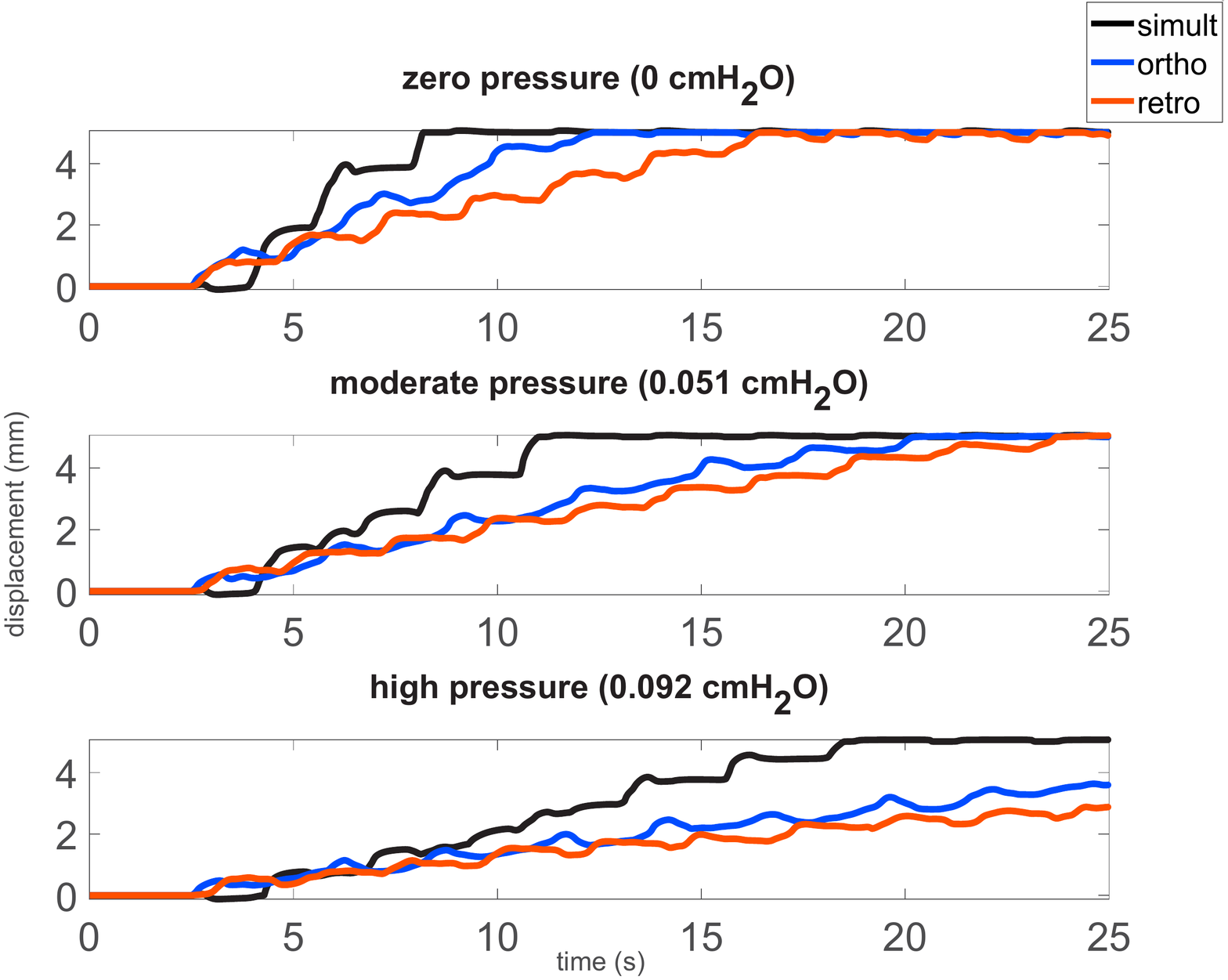}
%    \includegraphics[width=1\textwidth]{Fig14.eps}
\caption{Tracer 1 displacement in simultaneous, orthograde, and retrograde 4-lymphangion chains pumping at low, moderate, and high AAPDs.  Displacement timecourse plots are shown over cycles 2-10 (T = 2.5-25 s).  The displacement is the distance from the initial position of tracer 1 (see left-most target symbol in Fig. \ref{fig:2.2.1}); tracers remain along the vessel midline throughout each simulation}
\label{fig:3.6.3}       % Give a unique label
\end{figure*}

\subsubsection{Flow-rate and pressure comparison}
\label{sec:3.6.2}

Flow\hyp{}rate and pressure plots corroborate the video observations.  As shown in Fig. \ref{fig:3.6.1} panel (a) (reproduction of panel (a) in Fig. \ref{fig:3.3.1}), the simultaneous pump at a moderate pressure difference of 0.051 cm H$_2$O generates two significant periods of concerted forward flow each cycle with minimal back flow (associated with the closure of valves 1 and 5).  One period of forward flow corresponds to downstream ejection and one to refilling from upstream.  The largest flow rate occurs just upstream of valve 5 during systole; the second\hyp{}largest flow rate occurs just upstream of valve 1 during diastole.  

As seen in panels (b)\hyp{}(c), the flow rates generated by the orthograde and retrograde pumps are smaller than those in the simultaneous pump.  They are also unorganized with significant periods of forward and backward flow, consistent with the back\hyp{}and\hyp{}forth tracer movement.  On the other hand, the fast, non\hyp{}simultaneous\hyp{}pump flow rates (panels (d)\hyp{}(e)) exhibit features similar to those from the simultaneous pump, though they are lagged in time.  Namely, the fast\hyp{}orthograde pump has two organized periods of forward flow with the largest flow\hyp{}rates measured just upstream of valve 5 during the systolic period (e.g., T = 23.25\hyp{}24.25 s) and the second\hyp{}largest collection of flow rates measured just upstream of valve 1 during the diastolic period (T = 22.5\hyp{}23.25 s and 24.25\hyp{}25 s).  The fast\hyp{}retrograde pump similarly exhibits two organized periods of forward flow but with the largest flow rates measured just upstream of valve 1 during the diastolic period (T = 22.5\hyp{}23 s and 24.5\hyp{}25 s) and the second\hyp{}largest flow rates occurring just upstream of valve 5 during the systolic period (T=23\hyp{}24.5 s).  In the fast, non\hyp{}simultaneous pumps, two periods of backflow also occur each cycle prior to valve closure and are more pronounced than for the simultaneous pump.  Thus the simultaneous and fast, non\hyp{}simultaneous pumps exhibit two prominent periods of forward flow each cycle with comparatively little back flow; their CMFRs are larger than those for the orthograde and retrograde pumps.  

We next examine pressure plots in Fig. \ref{fig:3.6.2}.  As seen in panel (a) (same as panel (d) in Fig. \ref{fig:3.3.1}), the simultaneous\hyp{}pump pressures measured just upstream of each valve rise and fall in tandem.  The fast\hyp{}orthograde pressures similarly rise in tandem, but they fall in sequence moving down the chain from left to right; see panel (d).  The fast\hyp{}retrograde pressures rise in sequence from right to left and fall in tandem; see panel (e).  Thus, the fast\hyp{}orthograde pump resembles the simultaneous pump during the systolic period with its coordinated pressure rise, whereas the fast\hyp{}retrograde pump resembles the simultaneous pump during the diastolic period with its coordinated pressure drop.

Taken together, the flow\hyp{}rate and pressure information reveal that the fast\hyp{}orthograde pump quickly ejects fluid and has sustained, slower fluid intake, and the fast\hyp{}retrograde pump quickly pulls fluid in and has sustained, slower ejection.  The simultaneous pump has coordinated systolic and diastolic periods with prominent peak flow rates during both these times.  Like the fast\hyp{}orthograde pump, it also exhibits larger peak ejection flow rates than peak entry flow rates.  The orthograde pump pressures also rise in tandem and fall in sequence, but the sequential pressure drops occur while other pressures are still rising; see panel (b) of Fig. \ref{fig:3.6.2}.  \textcolor{myred}{Similar to the pressure patterns for our orthograde pump in Fig. \ref{fig:3.6.2}, \citet{Bertram2011} describe a staircase progression in pressures in their 0.5\hyp{}s orthograde 4\hyp{}lymphangion chain that overcame the outlet pressure each cycle.}  Our retrograde pump pressures rise in sequence and fall in tandem, though the partially coordinated pressure drops occur while other pressures are rising; see panel (c).  Attributable to the \textcolor{myred}{contraction propagation speed,} periods of rising and falling pressures overlap in a more pronounced way than they do for the fast, non\hyp{}simultaneous pumps.  

Flow\hyp{}rate plots at AAPDs of 0 cmH$_2$O and 0.092 cmH$_2$O are similar in spirit to those at 0.051 cmH$_2$O with the exception that the simultaneous pump at  0.092 cmH$_2$O has larger peak flow rates upstream of valve 1 rather than valve 5 (data not shown).  Pressure plots at 0 cmH$_2$O and 0.092 cmH$_2$O exhibit different shapes than at 0.051 cmH$_2$O, but the sequential and synchronized temporal patterns of the non\hyp{}simultaneous pump pressures persist (data not shown).  Interestingly, peak pressures increase with the AAPD for all pump types, though this is perhaps unsurprising since at 0.092 cmH$_2$O, all five pumps still yield positive CMFRs.  

\subsubsection{Tracer-transport comparison}
\label{sec:3.6.3}
Tracer\hyp{}transport plots showing displacement for tracer 1 (originating at the vessel midline near valve 1) for simultaneous, orthograde, and retrograde pumps operating at AAPDs of 0, 0.051, and 0.092 cmH$_2$O are provided in Fig. \ref{fig:3.6.3}.  We show only results for these three types of pumps because they exhibit the greatest differences in tracer transport.  As the downstream pressure increased, tracers took longer to reach the downstream source in all three pumps.  The simultaneous pump transported tracer 1 to the downstream source most rapidly at all three AAPDs.  Counter\hyp{}intuitively, the time gap between the tracer 1 arrival at the downstream source in the simultaneous vs. orthograde or retrograde pumps was longer at a pressure of 0.051 cmH$_2$O than at 0 despite CMFRs being more similar at 0.051 cmH$_2$O.  \textcolor{myred}{This likely stems from differences in velocities measured at the vessel midline (where the tracers are always located) and averages of the horizontal velocity component along the vessel diameter (used to compute CMFRs)}.  At 0.092 cmH$_2$O, only the simultaneous pump transported tracer 1 to the downstream source by the end of the simulation.  The largest cycle-mean velocities and the largest instantaneous velocities for tracer 1 attained in \textcolor{myred}{cycles 2-10} for the three pump types at the three AAPDs are displayed in Table \ref{tab:3.6.1}.  
 
\subsubsection{Summary}
\label{sec:3.6.4}     
The videos, flow\hyp{}rate plots, pressure plots, and tracer\hyp{}transport plots elucidate differences in pump characteristics that affect CMFRs.  Due to the synchronous contraction, the simultaneous, 4\hyp{}lymphangion chain essentially functioned as a single, long, pumping lymphangion with only its bounding valves opening and closing each cycle.  At all but the highest AAPDs, this yielded organized, concerted periods of forward flow with little backflow; the interior valves conferred low flow resistance due to their persistent openness.  Different behaviors were observed in the non\hyp{}simultaneous pumps and were associated with reductions in their CMFRs.  As the AAPD increased from 0, the CMFRs decreased linearly for all pumps though with different slopes.  The simultaneous pump CMFR was most responsive to changes in axial pressure (smallest\hyp{}magnitude slope); thus its marked advantage in generating large CMFRs at low AAPDs diminished as the downstream pressure rose.  When the AAPD was sufficiently large, all pumps failed to yield positive CMFR.  Also, at the three AAPDs of 0, 0.051, and 0.092 cmH$_2$O, valves 1 and 5 in the simultaneous pumps experienced the largest adverse, cycle\hyp{}mean trans\hyp{}valve pressures among all the pump types (results not shown).  In the high\hyp{}AAPD regime, large trans\hyp{}valve pressures are exceedingly difficult to overcome.  Unlike the simultaneous pump, the interior\hyp{}valve closure in the non\hyp{}simultaneous pumps shields valves 1 and 5 from prolonged exposure to high pressures.

\begin{table}
{\scriptsize
\caption{Tracer 1 velocities in simultaneous, orthograde, and retrograde 4-lymphangion pumps.  The largest cycle-mean velocities (avg. vel.) and the largest instantaneous velocities (inst. vel.) over \textcolor{myred}{cycles 2 through 10} are displayed for tracer 1 for each contraction style and each AAPD}
\label{tab:3.6.1}       % Give a unique label
\begin{tabular}{lrrr}
\hline\noalign{\smallskip}
style & AAPD (cmH$_2$O) & avg. vel. (mm/s) & inst. vel. (mm/s)   \\
\noalign{\smallskip}\hline\noalign{\smallskip}
simult & 0 & 0.79 & 7.91\\
\text{} & 0.051 & 0.57 & 9.69\\
\text{} & 0.092 & 0.33 & 6.50\\
\hline\noalign{\smallskip}
ortho & 0 & 0.74 & 3.91\\
\text{} & 0.051 & 0.42 & 2.99\\
\text{} & 0.092 & 0.25 & 2.55\\
\hline\noalign{\smallskip}
retro & 0 & 0.56 & 3.99\\
\text{} & 0.051 & 0.37 & 2.39\\
\text{} & 0.092 & 0.22 & 1.67\\
\noalign{\smallskip}\hline
\end{tabular}}
\end{table}      
\section{Conclusions}
\label{sec:4}
The \textcolor{myred}{two\hyp{dimensional}, fluid\hyp{}structure interaction} model and results presented herein provide new insight into lymphatic transport in chains of variously contracting lymphangions.  We simulated the full Navier\hyp{}Stokes fluid equations without restrictive assumptions on the flow and utilized the immersed boundary method for the two\hyp{}way FSI.  To our knowledge, no reports of fully coupled, two\hyp{}way, FSI models of contracting lymphangion chains with physical valves exist.  \textcolor{myred}{While modelling in 2D has some limitations, it also has an advantage of being more computationally tractable than modelling in 3D, particularly for the types of simulations we performed.  Limitations include the inability to explicitly include the valve and vessel 3D geometry and the assumption that quantities do not vary in the transverse direction.  We used the latter assumption (over a depth equal to the lymphangion diameter) to calculate volumetric flow rates.  Valves have complicated geometries in 3D; we implicitly model 3D features via the valve buttress forces.  However, the valves in our model exhibit hysteretic gap\hyp{}distance behaviors consistent with reports in the literature from 3D models \citep{Ballard2018,Wilson2018}.  Also, our open-valve resistance estimates are perhaps underestimated due to the 2D nature of the model.  Nonetheless, a 2D model has value in providing insight on lymphatic transport through various\hyp{}length chains of lymphangions with different contraction styles and various AAPDs.  It has the potential to suggest and complement biology experiments, to explore regimes that are impossible to consider in the experimental setting, to inform lower-dimensional models, and to refine the range of experiments one may be interested in probing with a 3D model.}

\textcolor{myred}{Comparing analytic Poiseuille flow and closed channel flow rates for a pipe (3D) and a channel of equal diameter and height, for a given pressure drop and the same viscous fluid, the flow rates in the channel are about 27\% larger than in the pipe.  We are comparing volumetric flow rates using the height of the channel as a characteristic depth.  To yield the same pressure\hyp{}driven flow rates in the channel as in the pipe, it would require a pressure reduction of about 21\%.  This flow\hyp{}rate and pressure\hyp{}drop comparison lends some support to the use of lower pressures in a 2D setting.  The adverse axial pressure differences (AAPDs) in our study are about an order of magnitude smaller than those typically used in isolated vessel experiments.  However, our 1\hyp{}lymphangion pumps fail at AAPDs roughly four times larger (0.097 cmH$_2$O vs. 0.025 cmH$_2$O) than those reported by \citet{Li2019} for their 2D computational model.  We interpret the AAPDs in the present study as physiologically relevant during transitions from adverse to favorable axial pressure difference when the AAPD would be small.  We are in the process of developing robust numerical methods to deal with higher pressures.}  We ignored gravitational effects which may become significant in longer chains \citep{Bertram2011}.  Also, contractions were prescribed, but in reality, contractile behavior exhibits adaptation to pressure and shear \citep{Mukherjee2019,Scallan2012}.   

In addition to reporting pressure and velocity plots from our model, we generated pump\hyp{}function data for many types of pumps.  No such data from higher\hyp{}dimensional models appear in the literature.  The simultaneous pumps generally yielded the largest CMFRs, and these pumps were the most responsive to pressure changes.  \textcolor{myred}{Lengthening the chain generally leads to increased CMFRs and higher AAPDs at which the pump fails to yield positive CMFR.}  Moreover, the first and last valves in the simultaneous pumps bore the largest trans\hyp{}valve pressures.  The interior valves in non\hyp{}simultaneous pumps generally opened and closed each cycle thus breaking up the pressure column and reducing the trans\hyp{}valve pressure burden.  In light of this, there may be an important role for \textcolor{myred}{slowly propagating} contractions in the high\hyp{}AAPD regime \textcolor{myred}{and quickly propagating contractions in the low\hyp{}AAPD regime}.  Also, in long chains of lymphangions, simultaneous contraction is not expected to occur along the entire length of the chain \citep{Bertram2016}.  Perhaps there are functional subgroups of lymphangions in series whose contractile behaviors are closely related (due to variable gap\hyp{}junctions, muscle\hyp{}cell coverage, or pacemaker sites), and these groups would be interesting to simulate.  Nonetheless, if therapeutics could be designed to selectively target and reduce the contraction propagation speed, to resemble the slow orthograde or retrograde pumps in the current investigation, it may be advantageous to regulate the CMFR and limit the trans\hyp{}valve pressures when a chain of lymphangions is facing a large adverse pressure difference.  \textcolor{myred}{Similarly, speeding up contraction propagation may boost CMFRs when there is a slightly adverse pressure difference.}    

Novel tracer\hyp{}transport plots were also reported in the present work, and simulation videos with tracers are helpful for visualizing transport.  Non\hyp{}simultaneous contractions propagated along the lymphangion chains, and various valve behaviors were observed and undoubtedly affected by the contraction style.  Reports from other higher\hyp{}dimensional models have been restricted to the case of a rigid vessel in a non\hyp{}contractile setting, a vessel without physical valves, or a vessel with only a single, contracting lymphangion.  Thus, the simulations with various\hyp{}style contractions in the valved, lymphangion chains in the present work are a novel contribution to the quantitative lymphatic biology field.   

%\section{Supplementary Information}
%See extra PDF (discretization equations too long for two-body template).  All the movies, which?!  Need to write captions/descriptions.  Check again formatting requirements.  

\begin{acknowledgements}
Simulations were run at the Center for High-Performance Computing (CHPC) at the University of Utah.  HE acknowledges partial support under NSF RTG-1148230.  ALF was supported in part by NSF grant DMS-1716898.  ALF and AB were supported in part by NHLBI grant 1U01HL143336.  VS was supported by grant NSF CCF 1714844.  
\end{acknowledgements}
\section*{Conflict of interest}
The authors declare that they have no conflict of interest.

\bibliographystyle{spbasic_cp2}      % basic style, author-year citations
\bibliography{library}   % name your BibTeX data base

%% Non-BibTeX users please use
%\begin{thebibliography}{}
%%
%% and use \bibitem to create references. Consult the Instructions
%% for authors for reference list style.
%\end{thebibliography}

\end{document}

\title{Supplementary Information%\thanks{Grants or other notes
}

%\titlerunning{Supplementary Information}        % if too long for running head


%\maketitle
%\onehalfspacing
%\doublespacing
\begin{center}
{\LARGE
\textbf{Supplementary Information}}
\bigskip

Pump efficacy in a \textcolor{myred}{two-dimensional}, fluid-structure interaction model of a chain of contracting lymphangions
\bigskip

Hallie Elich, Aaron Barrett, Varun Shankar, Aaron L. Fogelson
\bigskip

\textit{Biomechanics and Modeling in Mechanobiology}
\bigskip

Corresponding author: Hallie Elich, Department of Mathematics, University of Utah (elich@math.utah.edu)
\end{center}
  
\section{Discretization}
\label{sec:1} 
We provide additional details for the spatial discretizations of $\Omega$ and $\Gamma,$ the finite difference discretization of spatial differential operators in our model equations, the temporal discretization, and the numerical\hyp{}implementation scheme. 
\subsection{Spatial discretization}
\label{sec:1.1}  
\begin{figure}
 % \includegraphics[width=0.5\textwidth]{staggered_grid.pdf}
 %\includegraphics[width=0.5\textwidth]{Figures/Fig3.pdf}
  \includegraphics[width=0.5\textwidth]{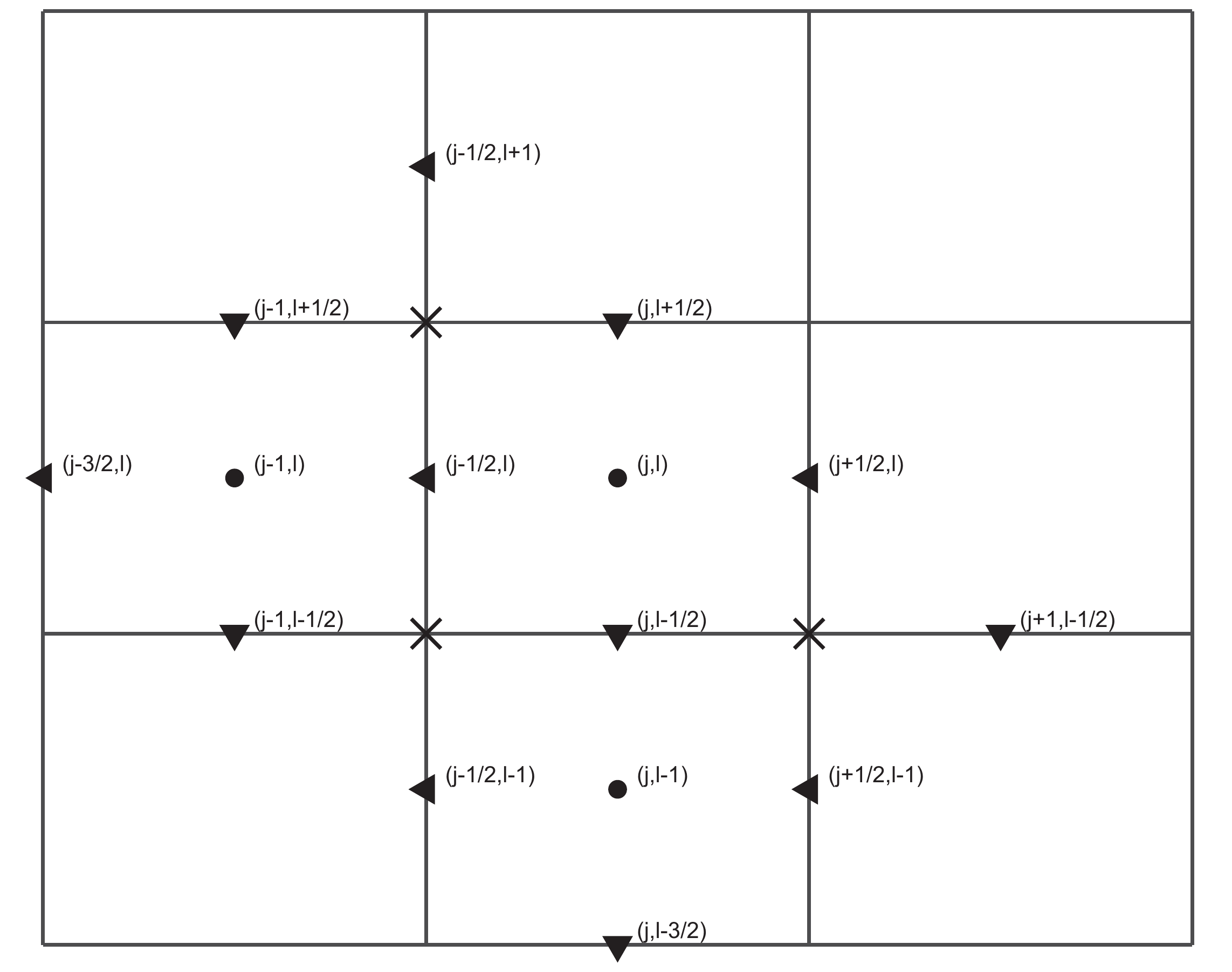}
%  \includegraphics[width=0.5\textwidth]{Fig1_SI.eps}
\caption{Portion of a MAC grid with select grid cell\hyp{}centers and grid edge\hyp{}centers labelled with the corresponding $\bvx$ subscripts.  Select cell centers are denoted with dots, left\hyp{}edge centers are denoted with left arrowheads, bottom\hyp{}edge centers are denoted with down arrowheads, grid vertices are denoted with x's}
\label{fig:1.1.1}       % Give a unique label
\end{figure}

Recall the 2D dimensionless system, i.e., dimensionless 2D versions of Eqs. 1\hyp{}6 from the main text and with Eq. 1 replaced with Eq. 14.  For convenience, that system is reproduced here (tildes are omitted):
\begin{equation}\label{eq:1.1.1}
\text{Re}\left(\bvu_t + \nabla \cdot \left(\bvu\bvu^T\right)\right) = -\nabla p + \Delta\bvu + \bvf - \kappa\sigma\bvu,
\end{equation}

\begin{equation}\label{eq:1.1.2}
\nabla \cdot \bvu = \sum_{i=1}^2 Q_i(t)\psi_i(\bvx),
\end{equation}

\begin{equation}\label{eq:1.1.3}
Q_i(t) = \dfrac{1}{R_i}\left\lbrace P_i^{\text{ext}}(t)-\int_{\Omega}p(\bvx,t)\psi_i(\bvx)\,d\bvx \right\rbrace,
\end{equation}

\begin{equation}\label{eq:1.1.4}
\bvf(\bvx,t) = \int_{\Gamma} \bvF(s,t)\delta(\bvx-\bvX(s,t))\,ds,
\end{equation}
\begin{equation}\label{eq:1.1.5}
\dfrac{\partial \bvX}{\partial t}(s,t) = \bvu(\bvX(s,t),t) = \int_{\Omega}\bvu(\bvx,t)\delta(\bvx-\bvX(s,t))\,d\bvx,
\end{equation}
\begin{equation}\label{eq:1.1.6}
\bvu(\bvx,0) \equiv \boldsymbol{0},
\end{equation} where $\Omega = \left[0, a\right]\times \left[0, b\right],$ and the IB configuration is $\bvX(s,t)$ with $0 \leq s\leq \dfrac{l}{L}.$
Let $\bvu(\bvx,t) = (u(\bvx,t),v(\bvx,t))^T$, $\bvf(\bvx,t) = (f(\bvx,t), g(\bvx,t))^T,$ $\bvF(s,t) = (F(s,t), G(s,t))^T,$ and $\bvx = (x,y)^T$.  In the main text, $\Omega$ is discretized into a uniform MAC grid of meshwidth $h$.  The $(j,l)$\hyp{}th MAC grid cell has center $\bvx_{j,l} = (x_j, y_l) = \left(\left(j - \frac{1}{2}\right)h, \left( l - \frac{1}{2}\right)h \right),$ left\hyp{}edge center $\bvx_{j-1/2,l}=(x_{j-1/2}, y_l)=((j-1)h,(l-\frac{1}{2})h),$ and bottom\hyp{}edge center $\bvx_{j,l-1/2}=(x_j,y_{l-1/2})=((j-\frac{1}{2})h,(l-1)h),$ for $j = 1,\ldots, n_x$ and $l = 1,\ldots,n_y;$  see Fig. \ref{fig:1.1.1}.  

We denote by $\Delta s$ the uniform increment in parameter space and index Lagrangian nodes with a subscript, $k,$ i.e., $s_k$ for $k = 1,\ldots N.$  The valve IB points have $\Delta s_v\approx \Delta s;$ each leaflet IB point is indexed $k = 1,\ldots,N_v.$ 

With $\Omega$ and $\Gamma$ spatially discretized, we introduce notation for the solution to the discrete approximation of the 2D dimensionless system Eqs. \ref{eq:1.1.1}-\ref{eq:1.1.6}.  Let \\
$U_{j-1/2,l}(t) \approx u(\bvx_{j-1/2,l},t),$\\
$V_{j,l-1/2}(t) \approx v(\bvx_{j,l-1/2},t),$\\
$\bvU_{j,l}=(U_{j-1/2,l},V_{j,l-1/2})^T,$\\
$P_{j,l}(t) \approx p(\bvx_{j,l},t),$  \\
$f_{j-1/2,l}(t) \approx f(\bvx_{j-1/2,l},t),$\\
$g_{j,l-1/2}(t) \approx g(\bvx_{j,l-1/2},t),$\\
$\bvf_{j,l}=(f_{j-1/2,l},g_{j,l-1/2})^T,$\\
$F_k(t) \approx F(s_k,t),$\\
$G_k(t) \approx G(s_k,t),$\\
$\bvF_k(t) = (F_k(t), G_k(t))^T,$ and\\
$\bvX_k(t) = \bvX(s_k,t).$   \\
Thus other than for $\bvf,$ $\bvF,$ and their components, we use capital letters to denote quantities corresponding to the discrete-approximation solution.  Whenever $\bvf$, $\bvF$, or their component functions appear with a subscript, usage of the corresponding discrete function is implied.  Since the components of $\bvU_{j,l}$ are defined at grid left\hyp{}edge centers and bottom\hyp{}edge centers, we define a separate Brinkman indicator for each component.  The first component indicator is 1 at left\hyp{}edge centers outside the structure and 0 at all other left\hyp{}edge centers and similarly for the second component indicator at bottom\hyp{}edge centers.  Let $\boldsymbol{\sigma}_{j,l} = (\sigma_{L_{j-1/2,l}},\sigma_{B_{j,l-1/2}})^T,$ with $\sigma_{L_{j-1/2,l}}(t) = \sigma(\bvx_{j-1/2,l},t),$ and $\sigma_{B_{j,l-1/2}}(t) = \sigma(\bvx_{j,l-1/2},t).$  Define $\Psi^r_{i_{j,l}} = \psi_i^r(\bvx_{j,l}),$ $\Psi^c_{j,l} = \psi^c(\bvx_{j,l}),$ and $\Psi_{i_{j,l}} = \Psi^r_{i_{j,l}} - \Psi^c_{j,l}.$  These functions satisfy        
\begin{equation}\label{eq:1.1.7}
\sum_{j,l} \Psi^r_{i_{j,l}} h^2 = \sum_{j,l} \Psi^c_{j,l}h^2 = 1,
\end{equation} 
which are discretized 2D integrals and analogues of the integrals in Eq. 7 of the main text (each dimensionless 3D integral features an integrand that is constant in $z$ and with $z$\hyp{}integration limits from 0 to 1). 

We write the momentum (Eq. \ref{eq:1.1.1}) and continuity (Eq. \ref{eq:1.1.2}) equations in component/expanded form:

\begin{equation}\label{eq:1.1.8}
\text{Re}\left(\dfrac{\partial u}{\partial t} + \dfrac{\partial}{\partial x}(u^2) + \dfrac{\partial}{\partial y}(uv)\right) = -\dfrac{\partial p}{\partial x}+ \dfrac{\partial ^2 u}{\partial x^2}+ \dfrac{\partial ^2 u}{\partial y^2} + f - \kappa\sigma u,
\end{equation}

\begin{equation}\label{eq:1.1.9}
\text{Re}\left(\dfrac{\partial v}{\partial t} + \dfrac{\partial}{\partial x}(uv) + \dfrac{\partial}{\partial y}(v^2) \right) = -\dfrac{\partial p}{\partial y}+ \dfrac{\partial ^2 v}{\partial x^2}+ \dfrac{\partial ^2 v}{\partial y^2} + g - \kappa\sigma v,
\end{equation}

\begin{equation}\label{eq:1.1.10}
\dfrac{\partial u}{\partial x} + \dfrac{\partial v}{\partial y} = \sum_{i=1}^2 Q_i\psi_i.
\end{equation}

We use centered differencing and averaging, when quantities do not live at required locations, to spatially discretize Eqs. \ref{eq:1.1.8}-\ref{eq:1.1.10}.  We focus on the $(j,l)$-th grid cell and use Fig. \ref{fig:1.1.1} to aid our work.  We use the following approximations for terms in Eq. \ref{eq:1.1.8}, each of which must live at the left-edge center, $\boldsymbol{x}_{j-1/2,l}$:

\begin{equation}\label{eq:1.1.11}
\dfrac{\partial}{\partial x}(u^2)\Big|_{\boldsymbol{x}_{j-1/2,l}} \approx \dfrac{U_{j+1/2,l}^2 - U_{j-3/2,l}^2}{2h},
\end{equation}

\begin{equation}\label{eq:1.1.12}
\begin{aligned}
&\dfrac{\partial}{\partial y}(uv)\Big|_{\boldsymbol{x}_{j-1/2,l}} \approx\\&\approx \dfrac{\left(\dfrac{U_{j-1/2,l+1}+U_{j-1/2,l}}{2}\right)\left(\dfrac{V_{j,l+1/2}+V_{j-1,l+1/2}}{2}\right) - \left(\dfrac{U_{j-1/2,l}+U_{j-1/2,l-1}}{2}\right)\left(\dfrac{V_{j,l-1/2}+V_{j-1,l-1/2}}{2}\right)}{h},
\end{aligned}
\end{equation}

\begin{equation}\label{eq:1.1.13}
\dfrac{\partial p}{\partial x}\Big|_{\boldsymbol{x}_{j-1/2,l}} \approx \dfrac{P_{j,l} - P_{j-1,l}}{h}=: (D_{h_x}^cP)_{j-1/2,l},
\end{equation}

\begin{equation}\label{eq:1.1.14}
\dfrac{\partial ^2 u}{\partial x^2}\Big|_{\boldsymbol{x}_{j-1/2,l}} \approx \dfrac{U_{j+1/2,l} - 2U_{j-1/2,l}+U_{j-3/2,l}}{h^2},
\end{equation}

and

\begin{equation}\label{eq:1.1.15}
\dfrac{\partial ^2 u}{\partial y^2}\Big|_{\boldsymbol{x}_{j-1/2,l}} \approx \dfrac{U_{j-1/2,l+1} - 2U_{j-1/2,l}+U_{j-1/2,l-1}}{h^2}.
\end{equation}

Similarly, we use the following approximations for terms in Eq. \ref{eq:1.1.9}, each of which must live at the bottom-edge center, $\boldsymbol{x}_{j,l-1/2}$:
\begin{equation}\label{eq:1.1.16}
\begin{aligned}
&\dfrac{\partial}{\partial x}(uv)\Big|_{\boldsymbol{x}_{j,l-1/2}} \approx\\&\approx \dfrac{\left(\dfrac{U_{j+1/2,l}+U_{j+1/2,l-1}}{2}\right)\left(\dfrac{V_{j+1,l-1/2}+V_{j,l-1/2}}{2}\right) - \left(\dfrac{U_{j-1/2,l}+U_{j-1/2,l-1}}{2}\right)\left(\dfrac{V_{j,l-1/2}+V_{j-1,l-1/2}}{2}\right)}{h},
\end{aligned}
\end{equation}

\begin{equation}\label{eq:1.1.17}
\dfrac{\partial}{\partial y}(v^2)\Big|_{\boldsymbol{x}_{j,l-1/2}} \approx \dfrac{V_{j,l+1/2}^2 - V_{j,l-3/2}^2}{2h},
\end{equation}

\begin{equation}\label{eq:1.1.18}
\dfrac{\partial p}{\partial y}\Big|_{\boldsymbol{x}_{j,l-1/2}} \approx \dfrac{P_{j,l} - P_{j,l-1}}{h} =: (D_{h_y}^cP)_{j,l-1/2},
\end{equation}

\begin{equation}\label{eq:1.1.19}
\dfrac{\partial ^2 v}{\partial x^2}\Big|_{\boldsymbol{x}_{j,l-1/2}} \approx \dfrac{V_{j+1,l-1/2} - 2V_{j,l-1/2}+V_{j-1,l-1/2}}{h^2},
\end{equation}

and

\begin{equation}\label{eq:1.1.20}
\dfrac{\partial ^2 v}{\partial y^2}\Big|_{\boldsymbol{x}_{j,l-1/2}} \approx \dfrac{V_{j,l+1/2} - 2V_{j,l-1/2}+V_{j,l-3/2}}{h^2}.
\end{equation}

Lastly, for Eq. \ref{eq:1.1.10} we have at the cell center $\boldsymbol{x}_{j,l}:$
\begin{equation}\label{eq:1.1.21}
\dfrac{\partial u}{\partial x}\Big|_{\boldsymbol{x}_{j,l}} \approx \dfrac{U_{j+1/2,l} - U_{j-1/2,l}}{h}=: (D_{h_x}^LU)_{j,l}
\end{equation}

and

\begin{equation}\label{eq:1.1.22}
\dfrac{\partial v}{\partial y}\Big|_{\boldsymbol{x}_{j,l}} \approx \dfrac{V_{j,l+1/2} - V_{j,l-1/2}}{h} =: (D_{h_y}^BV)_{j,l}.
\end{equation}
 
Note that Eqs. \ref{eq:1.1.12} and \ref{eq:1.1.16} involve approximations to the product $UV$ located at the vertical and horizontal pairs, respectively, of x's in Fig. \ref{fig:1.1.1}.  Also note the discrete spatial difference operators $D_{h_x}^c,$ $D_{h_y}^c,$ $D_{h_x}^L,$ and $D_{h_y}^B$ defined in Eqs. \ref{eq:1.1.13}, \ref{eq:1.1.18}, \ref{eq:1.1.21}, and \ref{eq:1.1.22}, respectively. 

We introduce additional spatial difference operators $\Delta_{h_x}^L,$ $\Delta_{h_y}^B,$ and $\boldsymbol{C_h}(\boldsymbol{U}),$ where
$(\Delta_{h_x}^LU)_{j-1/2,l}$ is given by the sum of the right-hand sides of Eqs. \ref{eq:1.1.14}\hyp{}\ref{eq:1.1.15}, $(\Delta_{h_y}^BV)_{j,l-1/2}$ is given by the sum of the right-hand sides of Eqs. \ref{eq:1.1.19}\hyp{}\ref{eq:1.1.20}, $(C_h^{(1)}(U))_{j-1/2,l}$ is given by the sum of the right-hand sides of Eqs. \ref{eq:1.1.11}\hyp{}\ref{eq:1.1.12}, and $(C_h^{(2)}(U))_{j,l-1/2}$ is given by the sum of the right-hand sides of Eqs. \ref{eq:1.1.16}\hyp{}\ref{eq:1.1.17}.    

Putting this all together, we define the discrete gradient operator $\boldsymbol{D_h^c}:= (D_{h_x}^c, D_{h_y}^c)^T,$
where $\boldsymbol{D_h^c}P = (D_{h_x}^cP, D_{h_y}^cP)^T;$ we define the discrete divergence operator 
$\boldsymbol{D_h^{L,B}} := (D_{h_x}^L, D_{h_y}^B)^T,$ where $\boldsymbol{D_h^{L,B}}\cdot \boldsymbol{U} = D_{h_x}^L U + D_{h_y}^B V;$ we define the discrete vector Laplacian $\boldsymbol{\Delta_h^{L,B}} = (\Delta_{h_x}^L, \Delta_{h_y}^B)^T,$
where $\boldsymbol{\Delta_h^{L,B}}\boldsymbol{U} = (\Delta_{h_x}^L U, \Delta_{h_y}^B V)^T;$
and we define the discrete conservative\hyp{}form, advective\hyp{}term operator $\boldsymbol{C_h}(\cdot) = (C_h^{(1)}(\cdot), C_h^{(2)}(\cdot))^T,$
where $\boldsymbol{C_h}(\boldsymbol{U}) = (C_h^{(1)}(U), C_h^{(2)}(V))^T.$  We also define $\Delta_h^c = \boldsymbol{D_h^{L,B}}\cdot \boldsymbol{D_h^c}$ which yields the 5-point Laplacian operator which maps discrete scalar functions (i.e., cell-centered quantities) to discrete scalar functions.  For example, $(\Delta_h^cP)_{j,l} = \dfrac{p_{j+1,l}-2p_{j,l}+p_{j-1,l}}{h^2} + \dfrac{p_{j,l+1}-2p_{j,l}+p_{j,l-1}}{h^2}.$

We have defined operators $\boldsymbol{D_h^c},$ $\boldsymbol{D_h^{L,B}},$ $\boldsymbol{\Delta_h^{L,B}},$ $\boldsymbol{C_h},$ and $\Delta_h^c.$  We streamline notation and use $\boldsymbol{D_h},$ $\boldsymbol{\Delta_h},$ $\boldsymbol{C_h},$ and $\Delta_h,$ thus omitting the $c$ and $L,B$ superscripts.  It will be clear from context precisely which operator is being used. 
\newpage

Spatial discretization yields these semi\hyp{}discrete approximations to Eqs. \ref{eq:1.1.1}\hyp{}\ref{eq:1.1.6}:
%2.5.2
\begin{equation}\label{eq:1.1.23}
\text{Re}\left(\bvU_{j,l}'(t)+ \boldsymbol{C_h}(\bvU_{j,l}(t))\right) =-\boldsymbol{D_h}P_{j,l}(t) + \boldsymbol{\Delta_h} \bvU_{j,l}(t) + \bvf_{j,l}(t) - \kappa\boldsymbol{\sigma}_{j,l}(t)\bvU_{j,l}(t),
\end{equation}
%\bvU_{j,l}'(t)+ \text{Re}\boldsymbol{C_h}(\bvU_{j,l}(t)) = \\
%=-\boldsymbol{D_h}P_{j,l}(t) + \boldsymbol{\Delta_h} \bvU_{j,l}(t) + \bvf_{j,l}(t) - \kappa\boldsymbol{\sigma}_{j,l}\bvU_{j,l}(t)
%\end{equation}
\begin{equation}\label{eq:1.1.24}
\boldsymbol{D_h}\cdot \bvU_{j,l}(t) = \sum_{i=1}^2 Q_i(t)\Psi_{i_{j,l}},
\end{equation}

\begin{equation}\label{eq:1.1.25}
Q_i(t) = \dfrac{1}{R_i}\left\lbrace P_i^{\text{ext}}(t)-\sum_{j,l}P_{j,l}(t)\Psi_{i_{j,l}}h^2 \right\rbrace,
\end{equation}

\begin{equation}\label{eq:1.1.26}
f_{j-1/2,l}(t) = \sum_k F_k(t)\delta_h(\bvx_{j-1/2,l}-\bvX_k(t))\Delta s,
\end{equation}

\begin{equation}\label{eq:1.1.27}
g_{j,l-1/2}(t) = \sum_k G_k(t)\delta_h(\bvx_{j,l-1/2}-\bvX_k(t))\Delta s,
\end{equation}

\begin{equation}\label{eq:1.1.28}
X_k'(t) = \sum_{j,l}U_{j-1/2,l}(t)\delta_h(\bvx_{j-1/2,l}-\bvX_k(t))h^2,
\end{equation}

\begin{equation}\label{eq:1.1.29}
Y_k'(t) = \sum_{j,l}V_{j,l-1/2}(t)\delta_h(\bvx_{j,l-1/2}-\bvX_k(t))h^2,
\end{equation}
\begin{equation}\label{eq:1.1.30}
\bvU_{j,l}(0) = 0.
\end{equation}
These equations hold for all $j = 1,\ldots, n_x;$ $l = 1,\ldots,n_y;$ $i = 1,2;$ and the boundary conditions are periodic.  The prime superscript indicates a derivative with respect to time.  The semi\hyp{}discrete versions of Eqs. \ref{eq:1.1.4} and \ref{eq:1.1.5} have been written in component form.  In Eqs. \ref{eq:1.1.26} and \ref{eq:1.1.28}, each $\delta_h$ is centered at an IB point and is evaluated at cell left\hyp{}edge centers; in Eqs. \ref{eq:1.1.27} and \ref{eq:1.1.29}, $\delta_h$ is evaluated at cell bottom edge\hyp{}centers.  In Eq. \ref{eq:1.1.23}, the corresponding components of $\boldsymbol{\sigma}_{j,l}$ and $\bvU_{j,l}$ are multiplied.  

\subsection{Temporal discretization}
\label{sec:1.2}
We numerically integrate Eqs. \ref{eq:1.1.23}\hyp{}\ref{eq:1.1.24} using a second\hyp{}order\hyp{}accurate Runge\hyp{}Kutta method based on the midpoint rule as in \citet{Peskin2002}.  For current purposes, we assume body forces are known.  In time\hyp{}stepping from level $n$ to level $n+1,$ there is first a preliminary substep from time\hyp{}level $n$ to time\hyp{}level $n + 1/2$ followed by a main step, which utilizes results of the preliminary substep, from time\hyp{}level $n$ to time\hyp{}level $n+1.$

The preliminary substep involves the fully discretized system (omitting $j,l$ subscripts):
\begin{equation}\label{eq:2.5.10}
\text{Re}\left(\dfrac{\bvU^{n+1/2} - \bvU^n}{\Delta t/2} + {\boldsymbol{C_h}}(\bvU^n)\right) = -\boldsymbol{D_h}\tilde{P}^{n+1/2} + \boldsymbol{\Delta_h} \bvU^{n+1/2} + \bvf^{n+1/2} - \kappa\boldsymbol{\sigma}^n\bvU^n,
\end{equation}
\begin{equation}\label{eq:2.5.11}
\boldsymbol{D_h}\cdot \bvU^{n+1/2} = \sum_{i=1}^2 Q_i^{n+1/2}\Psi_i,
\end{equation} and
\begin{equation}\label{eq:2.5.12}
Q_i^{n+1/2} = \dfrac{1}{R_i}\left\lbrace P_i^{\text{ext},n+1/2}-\sum\tilde{P}^{n+1/2}\Psi_ih^2 \right\rbrace,
\end{equation}
where $\tilde{P}^{n+1/2}$ is an approximation to the pressure at the half\hyp{}time level.
We rewrite Eq. \ref{eq:2.5.10} with only unknowns on the left\hyp{}hand side and condense the known quantities on the right\hyp{}hand side with new notation:
\begin{equation}\label{eq:2.5.13}
\left( \text{Re}I - \dfrac{\Delta t}{2}\boldsymbol{\Delta_h} \right)\bvU^{n+1/2}+\dfrac{\Delta t}{2}\boldsymbol{D_h}\tilde{P}^{n+1/2}=\text{Re}\bvU^n - \dfrac{\Delta t}{2}\text{Re}\boldsymbol{C_h}(\bvU^n) + \dfrac{\Delta t}{2}\bvf^{n+1/2} - \dfrac{\Delta t}{2}\kappa\boldsymbol{\sigma}^n\bvU^n=: \boldsymbol{R}^{n,n+1/2}_{pc}.
\end{equation}  
We next take the discrete divergence\footnote{The discrete spatial difference operators commute due to the periodic boundary conditions, i.e., $\boldsymbol{D_h}\cdot \boldsymbol{\Delta_h} = \Delta_h\boldsymbol{D_h}\cdot$ and $\boldsymbol{D_h}\cdot \boldsymbol{D_h} = \Delta_h.$} of Eq. \ref{eq:2.5.13} to obtain the following pressure\hyp{}Poisson equation:
\begin{equation}\label{eq:2.5.14}
\dfrac{\Delta t}{2}\Delta_h \tilde{P}^{n+1/2} = \boldsymbol{D_h} \cdot \boldsymbol{R}^{n,n+1/2}_{pc} - \sum_{i = 1}^2 Q_i^{n+1/2}\left(\text{Re}I - \dfrac{\Delta t}{2}\Delta_h\right)\Psi_i.
\end{equation}

Similar to what is done in \citet{Peskin1977} and \citet{ARTHURS1998402}, we consider three auxiliary problems:
\begin{equation}\label{eq:2.5.15}
\dfrac{\Delta t}{2}\Delta_h \phi_i = \left(\text{Re}I-\dfrac{\Delta t}{2}\Delta_h \right)\Psi_i \hspace{10pt} i = 1,2,
\end{equation}and
\begin{equation}\label{eq:2.5.16}
\dfrac{\Delta t}{2}\Delta_h\phi_0^{n+1/2} = \boldsymbol{D_h}\cdot \boldsymbol{R}^{n,n+1/2}_{pc}. 
\end{equation}Since the $\Psi_i$'s do not change with time, the $\phi_i$'s bear no time indices.  We use MATLAB's built-in, direct solver on Eqs. \ref{eq:2.5.15}\hyp{}\ref{eq:2.5.16}.  The problems are singular due to the periodic boundary conditions.  The discrete compatibility condition is that the right\hyp{}hand sides of Eqs. \ref{eq:2.5.15}\hyp{}\ref{eq:2.5.16} sum to 0 on the grid; this is satisfied due to the definition of $\Psi_i$ and since we have taken the discrete divergence of a periodic vector field.  Equipped with solutions to Eqs. \ref{eq:2.5.15}\hyp{}\ref{eq:2.5.16}, we define (with $Q_i^{n+1/2}$ to\hyp{}be\hyp{}determined):
\begin{equation}\label{eq:2.5.17}
\tilde{P_0}^{n+1/2} = \phi_0^{n+1/2} - \sum_{i=1}^2 Q_i^{n+1/2}\phi_i.
\end{equation}
This $\tilde{P_0}^{n+1/2}$ satisfies Eq. \ref{eq:2.5.14} for any pair $Q_i^{n+1/2}.$  Moreover, it is only defined up to an additive constant.  We define
\begin{equation}\label{eq:2.5.17a}
\tilde{P}^{n+1/2} = \tilde{P_0}^{n+1/2}- (\Psi^c,\tilde{P_0}^{n+1/2})_h,
\end{equation}
where\[(\alpha,\beta)_h = \sum_{j,l} \alpha(\bvx_{j,l})\beta(\bvx_{j,l})h^2\] denotes the discrete inner product.  Now, $\tilde{P}^{n+1/2}$ has average value 0 over the compensatory source, and we consider this 0 to be equal to the experimental reference pressure with respect to which $P_i^{\text{ext}}$ is measured (gauge pressure 0).  Substituting Eq. \ref{eq:2.5.17a} into Eq. \ref{eq:2.5.12} yields the matrix equation:
\begin{equation}\label{eq:2.5.18}
\left(\begin{array}{cc}
R_1 - (\phi_1,\Psi_1)_h  & -(\phi_2,\Psi_1)_h  \\
-(\phi_1,\Psi_2)_h & R_2 - (\phi_2,\Psi_2)_h  \\
\end{array}\right)\left(\begin{array}{c} Q_1^{n+1/2} \\ Q_2^{n+1/2} \end{array}\right) = \left(\begin{array}{c} P_1^{\text{ext},n+1/2}-(\phi_0^{n+1/2},\Psi_1)_h\\ P_2^{\text{ext},n+1/2}-(\phi_0^{n+1/2},\Psi_2)_h\end{array}\right),
\end{equation}  Solving this system yields $Q_i^{n+1/2}$ for $i = 1,2.$  Now $\tilde{P}^{n+1/2}$ is fully determined in Eq. \ref{eq:2.5.17a}.  We then return to Eq. \ref{eq:2.5.13} and obtain the Helmholtz problem:
\begin{equation}\label{eq:2.5.19}
\left( \text{Re}I - \dfrac{\Delta t}{2}\boldsymbol{\Delta_h} \right)\bvU^{n+1/2}= \boldsymbol{R}^{n,n+1/2}_{pc} - \dfrac{\Delta t}{2}\boldsymbol{D_h}\tilde{P}^{n+1/2}.
\end{equation}
This problem is nonsingular, solved directly in MATLAB, and decouples for the $U$ and $V$ components.  The main step involves the fully discretized system \footnote{We acknowledge the difference in time\hyp{}level for the pressure and flow\hyp{}rate in Eq. \ref{eq:2.5.22} but postpone consideration for future work.  We have linearly extrapolated $\boldsymbol{\sigma}^{n-1}$ and $\boldsymbol{\sigma}^{n}$ to approximate $\boldsymbol{\sigma}^{n+1/2}.$}:
\begin{equation}\label{eq:2.5.20}
\text{Re}\left(\dfrac{\bvU^{n+1} - \bvU^n}{\Delta t} + \boldsymbol{C_h}(\bvU^{n+1/2})\right) =-\boldsymbol{D_h}P^{n+1/2} + \boldsymbol{\Delta_h}\left(\dfrac{\bvU^n + \bvU^{n+1}}{2}\right)+ \bvf^{n+1/2} - \kappa\left(\dfrac{3\boldsymbol{\sigma}^n-\boldsymbol{\sigma}^{n-1}}{2}\right)\bvU^{n+1/2},
\end{equation}
\begin{equation}\label{eq:2.5.21}
\boldsymbol{D_h}\cdot \bvU^{n+1} = \sum_{i=1}^2 Q_i^{n+1}\Psi_i,
\end{equation} and
\begin{equation}\label{eq:2.5.22}
Q_i^{n+1} = \dfrac{1}{R_i}\left\lbrace P_i^{\text{ext},n+1}-\sum P^{n+1/2}\Psi_ih^2 \right\rbrace,
\end{equation}
where $P^{n+1/2}$ is another approximation to the pressure at the half\hyp{}time level.  This system is solved in the same way as that in the preliminary substep.  

\subsection{Numerical-implementation scheme}
\label{sec:1.3} 
We describe how to proceed from time\hyp{}level $n$ to time\hyp{}level $n+1.$  At time\hyp{}level $n,$ we know $\bvU^n$ and $\bvX^n.$  
\begin{enumerate}
\item Move the IB points to the half\hyp{}time level via:
\begin{equation}\label{eq:2.5.30}
\dfrac{X_k^{n+1/2}-X_k^n}{(\frac{\Delta t}{2})} = \sum_{j,l}U_{j-1/2,l}^n\delta_h(\bvx_{j-1/2,l}-\bvX_k^n)h^2,
\end{equation}
\begin{equation}\label{eq:2.5.31}
\dfrac{Y_k^{n+1/2}-Y_k^n}{(\frac{\Delta t}{2})} = \sum_{j,l}V_{j,l-1/2}^n\delta_h(\bvx_{j,l-1/2}-\bvX_k^n)h^2.
\end{equation}
\item Use the new boundary configuration $\bvX^{n+1/2}$ to compute the Lagrangian force density $\bvF^{n+1/2}$. 
\item Spread the Lagrangian forces to the Eulerian grid using the half time\hyp{}level $\delta_h$ values
\begin{equation}\label{eq:2.5.32}
f^{n+1/2}_{j-1/2,l} = \sum_k F_k^{n+1/2}\delta_h(\bvx_{j-1/2,l}-\bvX_k^{n+1/2})\Delta s,
\end{equation}
\begin{equation}\label{eq:2.5.33}
g^{n+1/2}_{j,l-1/2} = \sum_k G_k^{n+1/2}\delta_h(\bvx_{j,l-1/2}-\bvX_k^{n+1/2})\Delta s.
\end{equation}
\item Use $\bvf^{n+1/2}=(f^{n+1/2},g^{n+1/2})^T$ and the numerical integration described in Sect. \ref{sec:1.2} to obtain $\bvU^{n+1/2}$ and $\bvU^{n+1}.$  
\item Find the IB\hyp{}point locations at the full\hyp{}time level using the preliminary substep velocity and IB positions at the $n$ and $n+1/2$ time\hyp{}levels:
\begin{equation}\label{eq:2.5.34}
\dfrac{X_k^{n+1}-X_k^n}{\Delta t} = \sum_{j,l}U_{j-1/2,l}^{n+1/2}\delta_h(\bvx_{j-1/2,l}-\bvX_k^{n+1/2})h^2.
\end{equation}
\begin{equation}\label{eq:2.5.35}
\dfrac{Y_k^{n+1}-Y_k^n}{\Delta t} = \sum_{j,l}V_{j,l-1/2}^{n+1/2}\delta_h(\bvx_{j,l-1/2}-\bvX_k^{n+1/2})h^2.
\end{equation}
\end{enumerate} 
Thus, we obtain $\bvU^{n+1}$ and $\bvX^{n+1}.$  

We used an advective CFL condition $\dfrac{Ch}{\Delta t}\geq \text{max}_{j,l} ||\bvU_{j,l}^n||_2,$ to guide the size of the initial time step based on velocity estimates.  We used the Euclidean 2\hyp{}norm for the velocity at each grid cell and a CFL constant, $C = 0.1.$  The inequality was enforced at every time step.  To reduce machine error that otherwise contributed to a loss of horizontal (vessel\hyp{}midline) symmetry (also observed in \citet{ARTHURS1998402}), we symmetrized the vessel\hyp{}wall and valve IB points at half\hyp{} and full\hyp{}time levels about the midline by averaging corresponding diametrically opposed IB\hyp{}point $x$\hyp{}locations and reassigning the average as the new $x$\hyp{}locations for the corresponding top and bottom points.  Similarly, we averaged corresponding $y$\hyp{}coordinate distances to the midline and reset top and bottom $y$\hyp{}coordinates so each their distance to the midline was equal to the average.       

\subsection{\textcolor{myred}{Refinement study}}
\label{sec:1.4} 
\textcolor{myred}{Spatial and temporal refinement studies were performed on a simplified version of the model to assess convergence of the velocity field components and IB\hyp{}point locations.  The results shown in Tables \ref{tab:1.4.1}-\ref{tab:1.4.2} indicate approximate first-order convergence for both velocity components and the IB structure in space and time (in the grid 2-norm \citep{leveque2007finite}).  Grid 2-norms were computed as follows:}

\begin{equation}\label{eq:1.4.1}
||U^n||_2 = \left(\sum_{j,l}(U_{j-1/2,l}^n)^2 h^2\right)^{1/2},
\end{equation}

\begin{equation}\label{eq:1.4.2}
||V^n||_2 = \left(\sum_{j,l}(V_{j,l-1/2}^n)^2 h^2\right)^{1/2},
\end{equation}

\begin{equation}\label{eq:1.4.3}
||\bvX^n||_2 = \left(\sum_k((X_k^n)^2 + (Y_k^n)^2)\Delta s_v\right)^{1/2}.
\end{equation}

\textcolor{myred}{The order of convergence was estimated using the base-2 logarithm of the ratio of estimated errors in solution data from consecutively refined grids or from consecutively refined time-step sizes (as in \citet{leveque2007finite}).  Grid mesh widths were halved in spatial refinement, and time-step sizes were halved (using a fixed Eulerian grid) in temporal refinement.}
   
\begin{table}[H]
\caption{Spatial and temporal convergence of the velocity field components.  In spatial refinement (top half of table), the errors are estimated on each grid using results from the next\hyp{}finer grid as the reference solution as in \citet{leveque2007finite} and bicubic interpolation to transfer coarser-grid solution data to the next-finer grid.  In temporal refinement (bottom half of table), the estimated errors are obtained by subtracting the finer-timestep data from the coarser-timestep data (the grid is fixed as the medium grid).  These results were obtained from a simultaneous, 1-lymphangion pump operating at an AAPD of 0.051 cmH$_2$O simulated to $T = 3.8$ s.  Discrete $L^2$ norms of the horizontal velocity component were equal to 0.272 and 0.262 on the medium and fine grids, respectively; the same values for the vertical velocity component were 0.108 and 0.0845; all in mm/s.  In the table, $\Delta t$ is measured in s, the discrete $L^2$ estimated error is measured in mm/s, and the order is the estimated order of convergence \citep{leveque2007finite}.  We use notation e.g., e-5 to denote $\times 10^{-5}$.  Dashes such as ----- are included to indicate that data from the corresponding grid size listed for that table row were transferred to a different grid and used to estimate errors on that different grid.  In the case of temporal refinement, dashes have a similar meaning where data for the grid listed in that row were used but compared to other data generated using a different timestep with errors listed in the row for that other timestep} 
\label{tab:1.4.1}       
\begin{tabular}{lllllll}
\hline\noalign{\smallskip}
Grid size & $\Delta t$ & $L^2$ est. error in $U,V$ & order $U,V$\\
\noalign{\smallskip}\hline\noalign{\smallskip}
$96\times 32$ & 7.63 e-6  & ----- & -----\\
$192\times 64$  &  7.63 e-6  & 0.193, 0.126 & -----\\
$384\times 128$ & 7.63 e-6 & 0.0824, 0.0540 & 1.23, 1.22\\
\noalign{\smallskip}\hline
\noalign{\smallskip}

$192\times 64$ &   3.05 e-5 & 1.03 e-5, 7.03 e-6  & -----\\
$192\times 64$  &  1.53 e-5 & 5.48 e-6, 2.18 e-6 & 0.91, 1.69\\
$192\times 64$ &  7.63 e-6 & ----- & -----\\
\noalign{\smallskip}\hline
\end{tabular}
\end{table}

\begin{table}[H]
\caption{Spatial and temporal convergence of the IB-point locations.  Spatial errors (top nine rows of table) were estimated on each grid by subtracting locations of the corresponding subset of IB points from the next\hyp{}finer grid from locations of the IB points on the coarser grid.  Temporal errors (bottom nine rows of table) were estimated by subtracting the finer-timestep data from the coarser-timestep data (the grid is fixed as the medium grid).  These results were obtained from a simultaneous, 1-lymphangion pump operating at an AAPD of 0.051 cmH$_2$O simulated to $T = 3.8$ s.  Discrete $L^2$ norms of capsule IB-points were 4.47 on both the coarse and medium grids, respectively.  Discrete $L^2$ norms of the valve-IB points were 0.448 and 0.438 for the top leaflet of valve 1 on coarse and medium grids, respectively and 0.870 and 0.856 for the top leaflet of valve 2 on the coarse and medium grids, respectively; all discrete $L^2$ norm values in mm.  In the table, $N$ is the number of (capsule or valve top leaflet) IB points; $\Delta t$ is measured in s, type refers to the type of IB point: c denotes capsule, v1 denotes valve 1 leaflet, and v2 denotes valve 2 leaflet; the discrete $L^2$ estimated error is measured in mm; and the order is the estimated order of convergence \citep{leveque2007finite}.  Due to symmetry, only top-leaflet IB-point locations were considered in this refinement study.  We use notation e.g., e-5 to denote $\times 10^{-5}$.  Dashes have the same meaning as in Table \ref{tab:1.4.1}}
\label{tab:1.4.2}       
\begin{tabular}{llllll}
\hline\noalign{\smallskip}
Grid size & $N$ & $\Delta t$ & type & est. $L^2$ error & order\\
\noalign{\smallskip}\hline\noalign{\smallskip}
$96\times 32$ & 378 & 7.63 e-6  & c & 3.15 e-3 & -----\\
$192\times 64$ & 756 &  7.63 e-6  & c & 1.20 e-4 & 1.39\\
$384\times 128$ & 1512&  7.63 e-6& c & ----- & -----\\
\noalign{\smallskip}\hline
\noalign{\smallskip}

$96\times 32$ & 13 & 7.63 e-6 & v1 &  1.54 e-3 & -----\\
$192\times 64$ & 25 &  7.63 e-6 & v1 & 4.16 e-4 & 1.89\\
$384\times 128$ & 49 &  7.63 e-6& v1 & ----- & -----\\
\noalign{\smallskip}\hline
\noalign{\smallskip}

$96\times 32$ & 13 & 7.63 e-6 & v2 & 5.38 e-3 & -----\\
$192\times 64$ & 25 &  7.63 e-6 & v2 & 9.03 e-4 & 2.57\\
$384\times 128$ & 49 &  7.63 e-6& v2 & ----- & -----\\
\noalign{\smallskip}\hline
\noalign{\smallskip}

$192\times 64$ & 756 &  3.05 e-5 & c &  8.09 e-7 & -----\\
$192\times 64$ & 756 &  1.53 e-5 & c & 4.13 e-7 & 0.97\\
$192\times 64$ & 756 &  7.63 e-6 & c & ----- & ----- \\
\noalign{\smallskip}\hline
\noalign{\smallskip}

$192\times 64$ & 25 &  3.05 e-5 & v1 &  1.88 e-8 & -----\\
$192\times 64$ & 25 &  1.53 e-5 & v1 & 6.58 e-9 & 1.52\\
$192\times 64$ & 25 &  7.63 e-6 & v1 & ----- & ----- \\
\noalign{\smallskip}\hline
\noalign{\smallskip}

$192\times 64$ & 25 &  3.05 e-5 & v2 &  4.31 e-7 & -----\\
$192\times 64$ & 25 &  1.53 e-5 & v2 & 1.29 e-8 & 5.06\\
$192\times 64$ & 25 &  7.63 e-6 & v2 & ----- & ----- \\
\noalign{\smallskip}\hline
\end{tabular}
\end{table}

\section{Tension and bending Lagrangian force densities}
\label{sec:2}
We provide equations for the tension and bending Lagrangian force densities.  We present these equations in a dimensional setting, though they are non\hyp{}dimensionalized with scalings described in Sect. 2.4 of the main text for use in Eqs. \ref{eq:2.5.32}\hyp{}\ref{eq:2.5.33} of this supplement.  

Each valve leaflet originates at a vessel wall IB point and comprises an open curve.  We discretize the tension and bending elastic energy functionals in Eqs. 8\hyp{}9 of the main text first for the valve leaflets (thus with $k_t$ and $k_b$ replaced with $k_{t_v}$ and $k_{t_v},$ respectively).  As in \citet{PeskinNotes}, we discretize $E_T$ as 
 \begin{equation}\label{eq:2.6.3}
E_{T,D}(\bvX_1\cdots\bvX_{N_v}) := \dfrac{k_{t_v}}{2}\sum_{k = 1}^{N_v-1}\left(\dfrac{\left\Vert\bvX_{k+1}-\bvX_k\right\Vert}{\Delta s_v}-\dfrac{\left\Vert\bvX_{k+1}^0-\bvX_k^0\right\Vert}{\Delta s_v} \right)^2\Delta s_v.
\end{equation}  This discretized energy functional is differentiated with respect to each of the two components of $\boldsymbol{X}_k$ to yield Lagrangian forces at $\boldsymbol{X}_k$.  The force at the $l$\hyp{}th point is given by 
\begin{equation}\label{eq:2.6.4}
\bvF_l\Delta s_v = -\dfrac{\partial E_{T,D}}{\partial \bvX_l}= \left[(l \neq N_v)T_{l+1/2}\boldsymbol{\tau}_{l+1/2}-(l \neq 1)T_{l-1/2}\boldsymbol{\tau}_{l-1/2}\right],
\end{equation}
for $l = 1,\ldots,N_v,$\\
where $T_{l+1/2} = k_{t_v}\left(\dfrac{\left\Vert\bvX_{l+1}-\bvX_l\right\Vert}{\Delta s_v}-\dfrac{\left\Vert\bvX_{l+1}^0-\bvX_l^0\right\Vert}{\Delta s_v}\right),$\\
$\boldsymbol{\tau}_{l+1/2} = \dfrac{\bvX_{l+1}-\bvX_l}{\left\Vert\bvX_{l+1}-\bvX_l\right\Vert},$\\
$(l \neq N_v) = 1$ if $l \neq N_v$ and 0 if $l=N_v$,\\
 and $(l \neq 1) = 1$ if $l \neq 1$ and 0 if $l = 1$. The Lagrangian force density at the $l$\hyp{}th point is $\bvF_l.$  

For an open curve, we discretize $E_B$ as in \citet{PeskinNotes}
\begin{equation}\label{eq:2.6.6}
E_{B,D}(\bvX_1\cdots\bvX_{N_v}) := \dfrac{k_{b_v}}{2}\sum_{k = 2}^{N_v-1}|\boldsymbol{C}_k-\boldsymbol{C}_k^0|^2\Delta s_v,
\end{equation}

with 
\begin{equation}
\boldsymbol{C}_k = \dfrac{\bvX_{k+1}-2\bvX_k+\bvX_{k-1}}{(\Delta s_v)^2},\nonumber
\end{equation}

and 
\begin{equation}
\boldsymbol{C}_k^0 = \dfrac{\bvX_{k+1}^0-2\bvX_k^0+\bvX_{k-1}^0}{(\Delta s_v)^2},\nonumber
\end{equation}
for $k = 2,\ldots,N_v-1.$   
This discretized energy functional is differentiated with respect to each of the two components of $\bvX_k$ to yield Lagrangian forces.  The force at the $l$\hyp{}th point is given by 

\begin{align}\label{eq:2.6.7}
&\bvF_l\Delta s_v = -\dfrac{\partial E_{B,D}}{\partial \bvX_l}=\\
&= -k_{b_v}\left(\left\lbrace \dfrac{(l \geq 3)\boldsymbol{C}_{l-1}-(2\leq l \leq N_v-1)2\boldsymbol{C}_l+(l \leq N_v-2)\boldsymbol{C}_{l+1}}{(\Delta s_v)^2}\right\rbrace\right. \nonumber \\
&-\left.\left\lbrace \dfrac{(l \geq 3)\boldsymbol{C}_{l-1}^0-(2\leq l \leq N_v-1)2\boldsymbol{C}_l^0+(l \leq N_v-2)\boldsymbol{C}_{l+1}^0}{(\Delta s_v)^2}\right\rbrace\right)\Delta s_v,\nonumber
\end{align} using similar parenthetical logical expressions as in Eq. \eqref{eq:2.6.4} and with $\bvF_l$ the Lagrangian force density at the $l$\hyp{}th valve point.

For capsule IB points, the Lagrangian tension force density is the same as for leaflet points but without the parenthetical logical expressions $(l \neq N_v)$ and $(l \neq 1),$ it has $\Delta s_v$ replaced with $\Delta s,$ and it has tension constant $k_t.$  The bending force density at capsule IB points is the same as for leaflet points but without the parenthetical logical expressions, with bending coefficient $k_b$, and with $\Delta s_v$ replaced with $\Delta s.$       

In \textcolor{myred}{the} numerical implementation, all capsular and leaflet Lagrangian force densities are accumulated at each IB point and spread to the fluid via Eqs. \ref{eq:2.5.32}\hyp{}\ref{eq:2.5.33}.  In the case where a Lagrangian force (rather than force density is specified), it is interpreted as the product of a Lagrangian force density and the nodal increment, and it is spread to the fluid via the same equations.  

\section{Measuring flow rates, velocities, and pressures at flow meters}
\label{sec:3}

Flow meters are shown in cyan in Fig. 2 of the main text and are vertical line segments where we measure flow rates, velocities, and pressures.  In \textcolor{myred}{the} numerical implementation, at each full\hyp{}time level, a set of flow meter points is evenly distributed between the diametrically opposed IB points at each end of each flow meter with spacing approximately $h/2$.  At these ``auxiliary IB points,'' which do not affect the flow in any way, we interpolate velocity and pressure.  We can measure velocity or pressure at just the vessel midline or along all flow\hyp{}meter points, using measurements at the latter to estimate flow rates.  

We obtain the horizontal velocity component at each flow meter point at time level $n+1$ by using the right\hyp{}hand side of Eq. \ref{eq:2.5.34} but with $n+1/2$ replaced with $n+1$ and $\bvX_k$ replaced with the flow\hyp{}meter point.  If $\bvX_{\text{FM}_k}$ denotes a flow\hyp{}meter point, then the interpolated horizontal velocity component at that point at time level $n+1$, $U_{\text{FM}_k}^{n+1},$ is approximated 
\begin{equation}\label{eq:3.1}
U_{\text{FM}_k}^{n+1} \approx\sum_{j,l}U_{j-1/2,l}^{n+1}\delta_h(\bvx_{j-1/2,l}-\bvX_{\text{FM}_k}^{n+1})h^2.
\end{equation}

We approximate the flow rate using the midpoint rule to approximate the $y$\hyp{}integral of the horizontal velocity component along the flow meter and obtain
\begin{equation}\label{eq:3.2} 
\int_{y_l}^{y_u} u(x_0,y,(n+1)\Delta t)\,dy \approx \sum_k U_{\text{FM}_k}^{n+1}\Delta y_k,
\end{equation}
where $\Delta y_k \approx h/2$ is the uniform spacing of the flow\hyp{}meter points, $y_l$ and $y_u$ are lower and upper bounding $y$\hyp{}values corresponding to the bottom and top flow\hyp{}meter points, and $x_0$ is the $x$\hyp{}coordinate of the flow\hyp{}meter location.    

We similarly interpolate pressures at flow\hyp{}meter points via 
\begin{equation}\label{eq:3.3} 
P_{\text{FM}_k}^{n+1/2} \approx \left(\sum_{j,l}P_{j,l}^{n+1/2}\delta_h(\bvx_{j,l}-\bvX_{\text{FM}_k}^{n+1})h^2\right) - (P^{n+1/2},\Psi_c)_h,
\end{equation}
evaluating the discrete Delta function at cell centers and having subtracted the computational reference pressure.  
 
 \section{\textcolor{myred}{Contraction dynamics}}
\label{sec:4} 
\begin{figure}[H]
  \includegraphics[width=0.8\textwidth]{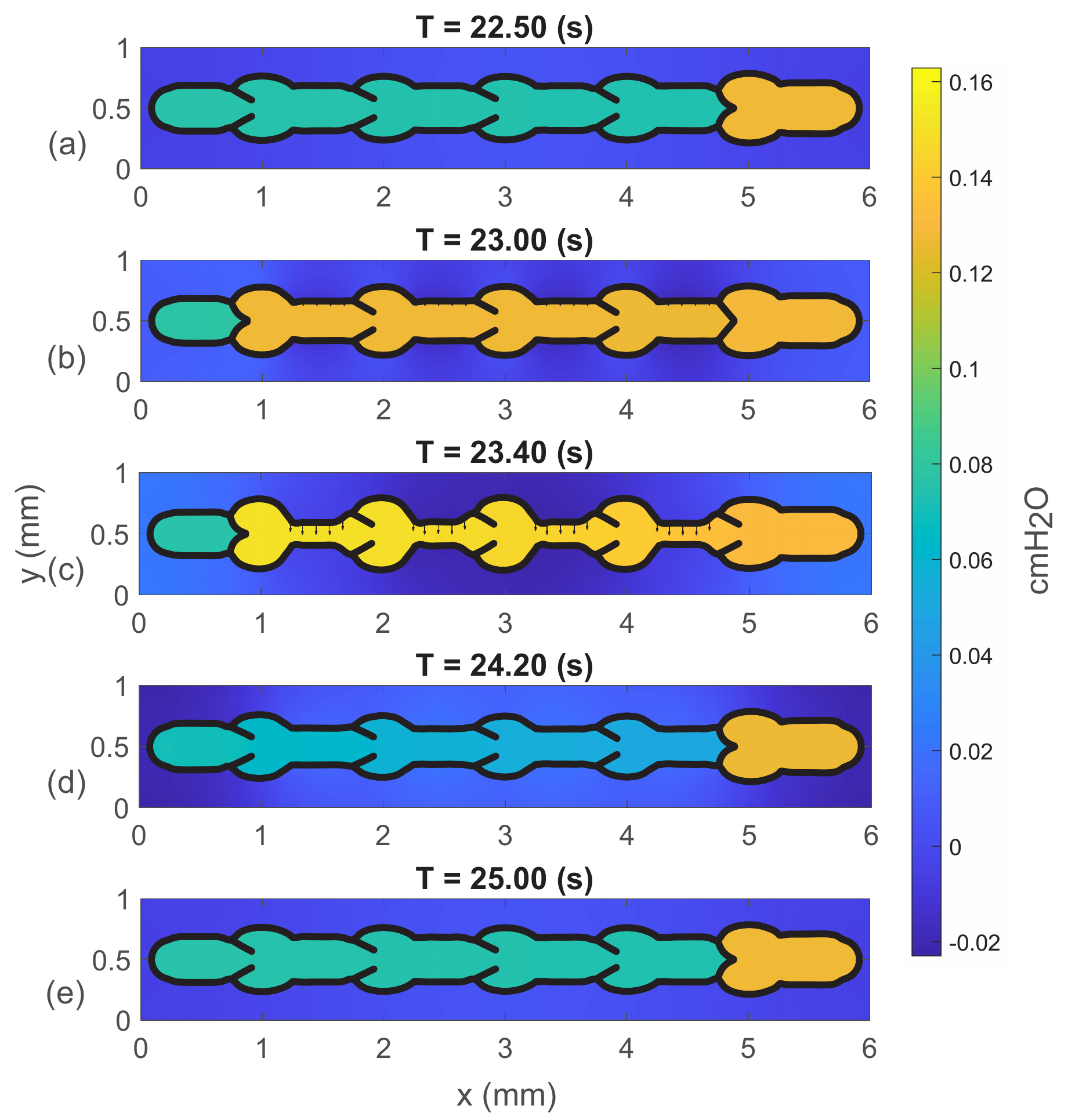}
%  \includegraphics[width=0.8\textwidth]{Fig2_SI.eps}
\caption{Contraction forces and pressure at select times during cycle 10 in a 4-lymphangion, simultaneous pump with an AAPD of 0.051 cmH$_2$O.  Four force vectors are provided along each upper contractile region.  The forces are uniform at all IB points within a contractile region contained in a single lymphangion; see Fig. 2 in the main text for contractile region locations.  The force vectors have been scaled (for all subplots) by a factor of 1/800; force magnitudes are measured in nano-Newtons.  If no force vectors appear, it is because they have small magnitude  }
\label{fig:4.1}       % Give a unique label
 \end{figure}

\begin{figure}[H]
  \includegraphics[width=0.8\textwidth]{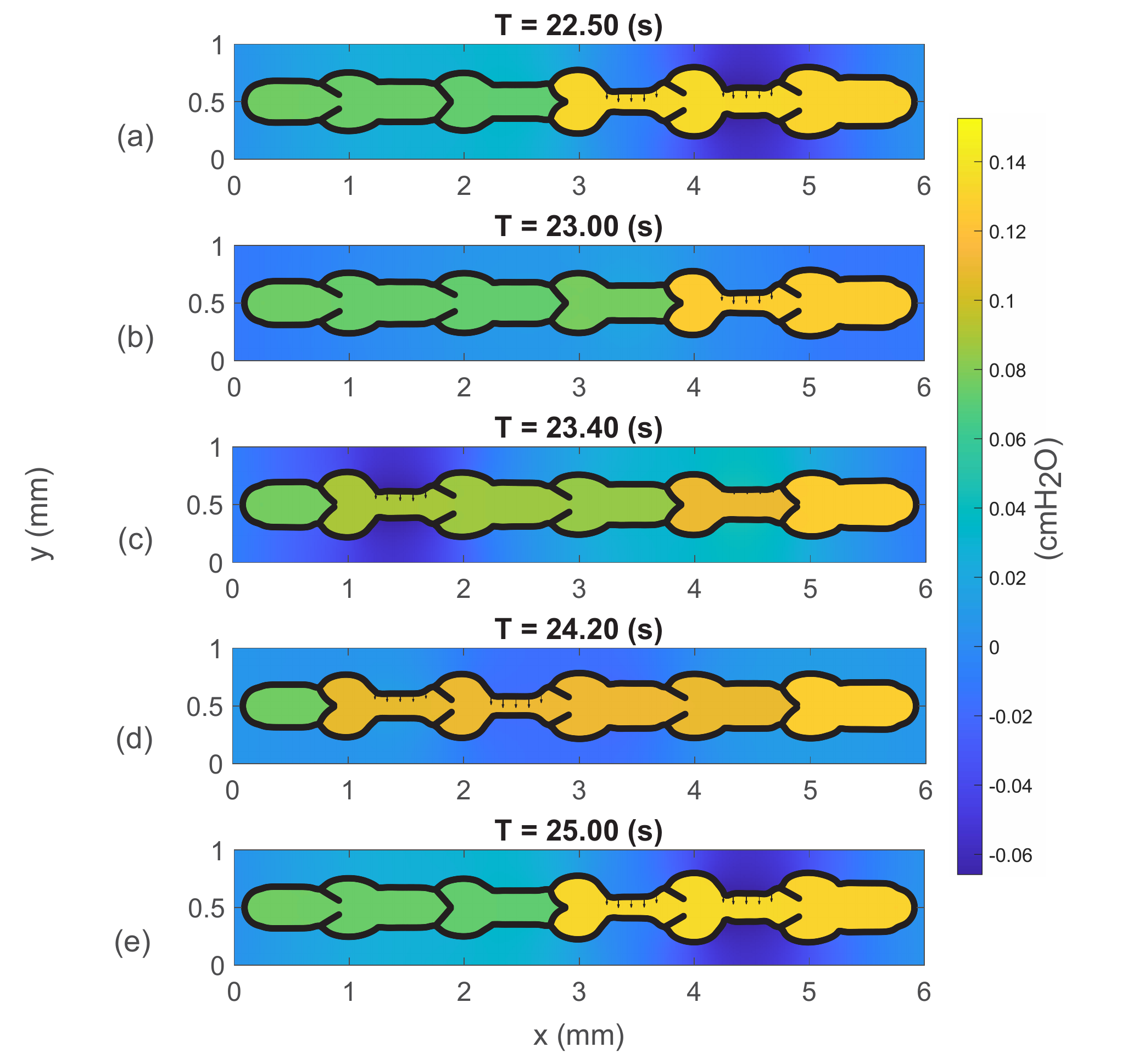}
%  \includegraphics[width=0.8\textwidth]{Fig3_SI.eps}
\caption{Contraction forces and pressure at select times during cycle 10 in a 4-lymphangion, orthograde pump with an AAPD of 0.051 cmH$_2$O.  Four force vectors are provided along each upper contractile region.  The forces are uniform at all IB points within a contractile region contained in a single lymphangion; see Fig. 2 in the main text for contractile region locations.  The force vectors have been scaled (for all subplots) by a factor of 1/800; force magnitudes are measured in nano-Newtons.  If no force vectors appear, it is because they have small magnitude  }
\label{fig:4.2}       % Give a unique label
 \end{figure}

\begin{figure}[H]
  \includegraphics[width=0.8\textwidth]{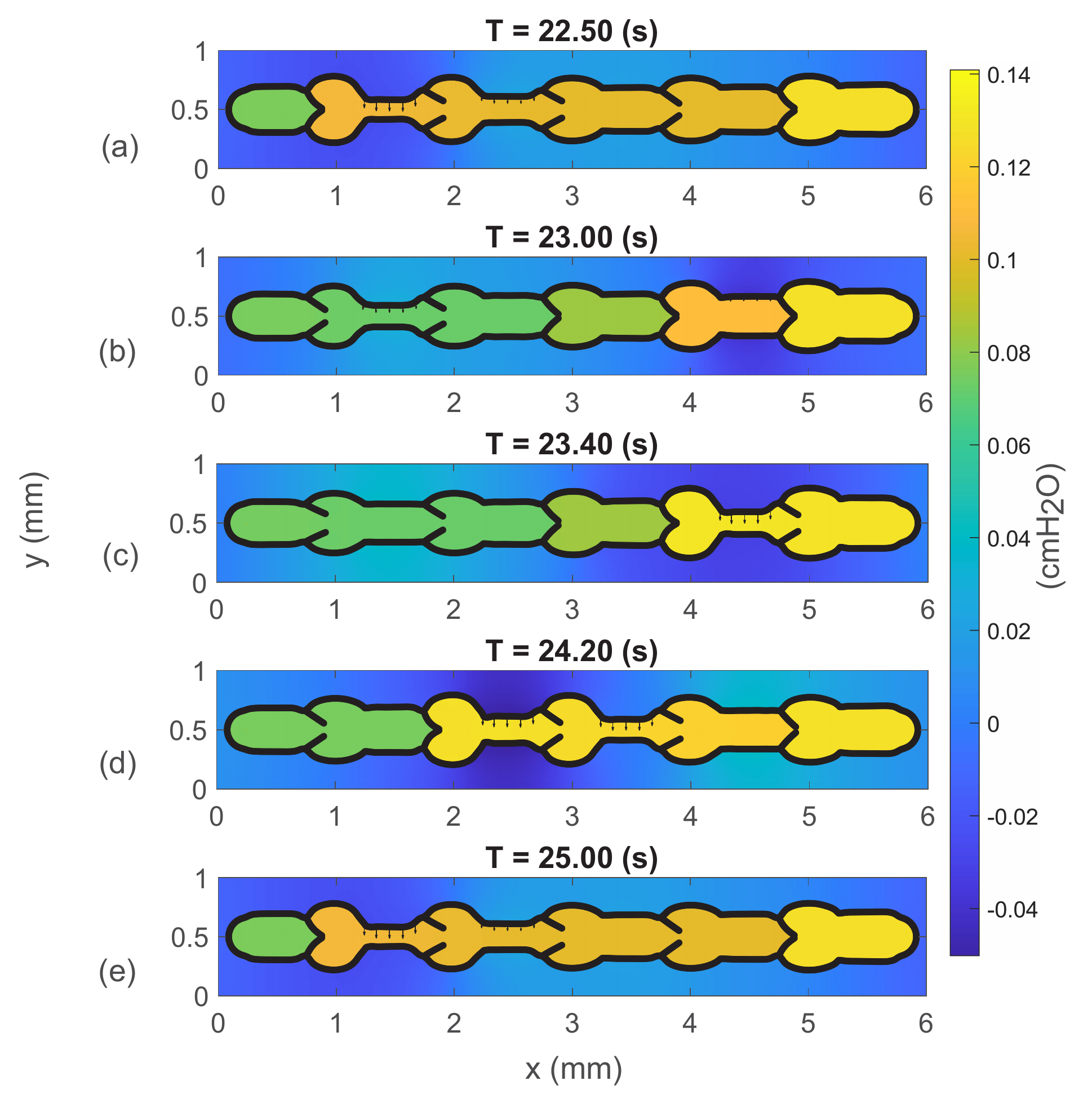}
%  \includegraphics[width=0.8\textwidth]{Fig4_SI.eps}
\caption{Contraction forces and pressure at select times during cycle 10 in a 4-lymphangion, retrograde pump with an AAPD of 0.051 cmH$_2$O.  Four force vectors are provided along each upper contractile region.  The forces are uniform at all IB points within a contractile region contained in a single lymphangion; see Fig. 2 in the main text for contractile region locations.  The force vectors have been scaled (for all subplots) by a factor of 1/800; force magnitudes are measured in nano-Newtons.  If no force vectors appear, it is because they have small magnitude  }
\label{fig:4.3}       % Give a unique label
 \end{figure}

\textcolor{myred}{Here we provide figures to illustrate the contraction forces at various times throughout a cycle in the simultaneous (Fig. \ref{fig:4.1}), orthograde (Fig. \ref{fig:4.2}), and retrograde (Fig. \ref{fig:4.3}) pumps.  The times we selected occur during cycle 10 and correspond to minimal contraction force for the simultaneous pump (T = 22.5 s, 25 s), maximal contraction force for the simultaneous pump ($T \approx 23.4$ s), and times that are half way between the times at which minimal and maximal simultaneous contraction forces occur (T = 23 s, 24.2 s).  In panels (a) and (b) of Fig. \ref{fig:4.2}, contraction is prominent in lymphangions 3 and 4 due to the previous contraction cycle.  In the same figure, lymphangion 1 is relaxed in panel (a) and beginning contraction in panel (b) with a small-magnitude contraction force.  In panel (c), lymphangion 1 is near end-systole.  Panels (c)-(e) particularly illustrate the contraction forces growing in magnitude moving down the chain as the orthograde contraction propagates to the right.  The observations in Fig. \ref{fig:4.3} are similar but with contraction propagating upstream (to the left).}          

 \section{\textcolor{myred}{Pump-function data for cycle-mean forward and cycle-mean backward flow rates}}
\label{sec:5} 

\textcolor{myred}{We decomposed the flow-rate data into positive and negative parts and averaged each over a contraction cycle and among the
interior flow meters to obtain cycle-mean forward and backward flow rates, CMFFRs and CMBFRs, respectively. We generated pump-function data plots analogous to those in
panels (c) and (d) of Fig. 10 of the main text but with CMFFRs and CMBFRs replacing CMFRs.}
\begin{figure}[H]
 % \includegraphics[width=0.5\textwidth]{staggered_grid.pdf}
 %\includegraphics[width=0.5\textwidth]{Figures/Fig3.pdf}
   \includegraphics[width=1\textwidth]{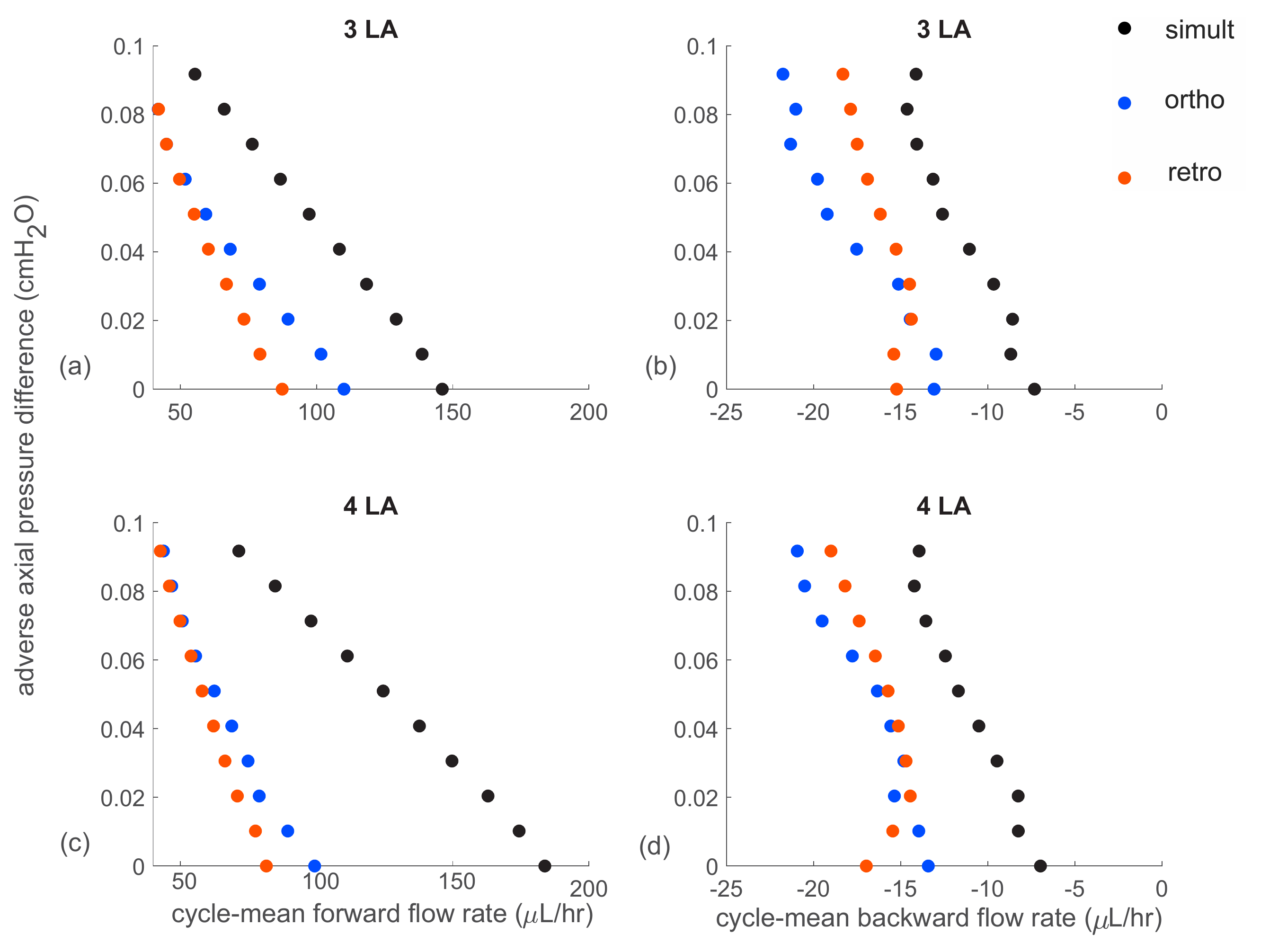}
%  \includegraphics[width=1\textwidth]{Fig5_SI.eps}
\caption{Cycle-mean forward flow rate (CMFFR) and cycle-mean backward flow rate (CMBFR) plots for 3- and 4-lymphangion simultaneous, orthograde, and retrograde pumps. Panels (a) and (b) include CMFFR and CMBFR data for the 3-lymphangion pumps, respectively; panels (c) and (d) are for the 4-lymphangion pumps.  Horizontal axis labels apply to both subplots in that column.  The vertical axis label applies to all subplots}
\label{fig:5.1}       % Give a unique label
 \end{figure}

\bibliographystyle{spbasic_cp2}      % basic style, author-year citations
\bibliography{library}   % name your BibTeX data base

%% Non-BibTeX users please use
%\begin{thebibliography}{}
%%
%% and use \bibitem to create references. Consult the Instructions
%% for authors for reference list style.
%\end{thebibliography}